\documentclass[
superscriptaddress,amsmath,amssymb,aps,longbibliography,twocolumn,showpacs,a4paper, notitlepage]{revtex4-2}

\usepackage{graphicx}
\usepackage{mathtools}
\usepackage{bm}
\usepackage{braket}
\usepackage{xcolor}

\usepackage{makecell}
\usepackage{graphicx}  
\usepackage{amsmath}   
\usepackage{hyperref}  
\usepackage{longtable} 

\usepackage[normalem]{ulem} 
\usepackage{xcolor}         

\usepackage{graphicx}    
\usepackage{amsmath}     
\usepackage{hyperref}    
\usepackage{array} 
\usepackage[longtable]{multirow}
\usepackage{ragged2e}
\usepackage{longtable}
\usepackage{mathtools}

\begin{document}

\title{Evaluating SCAN and r$^2$SCAN meta-GGA functionals for
predicting transition temperatures in antiferromagnetic materials}

\author{Nafise Rezaei}
\affiliation{Skolkovo Institute of Science and Technology, 121205, Bolshoy Boulevard 30, bld. 1, Moscow, Russia.}
\author{Mojtaba Alaei}
\affiliation{Skolkovo Institute of Science and Technology, 121205, Bolshoy Boulevard 30, bld. 1, Moscow, Russia.}
\affiliation{Department of Physics, Isfahan University of Technology, Isfahan 84156-83111, Iran.}
\author{Artem R. Oganov}
\affiliation{Skolkovo Institute of Science and Technology, 121205, Bolshoy Boulevard 30, bld. 1, Moscow, Russia.}

\date{\today}

\begin{abstract}
Recent advancements in exchange-correlation functionals within density functional theory  highlight the need for rigorous validation across diverse types of materials properties. In this study, we assess the performance of the newly developed meta-GGA r$^2$SCAN and its predecessor, SCAN, in predicting the Néel transition temperature of antiferromagnetic materials. Our analysis includes 48 magnetic materials, spanning both simple and complex systems. Using DFT, we compute the energies of various magnetic configurations and extract exchange interaction parameters through a least-squares fitting approach. These parameters are then used in classical Monte Carlo simulations to estimate the transition temperatures. 
Our results demonstrate that both SCAN and r$^2$SCAN greatly outperform standard GGA and GGA+$U$ methods, yielding predictions that closely align with experimental values. The Pearson correlation coefficients for SCAN and r$^2$SCAN are 0.97 and 0.98, respectively, when compared to experimental transition temperatures. Additionally, we calculate the energy differences between antiferromagnetic and ferromagnetic configurations to assess the performance of the hybrid HSE06 functional. We found that the HSE06 functional underestimates transition temperatures compared to the meta-GGA functionals and experimental values.  

\end{abstract}


\maketitle
\section{\label{sec:int1}INTRODUCTION}
The exchange-correlation (xc) functional is a crucial component of the Kohn-Sham density functional theory (DFT), as it encapsulates the complex many-body effects of electron-electron interactions. The accuracy of DFT calculations heavily depends on the choice of xc functional, especially for magnetic calculations. 
Magnetism serves as an ideal ground for testing and improving exchange-correlation (xc) functionals, as it directly arises from the presence of xc energy in the many-body Hamiltonian~\cite{DFT_MAG1,Yang2008}.

The local density approximation (LDA)~\cite{LDA} and the generalized gradient approximation (GGA)~\cite{GGA, PBE}, as commonly used xc functionals, both face challenges in describing systems with strong on-site Coulomb interactions, such as those involving $d$ and $f$ orbitals. These limitations, stemming from self-interaction errors~\cite{Blaha-meta}, lead to inaccuracies in predicting electronic structures and magnetic properties. For example, LDA often results in incorrect magnetic moments for strongly correlated materials and fails to predict the correct magnetic crystal structures~\cite{LDA_Fe}. GGA offers improved accuracy with respect to LDA, providing reasonable predictions for magnetic moments; nevertheless, GGA struggles to open sufficiently band gaps and fully capture strong electronic correlations.

To address these limitations, the GGA+$U$ method~\cite{GGAU1,GGAU2} was introduced, which adds an on-site Coulomb interaction term to better account for strong electronic correlations. However, the {accuracy of GGA+$U$ heavily depends on the $U$ parameter, which must be carefully tuned for each system. Hybrid functionals, which mix a portion of Hartree-Fock exchange with LDA or GGA functionals, offer further improvements in accuracy but are often computationally prohibitive, particularly for large or complex systems. 
Both GGA+$U$ and hybrid functionals fall into the category of orbital-dependent density functionals~\cite{ODDF}, where the direct use of orbitals in the Hamiltonian helps to partially correct self-interaction errors. Therefore, it is prudent to also consider other functionals in this category, such as meta-GGA, for studying magnetic materials.

Meta-generalized gradient approximation (meta-GGA) functionals, such as the Strongly Constrained and Appropriately Normed (SCAN) functional~\cite{SCAN}, represent a significant advancement over traditional functionals~\cite{meta-gga-mag}. 
These functionals improve the localization of \(d\)-electrons~\cite{delocal}, leading to more realistic band gap predictions and addressing many of the limitations associated with both LDA and GGA.  
The improved \(d\)-electron localization enables meta-GGA functionals to better describe magnetic materials that are Mott insulators. However, this feature can also result in an overestimation of magnetic moments in itinerant ferromagnets, such as iron (Fe)~\cite{Blaha-meta}. 

SCAN demonstrates superior performance in magnetic and non-magnetic materials~\cite{meta-gga-mag, meta-gga-gen} but demands dense real-space grids to mitigate numerical instabilities~\cite{r2SCAN}. This computational demand can pose challenges.   

To overcome the instability challenges, the restored regularized SCAN (r$^2$SCAN) functional~\cite{r2SCAN} was developed. This functional retains the accuracy of SCAN while improving numerical stability and computational efficiency~\cite{scan-r2scan}. In transition metal oxides, r$^2$SCAN has demonstrated superior performance, balancing accuracy and efficiency, making it an attractive option for studying magnetic systems.

Numerous benchmarks are available for assessing xc functionals in predicting properties such as bond lengths, band gaps, and energy barriers~\cite{Yang2012}. However, in the field of magnetism, where xc functionals play a critical role, benchmarking efforts remain limited, particularly for magnetic interactions. Studies on meta-GGA functionals for magnetic materials have highlighted limitations, such as their tendency to overestimate magnetic moments in itinerant ferromagnets, where GGA often yields more accurate results~\cite{SCAN_Fe1, SCAN_Fe2, Blaha-meta}. Other investigations have examined such properties as equilibrium volumes, band gaps for transition metals (3\(d\), 4\(d\), and 5\(d\))\cite{r2scan_3d_4d_5d}, Heusler alloys\cite{meta-gga-mag}, and transition metal oxides~\cite{SCAN_TMO}, showcasing certain advantages of meta-GGA over GGA. Despite these efforts, a comprehensive study on the prediction of magnetic thermodynamic properties is still lacking. Establishing robust benchmark sets for magnetic materials is crucial to addressing the challenges in DFT calculations for these systems.

In our previous work, we benchmarked GGA and GGA+$U$ functionals for predicting the transition temperatures of antiferromagnetic (AFM) materials. The results showed that GGA tends to overestimate, while GGA+$U$ tends to underestimate, the exchange coupling interactions~\cite{Mosleh2023}, leading to correspondingly higher and lower estimates of the Néel transition temperature. These findings highlight the need for improved xc functionals to achieve more accurate predictions of magnetic coupling.

In this study, we aim to evaluate the accuracy of the SCAN and r$^2$SCAN functionals in predicting the transition temperature of insulating AFM materials. To achieve this, we calculate magnetic exchange couplings using total energy calculations for various magnetic configurations with these functionals. These exchange interactions are then used to construct a model Hamiltonian, which is analyzed through Monte Carlo (MC) simulations.
We also evaluate how the predicted transition temperature from DFT changes as we advance to higher-level functionals on Jacob's ladder~\cite{Jacob-ladder}, ascending from meta-GGA to hybrid functionals like HSE06.

The paper is organized as follows: In Section~\ref{sec:comp}, we describe the computational details for obtaining transition temperatures. Section~\ref{sec:results} presents an analysis of the predicted transition temperatures using SCAN and r$^2$SCAN, including a comparison between the two functionals and with HSE06. The paper concludes with a summary of findings and implications.  
In the appendix, we provide a table listing the transition temperatures predicted by SCAN and r$^2$SCAN alongside their experimental values. Additionally, the supplementary materials include all exchange parameters obtained from SCAN and r$^2$SCAN for each compound.  

\section{\label{sec:comp}COMPUTATIONAL Details and Materials}
\subsection{DFT and MC}
The experimental structures, including both atomic positions and lattice vectors, are directly used for all \textit{ab initio} calculations. No \textit{ab initio} geometrical optimization is performed on these structures.
To determine up to which \(n\)th nearest neighbor to include in the calculations, we consider both the distance and the bonding connections between magnetic sites. Distances greater than 7 {\AA} are generally excluded unless there are atomic bond connections with angles close to 180° at such distances. For layered structures, we ensure that the interactions considered include at least one exchange interaction between adjacent layers.

Calculating exchange parameters up to the \(n\)th nearest neighbor requires a supercell structure that is mathematically capable of supporting such calculations. This is achieved using the SUPERHEX code~\cite{superhex}, which generates optimized supercells specifically designed for determining exchange interactions up to the desired \(n\)th nearest neighbor. For most compounds, the generated supercells contain 32 to 84 atoms, although for some structures, the number of atoms was as large as 112.

We perform spin-polarized density functional theory calculations using the Vienna Ab-initio Simulation Package (VASP)~\cite{vasp}, employing a plane-wave basis set with an energy cutoff of 600 eV. The Brillouin zone is sampled with a Monkhorst-Pack grid, maintaining a k-point spacing of 0.15~\AA$^{-1}$. 
To address convergence challenges in meta-GGA calculations, we initialize each magnetic configuration using wave functions and charge densities obtained from converged GGA or GGA+$U$ calculations. 

To estimate exchange interactions, we map the DFT results onto the Heisenberg Hamiltonian~\cite{Mosleh2023, Nafise2019}:

\[
\hat{\mathcal{H}} = -\frac{1}{2} \sum_{i,j} J_{ij} \hat{\mathbf{S}}_i \cdot \hat{\mathbf{S}}_j,
\]

where \(\hat{\mathbf{S}}_i\) and \(\hat{\mathbf{S}}_j\) are unit magnetic vectors at sites \(i\) and \(j\), and \(J_{ij}\) represents the coupling constants. 

To determine the Heisenberg exchange interactions using the total energy~\cite{Mosleh2023, Nafise2019}, at least \(n+1\) magnetic configurations are required to calculate coupling constants up to the {\it nth} nearest neighbors. However, small induced magnetic moments often appear on non-magnetic atoms, which can affect the results. To improve accuracy, it is recommended to use more magnetic configurations than the minimal requirement~\cite{Alaei2023}. The least-squares method is applied to fit the energy differences between magnetic configurations to evaluate exchange interactions.

For each compound, we use at least three times the minimal number of magnetic configurations and monitor the convergence of the coupling constants as the number of configurations increases. If convergence is unsatisfactory, we add more configurations until a satisfactory result is achieved.

To compute the Néel temperature, we utilize the ESpinS code~\cite{ESpinS}, which performs classical MC calculations based on the Heisenberg Hamiltonian. We select supercells containing at least 2000 magnetic atoms per cell. This choice ensures adequate sampling of magnetic interactions and minimizes finite-size effects.
We execute $10^6$ MC steps for thermalization and $10^6$ steps for sampling, collecting data every 5 steps. When large fluctuations are observed, the thermalization steps are increased to $2 \times 10^6$ and the sampling steps to $3 \times 10^6$ to enhance accuracy. To accelerate convergence to stable configurations, we employ the parallel tempering method, exchanging configurations every 10 MC steps.
\subsection{ MATERIALS}
In this study, we investigate a total of 48 AFM materials, including 27 of the 29 compounds from our previous work using GGA and GGA+$U$ methods~\cite{Mosleh2023}. Among the previously selected compounds, we encountered convergence issues in the total energy calculations for \(\mathrm{LiCoPO_4}\) when using the SCAN and r$^2$SCAN functionals.
Also, we exclude KMnSb from this study, which was included in our previous work, because we found no definitive experimental reports on its transition temperature.

We limit our selection to 3\textit{d} magnetic materials to avoid additional complexities arising from spin-orbit coupling. Additionally, we focus on insulating magnetic materials to avoid itinerant magnetism, which typically requires accounting for exchange interactions over very long distances.
The selected materials, along with references to their experimental structures, are as follows:
$\mathrm{YVO_3}$~\cite{YVO3-S}, $\mathrm{CrSb_2}$~\cite{CrSb2-S}, $\mathrm{CrCl_2}$~\cite{CrCl2-S}, $\mathrm{CrF_2}$~\cite{CrF2-S}, $\mathrm{Cr_2O_3}$~\cite{Cr2O3-S}, $\mathrm{CrSBr}$~\cite{CrSBr-ST}, $\mathrm{Cr_2TeO_6}$~\cite{Cr(Fe)2Te(W)O6-S}, $\mathrm{Cr_2WO_6}$~\cite{Cr(Fe)2Te(W)O6-S}, 
$\mathrm{MnO}$~\cite{MnO-S}, $\mathrm{MnS}$~\cite{MnS-ST}, $\mathrm{MnSe}$~\cite{MnSe-S}, $\mathrm{MnTe}$~\cite{MnTe-S},
$\mathrm{MnO_2}$~\cite{MnO2-S}, $\mathrm{MnF_2}$~\cite{MnF2-S}, $\mathrm{MnS_2}$~\cite{MnS2-S},$\mathrm{MnTe_2}$~\cite{MnTe2-ST1},
$\mathrm{LiMnO_2}$~\cite{LiMnO2-S}, $\mathrm{SrMnO_3}$~\cite{SrMnO3-ST}, $\mathrm{KMnF_3}$~\cite{KMnF3-ST1}, $\mathrm{MnPS_3}$~\cite{MnPS3-S},
$\mathrm{MnPSe_3}$~\cite{MnPSe3-S},
$\mathrm{MnWO_4}$~\cite{MnWO4-ST}, $\mathrm{Li_2MnO_3}$~\cite{Li2MnO3-ST}, 
$\mathrm{LiMnPO_4}$~\cite{LiMnPO4-S}, $\mathrm{Fe_2O_3}$~\cite{Fe2O3-S}, $\mathrm{SrFeO_2}$~\cite{SrFeO2-ST}, $\mathrm{BiFeO_3}$~\cite{BiFeO3-S}, $\mathrm{LaFeO_3}$~\cite{LaFeO3-S}, $\mathrm{YFeO_3}$~\cite{YFeO3-S}, $\mathrm{FePS_3}$~\cite{FePS3-S}, $\mathrm{Fe_2TeO_6}$~\cite{Cr(Fe)2Te(W)O6-S}, $\mathrm{SrFe_2S_2O}$~\cite{SrFe2S2O-S}, $\mathrm{CoWO_4}$~\cite{CoWO4-S}, 
$\mathrm{NiO}$~\cite{NiO-S}, $\mathrm{NiF_2}$~\cite{NiF2-S}, $\mathrm{NiBr_2}$~\cite{NiBr2-S}, $\mathrm{NiS_2}$~\cite{NiS2-ST}, $\mathrm{NiCl_2}$~\cite{NiCl2-S}, $\mathrm{NiPS_3}$~\cite{NiPS3-S},
$\mathrm{NiPSe_3}$~\cite{NiPSe3-S1},
$\mathrm{KNiF_3}$~\cite{KNiF3-ST1}, $\mathrm{NiWO_4}$~\cite{NiWO4-S}, $\mathrm{La_2NiO_4}$~\cite{La2NiO4-ST}, $\mathrm{K_2NiF_4}$~\cite{K2NiF4-S}, 
$\mathrm{KNiPO_4}$~\cite{KNiPO4-ST1}, $\mathrm{LiNiPO_4}$~\cite{LiNiPO4-S}, $\mathrm{CuO}$~\cite{CuO-ST}, $\mathrm{CuF_2}$~\cite{CuF2-S}.

\begin{table}[b]
\centering
\begin{ruledtabular}
\begin{tabular}{ccc}
Compound  & $J_{\text{SCAN(VASP)}}$ (meV) &  $J_{\text{SCAN(FHI-aims)}}$ (meV)  \\
\hline
\hline
CrCl$_2$& $\begin{array}{l} J_1 = -5.15 \\ J_2 =  0.08   \end{array}$
        & $\begin{array}{l} J_1 = -5.30 \\ J_2 = -0.27   \end{array}$ \\
        \hline
Cr$_2$O$_3$& $\begin{array}{l} J_1 = -29.21 \\ J_2 =  -17.63   \end{array}$
           & $\begin{array}{l} J_1 = -28.04 \\ J_2 = -18.39   \end{array}$ \\   
         \hline
MnO	& $\begin{array}{l} J_1 = -7.85 \\ J_2 = -10.13  \end{array}$ 
	& $\begin{array}{l} J_1 = -7.27 \\ J_2 = -10.28  \end{array}$ \\ 
	\hline
MnS	& $\begin{array}{l} J_1 = -2.00 \\ J_2 = -13.79  \end{array}$ 
	& $\begin{array}{l} J_1 = -1.49 \\ J_2 = -13.51  \end{array}$ \\ 
	\hline
MnSe    & $\begin{array}{l} J_2 = -13.31  \\ J_4 = -2.64 \end{array}$ 
	& $\begin{array}{l} J_2 = -11.21  \\ J_4 = -1.89 \end{array}$ \\ 
	\hline
MnF$_2$    & $\begin{array}{l} J_1 =  0.34   \\ J_2 = -4.07    \end{array}$ 
        & $\begin{array}{l} J_1 = -0.45  \\ J_2 = -3.96   \end{array}$ \\
	\hline
NiO     & $\begin{array}{l} J_1 =  2.29 \\ J_2 = -28.30  \end{array}$
        & $\begin{array}{l} J_1 =  2.40 \\ J_2 = -27.37  \end{array}$ \\
        \hline
NiF$_2$ & $\begin{array}{l} J_1 = -0.80 \\ J_2 = -3.84   \end{array}$ 
        & $\begin{array}{l} J_1 = -0.40 \\ J_2 = -3.77   \end{array}$ \\
 
\end{tabular}
	\caption{\label{tab:fhi-vasp} Comparison of the two largest exchange coupling interactions for each material calculated using VASP and FHI-aims packages.}
\end{ruledtabular}
\end{table}

\section{\label{sec:results}RESULTS and DISCUSSIONS}
\subsection{Exchange parameters}
\begin{figure}[t]
\includegraphics[scale=0.5]{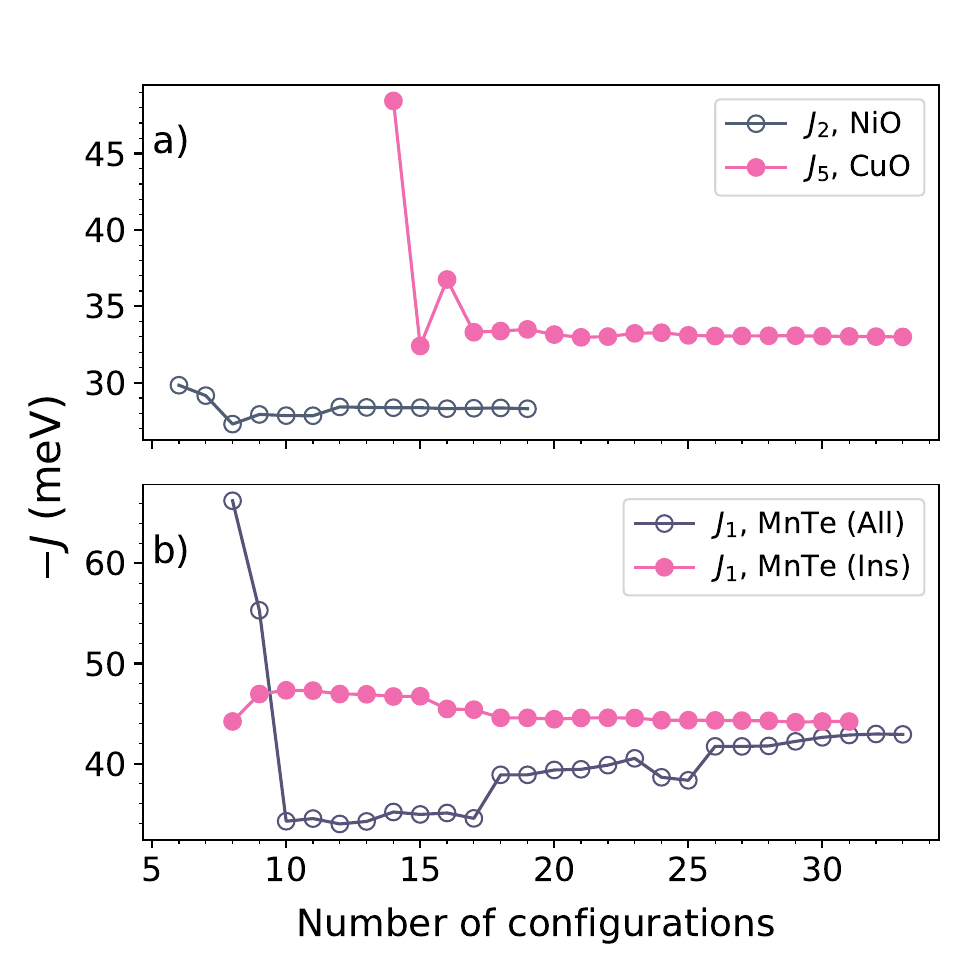}
\caption{\label{fig:J_conv} 
The plots illustrate the convergence behavior of exchange parameters as a function of the number of magnetic configurations. Plot (a) shows the convergence of the largest exchange parameters for NiO ($J_2$) and CuO ($J_5$). Plot (b) highlights the impact of including versus excluding metallic configurations on the convergence of $J_1$ for MnTe. "All" refers to data including both metallic and insulating configurations, while "Ins" refers to data derived exclusively from insulating configurations.
}
\end{figure}

\begin{figure}[t]
\includegraphics[scale=0.6]{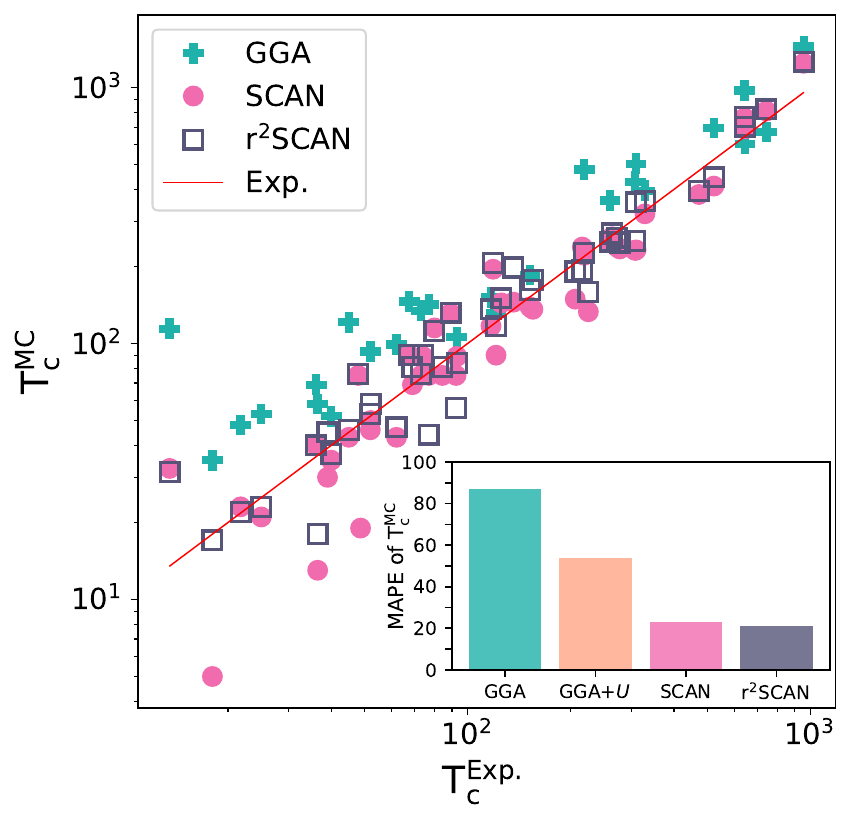}
\caption{\label{fig:tem} A comparison of computed Néel transition temperatures in SCAN and r$^2$SCAN functional 
	from MC simulations with experimental transition temperatures.  Inset figure represents the total mean absolute percentage error (MAPE) of transition temperatures in different functional. The GGA and GGA+$U$  data are taken from ref~\cite{Mosleh2023}.
    }
\end{figure}

We initially validated the reliability of our calculations by employing the Perdew-Burke-Ernzerhof PAW potentials in conjunction with the SCAN functional. 
To ensure accuracy, we selected a series of simple materials, including CrCl$_2$, Cr$_2$O$_3$, MnO, MnS, MnSe, MnF$_2$, NiO, and NiF$_2$,  computed their magnetic 
exchange couplings using the FHI-aims program package~\cite{FHI-aims}. A comparison between the results obtained from FHI-aims and VASP packages, presented in Table~\ref{tab:fhi-vasp}, 
demonstrates compatibility between the two codes, confirming that the meta-GGA functional can be reliably used with these pseudo-potentials.

\begin{figure*}
   \includegraphics[scale=0.65]{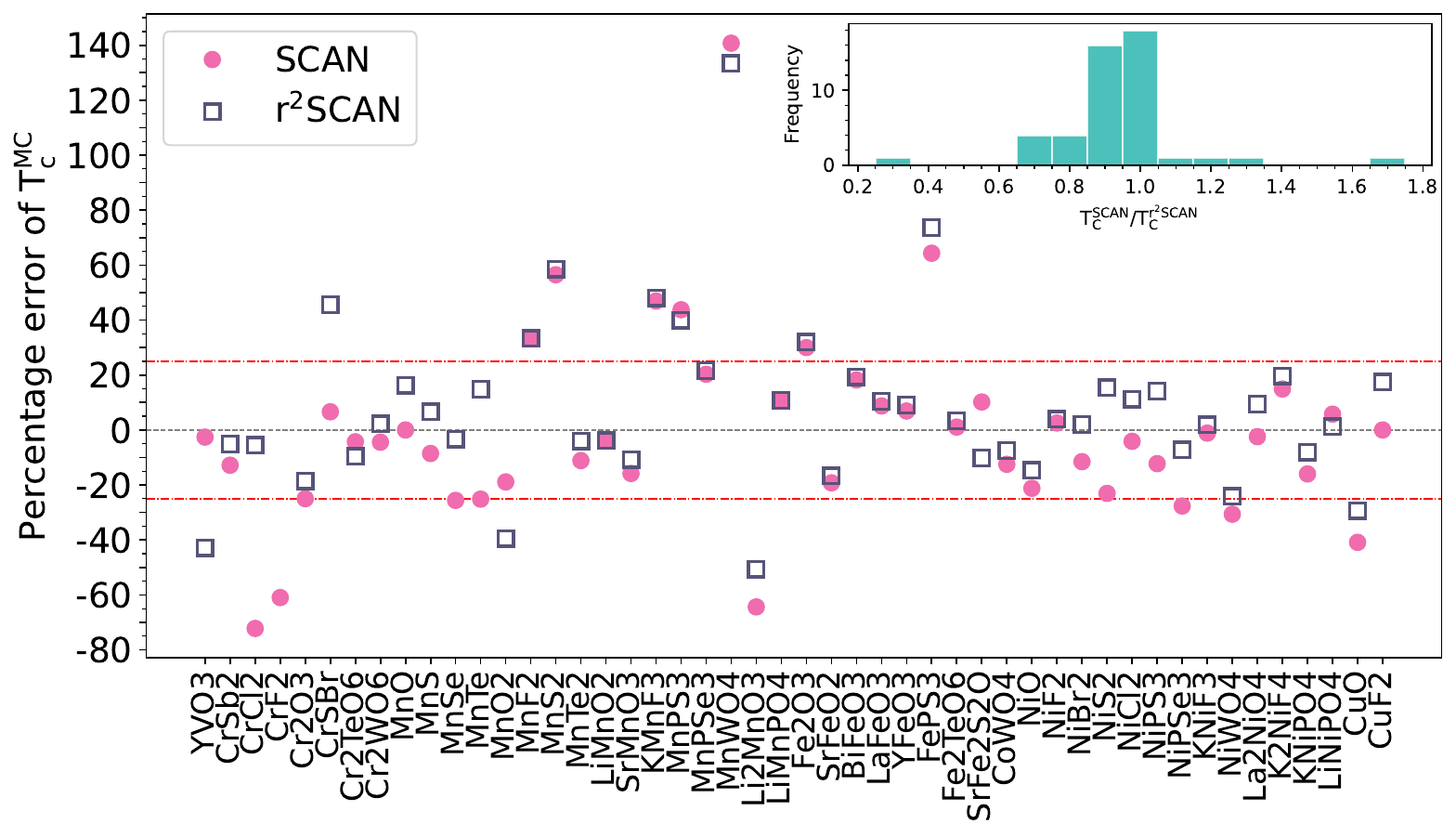}
\caption{\label{fig:error}Percentage error of transition temperatures calculated using MC simulations ($\mathrm{T_c^{MC}}$) with SCAN and r$^2$SCAN functionals across various compounds.  For CrF$_2$, r$^2$SCAN data is excluded as it incorrectly predicts CrF$_2$ to be FM.
Inset of figure represents the frequency distribution of ratio  $\mathrm{T_C^{SCAN}/T_C^{r^2SCAN}}$.
}
\end{figure*}
\begin{figure*}
   \includegraphics[scale=0.65]{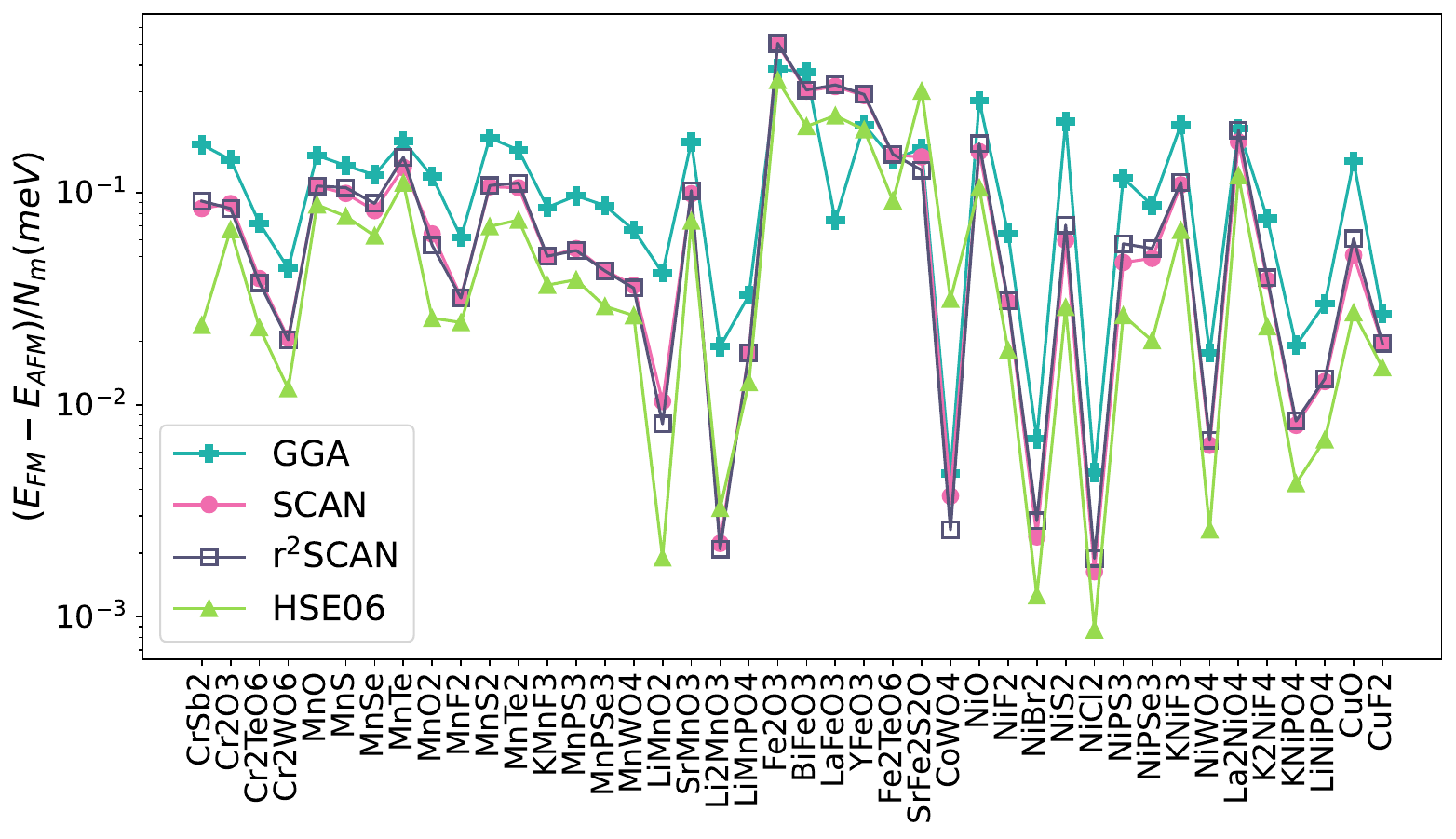}
\caption{\label{fig:dE}
Energy differences per magnetic atom between AFM and FM configurations for various compounds, calculated using four exchange-correlation functionals: PBE, HSE06, SCAN, and r$^2$SCAN. } 
\end{figure*}

Theoretically, obtaining Heisenberg exchange interactions up to the \(n\)th nearest neighbor requires a minimum of \(n+1\) unique magnetic configurations. However, practical considerations introduce additional challenges, such as the induced magnetic moments of anions like oxygen atoms. 
To address these complexities, it is advisable to utilize a greater number of magnetic configurations and to determine the Heisenberg exchanges using the least-squares method, as suggested in our previous work~\cite{Alaei2023}. 
Consequently, we employ approximately three times the minimum required number of magnetic configurations to account for these additional factors and enhance the accuracy of our results. 

The exchange constants for most compounds converge when only a few additional configurations beyond the minimal set are included. Figure ~\ref{fig:J_conv} (a) illustrates the variation of $J_2$ for NiO and $J_5$ for CuO as a function of the number of magnetic configurations. For both compounds, adding just 3 to 4 configurations beyond the minimal set is generally sufficient to achieve convergence.

For NiO, the minimal set of configurations yields a $J_2$ value that is reasonably close to the converged result. In contrast, for CuO, the minimal configurations result in a $J_5$ value that deviates significantly from the converged value. This demonstrates that relying solely on the minimal configuration set to calculate exchange interactions can lead to substantial inaccuracies.

We encounter convergence challenges when studying compounds that exhibit metallic band structure in certain magnetic configurations. In contrast, insulating configurations tend to converge more readily. To ensure reliable convergence of exchange interactions, we exclude metallic magnetic configurations from our analysis. Figure ~\ref{fig:J_conv} (b) illustrates how $J_1$ for MnTe fails to converge properly when all 34 magnetic configurations, including metallic ones, are considered. However, by excluding two metallic configurations, $J_1$ converges significantly faster to a stable value.

AFM ordering is observed in the exchange interaction couplings of all studied compounds, except for CrF$_2$, where the r$^2$SCAN functional predicts ferromagnetic (FM) ordering.
The exchange parameters obtained from SCAN and r$^2$SCAN functionals are given in the Supplemental material~\cite{SM}.

\subsection{SCAN and r$^2$SCAN transition temperature}
Figure \ref{fig:tem} compares the transition temperatures obtained from MC simulations using the SCAN and r$^2$SCAN meta-GGA functionals with experimental transition temperatures. The predictions from both SCAN and r$^2$SCAN functionals show variations, sometimes overestimating and sometimes underestimating the experimental values. However, overall, the transition temperatures computed with these meta-GGA functionals align significantly better with experimental data compared to those obtained using GGA and GGA+$U$ methods.

The mean absolute percentage error (MAPE) is 23\% for SCAN functional  and 22\% for r$^2$SCAN functional, demonstrating their superior accuracy. In comparison, GGA and GGA+$U$ functionals exhibit significantly higher MAPE values of 87\% and 54\% , respectively, as shown in the inset of Figure \ref{fig:tem} (the GGA and GGA+$U$ data are sourced from our previous work~\cite{Mosleh2023}).
The high Pearson correlation coefficients of 97\% for SCAN and 98\% for r$^2$SCAN suggest that r$^2$SCAN functional is well-suited for high-throughput calculations. When combined with machine learning techniques, it can enhance the prediction of transition temperatures.

Errors of the calculated transition temperatures for each material is illustrated in Figure \ref{fig:error}.
A total of 32 materials in the SCAN functional and 35 materials in the r$^2$SCAN functional exhibit errors of less than 25\%. 
The maximum errors are observed for MnWO$_4$, FePS$_3$, CrF$_4$, and MnS$_2$, with values of approximately 141 (133)\%, 64 (73)\%, 61 (not applicable for r$^2$SCAN functional due to incorrect prediction of ferromagnetic (FM) ordering instead of AFM), and 56 (59)\%, respectively, for SCAN (r$^2$SCAN) functional.

To evaluate the consistency between the SCAN and r$^2$SCAN functionals in predicting Néel transition temperatures, we analyze the ratio $\mathrm{T_C^{SCAN}/T_C^{r^2SCAN}}$. The frequency distribution of these ratios is shown in the inset of Figure~\ref{fig:error}. A ratio close to 1 indicates strong agreement between the two functionals.  
For most compounds (~77\%), the transition temperatures predicted by SCAN are lower than those predicted by r$^2$SCAN. However, the distribution is centered around 1, indicating that the two functionals generally provide comparable predictions. Deviations from this ratio reflect differences in how the functionals treat specific magnetic systems, likely due to their distinct parameterizations and underlying approximations.  In some cases, such as MnTe, the deviation is significant. SCAN functional predicts a transition temperature of 232 K, while r$^2$SCAN functional predicts 356 K, resulting in a difference of 124 K (see Table~\ref{tab:table2}).  

\subsection{ HSE06 functional versus  SCAN and r$^2$SCAN functionals}
We analyze the energy differences between FM and AFM configurations using the SCAN and r$^2$SCAN functionals, comparing them with results from the GGA and HSE06 functionals. These energy differences reflect the strength of exchange interactions and provide insights into how various \textit{ab initio} methods predict magnetic transition temperatures.  

Calculations are conducted on supercells containing two to four magnetic sites. While many of these supercells represent magnetic structures, some compounds require magnetic structures with more than four magnetic sites, rendering HSE06 calculations computationally prohibitive. Consequently, using these smaller supercells (with two to four magnetic sites) instead of magnetic unit cells led some calculations to incorrectly predict ferromagnetism as the stable configuration for certain compounds. To maintain consistency and accuracy, we exclude these cases from the analysis in this subsection. For supercells with four magnetic sites, we select the AFM configurations with the lowest energy among all possible AFM arrangements.

To reduce the computational cost of HSE06 calculations, the energy cutoff was set to 550 eV.  
As shown in Figure~\ref{fig:dE}, for most compounds, the energy differences per magnetic site predicted by SCAN and r$^2$SCAN functionals fall between those predicted by GGA and HSE06 functionals. GGA functional predicts larger energy differences, indicating stronger exchange coupling and higher transition temperatures for AFM materials. In contrast, HSE06 functional predicts smaller energy differences, suggesting weaker exchange interactions and lower transition temperatures.  
Since SCAN functional underestimates transition temperatures for nearly half of the compounds, HSE06 is expected to underestimate them even further in these cases. This highlights potential limitations of HSE06 functional for certain materials and underscores the need for continued development of DFT for magnetic systems.  

The energy difference between AFM and FM configurations is a key metric for understanding exchange interactions. To improve the reliability of this metric, advanced wavefunction-based methods, such as coupled-cluster singles and doubles (CCSD) or quantum MC simulations, are needed. Data from these methods would be invaluable for refining DFT functionals and enhancing their predictive accuracy for magnetic materials.  

\section{CONCLUSIONS}
In this study, we evaluated the performance of SCAN and r$^2$SCAN meta-GGA functionals in predicting the Néel transition temperatures of 48 AFM materials. Using a combination of DFT calculations, exchange parameter fitting, and MC simulations, we showed that both SCAN and r$^2$SCAN greatly outperform standard GGA and GGA+$U$ methods in aligning theoretical predictions with experimental data. These findings underscore the potential of meta-GGA functionals as reliable tools for studying magnetic phase transitions, marking a significant step toward more accurate modeling of complex magnetic systems. 
Although SCAN and r$^2$SCAN generally produce very similar results, our study reveals a significant discrepancy for CrF$_2$. SCAN correctly identifies the system as AFM, whereas r$^2$SCAN incorrectly predicts it to be FM. Additionally, for cases like MnTe, there is a notable difference between SCAN and r$^2$SCAN in predicting the transition temperature.
Although SCAN and r$^2$SCAN have shown significant success in predicting transition temperatures compared to GGA, our calculations faced a major challenge in achieving convergence during the DFT self-consistent cycles when applied to supercell structures. 
To address this issue, we initialized all calculations with results from GGA or GGA+$U$. While this approach partially mitigates the problem, it highlights a critical limitation that could hinder the broader application of meta-GGA functionals in magnetic material studies.  
We also presented a method for analyzing how the hybrid functional HSE06 predicts the transition temperature in comparison to meta-GGA functionals. Our findings suggest that HSE06 underestimates the transition temperature relative to meta-GGA functionals and to experiment.
\section{Acknowledgment}
This work has been supported by the Russian Science Foundation (Grant 19-72-30043).

\appendix
\section{Table of transition temperatures}

\begin{table}[h]
\caption{\label{tab:table2}
Néel temperature predicted by MC simulations using exchange parameters from SCAN and r$^2$SCAN functionals compared to experimental values.}
\begin{ruledtabular}
\scalebox{0.94}{%
\begin{tabular}{llll}
Compound & $\mathrm{T_C^{SCAN}}$ & $\mathrm{T_C^{r^2SCAN}}$ & $\mathrm{T_C^{exp.}}$ \\
       & (K)                  & (K)                      & (K)                  \\
\hline
    $\mathrm{YVO_3}$     &  75 &  44 & 77~\cite{YVO3-T}                                             \\   
    $\mathrm{CrSb_2}$    & 238 & 259 & 273~\cite{CrSb2-T1, CrSb2-T2}                                 \\
    $\mathrm{CrCl_2}$    &   5 &  17 & 16, 20~\cite{CrCl2-T1, CrCl2-T2, CrCl2-T3}                    \\  
    $\mathrm{CrF_2}$     &  19 &  -- & 48.7~\cite{CrF2-T}                                            \\
    $\mathrm{Cr_2O_3}$   & 231 & 251 & 308~\cite{Cr2O3-T1, Cr2O3-T2, Cr2O3-T3}                       \\
     $\mathrm{CrSBr}$     & 145 & 198 & 140, 132~\cite{CrSBr-ST, CrSBr-T}                             \\
    $\mathrm{Cr_2TeO_6}$ &  89 &  84 & 93~\cite{Cr2Te(W)O6-T}                                        \\
    $\mathrm{Cr_2WO_6}$  &  43 &  46 & 45~\cite{Cr2Te(W)O6-T}                                        \\
    MnO                  & 117 & 136 & 117~\cite{MnO-T}                                              \\
    MnS                  & 139 & 162 & 152~\cite{MnS-ST}                                             \\
    MnSe                 &  90 & 117 & 120, 122~\cite{MnSe-T1, MnSe-T2}                              \\  
    MnTe                 & 232 & 356 & 310~\cite{MnTe-T1, MnTe-T2}                                   \\
    $\mathrm{MnO_2}$     &  75 &  56 & 92, 93~\cite{MnO2-T1, MnO2-T2}                                \\
    $\mathrm{MnF_2}$     &  90 &  90 & 67.3, 67.7~\cite{MnF2-T, MnF2-KMnF3-T2}                       \\
    $\mathrm{MnS_2}$     &  75 &  76 & 48.2, 47.7, 47.9~\cite{MnS2-T1, MnS2-T2, MnS(Te)2-T3}         \\
    $\mathrm{MnTe_2}$    &  75 &  81 & 86.5, 83.8, 83~\cite{MnTe2-ST1, MnTe2-T2, MnS(Te)2-T3}         \\
    $\mathrm{LiMnO_2}$   & 250 & 250 & 261.5, 259~\cite{LiMnO2-T1, LiMnO2-T2}                        \\
    $\mathrm{SrMnO_3}$   & 234 & 248 & 278~\cite{SrMnO3-ST, SrMnO3-T}                                \\
    $\mathrm{KMnF_3}$    & 131 & 132 & 86.8, 88, 88.2, 89, 95~\cite{KMnF3-ST1, KNi(Mn)F3-T2, KNi(Mn)F3-T3, KMn(Ni)F3-T3, MnF2-KMnF3-T2, KMnF3-T4}   \\
    $\mathrm{MnPS_3}$    & 115 & 112 & 82, 78~\cite{MPX3-T, MPS3-T}                                  \\
    $\mathrm{MnPSe_3}$   &  89 &  90 & 74~\cite{MPX3-T, MnPSe3-T}                                    \\
    $\mathrm{MnWO_4}$    &32.5 & 31.5& 13.5~\cite{MnWO4-ST}                                          \\
    $\mathrm{Li_2MnO_3}$ &  13 &  18 & 36.5~\cite{Li2MnO3-ST}                                        \\
    $\mathrm{LiMnPO_4}$  &  40 &  40 & 34.8, 34, 36, 42, 33.8~\cite{LiMnPO4-T1, LiMnPO4-T2, LiMnPO4-T3, LiMnPO4-T4}\\
    $\mathrm{Fe_2O_3}$   &1243 &1262 & 946, 953, 960, 966~\cite{Fe2O30-T1, Fe2O3-T3, Fe2O3-T4, FE2O3-T5, Fe2O3-T6} \\   
    $\mathrm{SrFeO_2}$   & 382 & 394 & 473~\cite{SrFeO2-ST}                                              \\                                      
    $\mathrm{BiFeO_3}$   & 760 & 767 & 643~\cite{BiFeO3-T1, BiFeO3-T2}    \\      
    $\mathrm{LaFeO_3}$   & 809 & 822 & 738, 750~\cite{LaFeO3-T1, LaFeO3-T2}    \\          
    $\mathrm{YFeO_3}$    & 689 & 703 & 644.5~\cite{YFeO3-T}  \\
    $\mathrm{FePS_3}$    & 195 & 206 & 117, 116, 123~\cite{FePS3-T, MPX3-T, MPS3-T}  \\
    $\mathrm{Fe_2TeO_6}$ & 221 & 226 & 201, 233, 206.5, 244, 209~\cite{Fe2TeO6-1, Fe2TeO6-2, Fe2TeO6-3}    \\  
    $\mathrm{SrFe_2S_2O}$& 238 & 194 & 216~\cite{SrFe2S2O-T}    \\
    $\mathrm{CoWO_4}$    &  35 &  37 & 40~\cite{CoWO4-T}    \\  
    NiO                 & 412 & 446 & 523~\cite{NiO-T}    \\   
    $\mathrm{NiF_2}$     &  75 &  76 & 73, 73.3~\cite{NiF2-T1, NiF2-T2}  \\  
    $\mathrm{NiBr_2}$    &  46 &  53 & 52~\cite{NiBr2-T}    \\  
    $\mathrm{NiS_2}$     &  30 &  45 & 39~\cite{NiS2-ST}    \\  
    $\mathrm{NiCl_2}$    &  50 &  58 & 52, 52.3 ~\cite{NiCl2-T1, NiCl2-T2}    \\
    $\mathrm{NiPS_3}$    & 136 & 177 &  155~\cite{MPX3-T, MPS3-T}   \\  
    $\mathrm{NiPSe_3}$   & 149 & 191 & 206~\cite{MPX3-T}    \\
    $\mathrm{KNiF_3}$    & 261 & 269 & 275, 253~\cite{KNiF3-ST1,  KMn(Ni)F3-T3, KNi(Mn)F3-T2, KNi(Mn)F3-T3}   \\  
    $\mathrm{NiWO_4}$    &  43 &  47 & 62~\cite{NiWO4-T}    \\  
    $\mathrm{La_2NiO_4}$ & 321 & 360 & 330, 328~\cite{La2NiO4-ST, La2NiO4-T2}    \\  
    $\mathrm{K_2NiF_4}$  & 144 & 150 & 97.23, 98.7, 180~\cite{K2NiF4-T1, K2NiF4-T2, K2NiF4-T3}   \\
    $\mathrm{KNiPO_4}$   &  21 &  23 & 25~\cite{KNiPO4-ST1, KNiPO4-T2}    \\  
    $\mathrm{LiNiPO_4}$  &  23 &  22 & 21.8, 21.7~\cite{LiNiPO4-T1, LiNiPO4-T2}  \\  
    CuO                  & 133 & 159 & 220,225,230~\cite{CuO-ST, CuO-T(220), CuO-T(225)}    \\
    $\mathrm{CuF_2}$     &  69 &  81 & 69~\cite{CuF2-T}     \\                                              
\end{tabular}
}
\end{ruledtabular}
\end{table}

\bibliography{ref.bib}

\begin{thebibliography}{148}%
\makeatletter
\providecommand \@ifxundefined [1]{%
 \@ifx{#1\undefined}
}%
\providecommand \@ifnum [1]{%
 \ifnum #1\expandafter \@firstoftwo
 \else \expandafter \@secondoftwo
 \fi
}%
\providecommand \@ifx [1]{%
 \ifx #1\expandafter \@firstoftwo
 \else \expandafter \@secondoftwo
 \fi
}%
\providecommand \natexlab [1]{#1}%
\providecommand \enquote  [1]{``#1''}%
\providecommand \bibnamefont  [1]{#1}%
\providecommand \bibfnamefont [1]{#1}%
\providecommand \citenamefont [1]{#1}%
\providecommand \href@noop [0]{\@secondoftwo}%
\providecommand \href [0]{\begingroup \@sanitize@url \@href}%
\providecommand \@href[1]{\@@startlink{#1}\@@href}%
\providecommand \@@href[1]{\endgroup#1\@@endlink}%
\providecommand \@sanitize@url [0]{\catcode `\\12\catcode `\$12\catcode
  `\&12\catcode `\#12\catcode `\^12\catcode `\_12\catcode `\%12\relax}%
\providecommand \@@startlink[1]{}%
\providecommand \@@endlink[0]{}%
\providecommand \url  [0]{\begingroup\@sanitize@url \@url }%
\providecommand \@url [1]{\endgroup\@href {#1}{\urlprefix }}%
\providecommand \urlprefix  [0]{URL }%
\providecommand \Eprint [0]{\href }%
\providecommand \doibase [0]{https://doi.org/}%
\providecommand \selectlanguage [0]{\@gobble}%
\providecommand \bibinfo  [0]{\@secondoftwo}%
\providecommand \bibfield  [0]{\@secondoftwo}%
\providecommand \translation [1]{[#1]}%
\providecommand \BibitemOpen [0]{}%
\providecommand \bibitemStop [0]{}%
\providecommand \bibitemNoStop [0]{.\EOS\space}%
\providecommand \EOS [0]{\spacefactor3000\relax}%
\providecommand \BibitemShut  [1]{\csname bibitem#1\endcsname}%
\let\auto@bib@innerbib\@empty
\bibitem [{\citenamefont {Illas}\ \emph {et~al.}(2004)\citenamefont {Illas},
  \citenamefont {de~P.~R.~Moreira}, \citenamefont {Bofill},\ and\ \citenamefont
  {Filatov}}]{DFT_MAG1}%
  \BibitemOpen
  \bibfield  {author} {\bibinfo {author} {\bibfnamefont {F.}~\bibnamefont
  {Illas}}, \bibinfo {author} {\bibfnamefont {I.}~\bibnamefont
  {de~P.~R.~Moreira}}, \bibinfo {author} {\bibfnamefont {J.~M.}\ \bibnamefont
  {Bofill}},\ and\ \bibinfo {author} {\bibfnamefont {M.}~\bibnamefont
  {Filatov}},\ }\bibfield  {title} {\bibinfo {title} {{Extent and limitations
  of density-functional theory in describing magnetic systems}},\ }\href
  {https://doi.org/10.1103/PhysRevB.70.132414} {\bibfield  {journal} {\bibinfo
  {journal} {Phys. Rev. B}\ }\textbf {\bibinfo {volume} {70}},\ \bibinfo
  {pages} {132414} (\bibinfo {year} {2004})}\BibitemShut {NoStop}%
\bibitem [{\citenamefont {Cohen}\ \emph {et~al.}(2008)\citenamefont {Cohen},
  \citenamefont {Mori-S{\'a}nchez},\ and\ \citenamefont {Yang}}]{Yang2008}%
  \BibitemOpen
  \bibfield  {author} {\bibinfo {author} {\bibfnamefont {A.~J.}\ \bibnamefont
  {Cohen}}, \bibinfo {author} {\bibfnamefont {P.}~\bibnamefont
  {Mori-S{\'a}nchez}},\ and\ \bibinfo {author} {\bibfnamefont {W.}~\bibnamefont
  {Yang}},\ }\bibfield  {title} {\bibinfo {title} {{Insights into current
  limitations of density functional theory}},\ }\href
  {https://doi.org/10.1126/science.1158722} {\bibfield  {journal} {\bibinfo
  {journal} {Science}\ }\textbf {\bibinfo {volume} {321}},\ \bibinfo {pages}
  {792} (\bibinfo {year} {2008})}\BibitemShut {NoStop}%
\bibitem [{\citenamefont {Perdew}\ and\ \citenamefont {Zunger}(1981)}]{LDA}%
  \BibitemOpen
  \bibfield  {author} {\bibinfo {author} {\bibfnamefont {J.~P.}\ \bibnamefont
  {Perdew}}\ and\ \bibinfo {author} {\bibfnamefont {A.}~\bibnamefont
  {Zunger}},\ }\bibfield  {title} {\bibinfo {title} {{Self-interaction
  correction to density-functional approximations for many-electron systems}},\
  }\href {https://doi.org/10.1103/PhysRevB.23.5048} {\bibfield  {journal}
  {\bibinfo  {journal} {Phys. Rev. B}\ }\textbf {\bibinfo {volume} {23}},\
  \bibinfo {pages} {5048} (\bibinfo {year} {1981})}\BibitemShut {NoStop}%
\bibitem [{\citenamefont {Langreth}\ and\ \citenamefont {Perdew}(1980)}]{GGA}%
  \BibitemOpen
  \bibfield  {author} {\bibinfo {author} {\bibfnamefont {D.~C.}\ \bibnamefont
  {Langreth}}\ and\ \bibinfo {author} {\bibfnamefont {J.~P.}\ \bibnamefont
  {Perdew}},\ }\bibfield  {title} {\bibinfo {title} {{Theory of nonuniform
  electronic systems. I. Analysis of the gradient approximation and a
  generalization that works}},\ }\href
  {https://doi.org/10.1103/PhysRevB.21.5469} {\bibfield  {journal} {\bibinfo
  {journal} {Phys. Rev. B}\ }\textbf {\bibinfo {volume} {21}},\ \bibinfo
  {pages} {5469} (\bibinfo {year} {1980})}\BibitemShut {NoStop}%
\bibitem [{\citenamefont {Perdew}\ \emph {et~al.}(1996)\citenamefont {Perdew},
  \citenamefont {Burke},\ and\ \citenamefont {Ernzerhof}}]{PBE}%
  \BibitemOpen
  \bibfield  {author} {\bibinfo {author} {\bibfnamefont {J.~P.}\ \bibnamefont
  {Perdew}}, \bibinfo {author} {\bibfnamefont {K.}~\bibnamefont {Burke}},\ and\
  \bibinfo {author} {\bibfnamefont {M.}~\bibnamefont {Ernzerhof}},\ }\bibfield
  {title} {\bibinfo {title} {{Generalized Gradient Approximation Made
  Simple}},\ }\href {https://doi.org/10.1103/PhysRevLett.77.3865} {\bibfield
  {journal} {\bibinfo  {journal} {Phys. Rev. Lett.}\ }\textbf {\bibinfo
  {volume} {77}},\ \bibinfo {pages} {3865} (\bibinfo {year}
  {1996})}\BibitemShut {NoStop}%
\bibitem [{\citenamefont {Tran}\ \emph {et~al.}(2020)\citenamefont {Tran},
  \citenamefont {Baudesson}, \citenamefont {Carrete}, \citenamefont {Madsen},
  \citenamefont {Blaha}, \citenamefont {Schwarz},\ and\ \citenamefont
  {Singh}}]{Blaha-meta}%
  \BibitemOpen
  \bibfield  {author} {\bibinfo {author} {\bibfnamefont {F.}~\bibnamefont
  {Tran}}, \bibinfo {author} {\bibfnamefont {G.}~\bibnamefont {Baudesson}},
  \bibinfo {author} {\bibfnamefont {J.}~\bibnamefont {Carrete}}, \bibinfo
  {author} {\bibfnamefont {G.~K.~H.}\ \bibnamefont {Madsen}}, \bibinfo {author}
  {\bibfnamefont {P.}~\bibnamefont {Blaha}}, \bibinfo {author} {\bibfnamefont
  {K.}~\bibnamefont {Schwarz}},\ and\ \bibinfo {author} {\bibfnamefont {D.~J.}\
  \bibnamefont {Singh}},\ }\bibfield  {title} {\bibinfo {title} {{Shortcomings
  of meta-GGA functionals when describing magnetism}},\ }\href
  {https://doi.org/10.1103/PhysRevB.102.024407} {\bibfield  {journal} {\bibinfo
   {journal} {Phys. Rev. B}\ }\textbf {\bibinfo {volume} {102}},\ \bibinfo
  {pages} {024407} (\bibinfo {year} {2020})}\BibitemShut {NoStop}%
\bibitem [{\citenamefont {Cho}\ and\ \citenamefont {Scheffler}(1996)}]{LDA_Fe}%
  \BibitemOpen
  \bibfield  {author} {\bibinfo {author} {\bibfnamefont {J.-H.}\ \bibnamefont
  {Cho}}\ and\ \bibinfo {author} {\bibfnamefont {M.}~\bibnamefont
  {Scheffler}},\ }\bibfield  {title} {\bibinfo {title} {Ab initio
  pseudopotential study of fe, co, and ni employing the spin-polarized lapw
  approach},\ }\href {https://doi.org/10.1103/PhysRevB.53.10685} {\bibfield
  {journal} {\bibinfo  {journal} {Phys. Rev. B}\ }\textbf {\bibinfo {volume}
  {53}},\ \bibinfo {pages} {10685} (\bibinfo {year} {1996})}\BibitemShut
  {NoStop}%
\bibitem [{\citenamefont {Anisimov}\ \emph {et~al.}(1991)\citenamefont
  {Anisimov}, \citenamefont {Zaanen},\ and\ \citenamefont {Andersen}}]{GGAU1}%
  \BibitemOpen
  \bibfield  {author} {\bibinfo {author} {\bibfnamefont {V.~I.}\ \bibnamefont
  {Anisimov}}, \bibinfo {author} {\bibfnamefont {J.}~\bibnamefont {Zaanen}},\
  and\ \bibinfo {author} {\bibfnamefont {O.~K.}\ \bibnamefont {Andersen}},\
  }\bibfield  {title} {\bibinfo {title} {{Band theory and Mott insulators:
  Hubbard U instead of Stoner I}},\ }\href
  {https://doi.org/10.1103/PhysRevB.44.943} {\bibfield  {journal} {\bibinfo
  {journal} {Phys. Rev. B}\ }\textbf {\bibinfo {volume} {44}},\ \bibinfo
  {pages} {943} (\bibinfo {year} {1991})}\BibitemShut {NoStop}%
\bibitem [{\citenamefont {Anisimov}\ \emph {et~al.}(1997)\citenamefont
  {Anisimov}, \citenamefont {Aryasetiawan},\ and\ \citenamefont
  {Lichtenstein}}]{GGAU2}%
  \BibitemOpen
  \bibfield  {author} {\bibinfo {author} {\bibfnamefont {V.~I.}\ \bibnamefont
  {Anisimov}}, \bibinfo {author} {\bibfnamefont {F.}~\bibnamefont
  {Aryasetiawan}},\ and\ \bibinfo {author} {\bibfnamefont {A.~I.}\ \bibnamefont
  {Lichtenstein}},\ }\bibfield  {title} {\bibinfo {title} {{First-principles
  calculations of the electronic structure and spectra of strongly correlated
  systems: the LDA+ U method}},\ }\href
  {https://doi.org/10.1088/0953-8984/9/4/002} {\bibfield  {journal} {\bibinfo
  {journal} {Journal of Physics: Condensed Matter}\ }\textbf {\bibinfo {volume}
  {9}},\ \bibinfo {pages} {767} (\bibinfo {year} {1997})}\BibitemShut {NoStop}%
\bibitem [{\citenamefont {K\"ummel}\ and\ \citenamefont {Kronik}(2008)}]{ODDF}%
  \BibitemOpen
  \bibfield  {author} {\bibinfo {author} {\bibfnamefont {S.}~\bibnamefont
  {K\"ummel}}\ and\ \bibinfo {author} {\bibfnamefont {L.}~\bibnamefont
  {Kronik}},\ }\bibfield  {title} {\bibinfo {title} {{Orbital-dependent density
  functionals: Theory and applications}},\ }\href
  {https://doi.org/10.1103/RevModPhys.80.3} {\bibfield  {journal} {\bibinfo
  {journal} {Rev. Mod. Phys.}\ }\textbf {\bibinfo {volume} {80}},\ \bibinfo
  {pages} {3} (\bibinfo {year} {2008})}\BibitemShut {NoStop}%
\bibitem [{\citenamefont {Sun}\ \emph {et~al.}(2015)\citenamefont {Sun},
  \citenamefont {Ruzsinszky},\ and\ \citenamefont {Perdew}}]{SCAN}%
  \BibitemOpen
  \bibfield  {author} {\bibinfo {author} {\bibfnamefont {J.}~\bibnamefont
  {Sun}}, \bibinfo {author} {\bibfnamefont {A.}~\bibnamefont {Ruzsinszky}},\
  and\ \bibinfo {author} {\bibfnamefont {J.~P.}\ \bibnamefont {Perdew}},\
  }\bibfield  {title} {\bibinfo {title} {{Strongly Constrained and
  Appropriately Normed Semilocal Density Functional}},\ }\href
  {https://doi.org/10.1103/PhysRevLett.115.036402} {\bibfield  {journal}
  {\bibinfo  {journal} {Phys. Rev. Lett.}\ }\textbf {\bibinfo {volume} {115}},\
  \bibinfo {pages} {036402} (\bibinfo {year} {2015})}\BibitemShut {NoStop}%
\bibitem [{\citenamefont {Sokolovskiy}\ \emph {et~al.}(2023)\citenamefont
  {Sokolovskiy}, \citenamefont {Baigutlin}, \citenamefont {Miroshkina},\ and\
  \citenamefont {Buchelnikov}}]{meta-gga-mag}%
  \BibitemOpen
  \bibfield  {author} {\bibinfo {author} {\bibfnamefont {V.}~\bibnamefont
  {Sokolovskiy}}, \bibinfo {author} {\bibfnamefont {D.}~\bibnamefont
  {Baigutlin}}, \bibinfo {author} {\bibfnamefont {O.}~\bibnamefont
  {Miroshkina}},\ and\ \bibinfo {author} {\bibfnamefont {V.}~\bibnamefont
  {Buchelnikov}},\ }\bibfield  {title} {\bibinfo {title} {{Meta-GGA SCAN
  Functional in the Prediction of Ground State Properties of Magnetic
  Materials: Review of the Current State}},\ }\bibfield  {journal} {\bibinfo
  {journal} {Metals}\ }\textbf {\bibinfo {volume} {13}},\ \href
  {https://doi.org/10.3390/met13040728} {10.3390/met13040728} (\bibinfo {year}
  {2023})\BibitemShut {NoStop}%
\bibitem [{\citenamefont {Hait}\ and\ \citenamefont
  {Head-Gordon}(2018)}]{delocal}%
  \BibitemOpen
  \bibfield  {author} {\bibinfo {author} {\bibfnamefont {D.}~\bibnamefont
  {Hait}}\ and\ \bibinfo {author} {\bibfnamefont {M.}~\bibnamefont
  {Head-Gordon}},\ }\bibfield  {title} {\bibinfo {title} {{Delocalization
  Errors in Density Functional Theory Are Essentially Quadratic in Fractional
  Occupation Number}},\ }\href {https://doi.org/10.1021/acs.jpclett.8b02417}
  {\bibfield  {journal} {\bibinfo  {journal} {The Journal of Physical Chemistry
  Letters}\ }\textbf {\bibinfo {volume} {9}},\ \bibinfo {pages} {6280}
  (\bibinfo {year} {2018})}\BibitemShut {NoStop}%
\bibitem [{\citenamefont {Kaplan}\ \emph {et~al.}(2023)\citenamefont {Kaplan},
  \citenamefont {Levy},\ and\ \citenamefont {Perdew}}]{meta-gga-gen}%
  \BibitemOpen
  \bibfield  {author} {\bibinfo {author} {\bibfnamefont {A.~D.}\ \bibnamefont
  {Kaplan}}, \bibinfo {author} {\bibfnamefont {M.}~\bibnamefont {Levy}},\ and\
  \bibinfo {author} {\bibfnamefont {J.~P.}\ \bibnamefont {Perdew}},\ }\bibfield
   {title} {\bibinfo {title} {{The Predictive Power of Exact Constraints and
  Appropriate Norms in Density Functional Theory}},\ }\href
  {https://doi.org/https://doi.org/10.1146/annurev-physchem-062422-013259}
  {\bibfield  {journal} {\bibinfo  {journal} {Annual Review of Physical
  Chemistry}\ }\textbf {\bibinfo {volume} {74}},\ \bibinfo {pages} {193}
  (\bibinfo {year} {2023})}\BibitemShut {NoStop}%
\bibitem [{\citenamefont {Furness}\ \emph {et~al.}(2020)\citenamefont
  {Furness}, \citenamefont {Kaplan}, \citenamefont {Ning}, \citenamefont
  {Perdew},\ and\ \citenamefont {Sun}}]{r2SCAN}%
  \BibitemOpen
  \bibfield  {author} {\bibinfo {author} {\bibfnamefont {J.~W.}\ \bibnamefont
  {Furness}}, \bibinfo {author} {\bibfnamefont {A.~D.}\ \bibnamefont {Kaplan}},
  \bibinfo {author} {\bibfnamefont {J.}~\bibnamefont {Ning}}, \bibinfo {author}
  {\bibfnamefont {J.~P.}\ \bibnamefont {Perdew}},\ and\ \bibinfo {author}
  {\bibfnamefont {J.}~\bibnamefont {Sun}},\ }\bibfield  {title} {\bibinfo
  {title} {{Accurate and Numerically Efficient r2SCAN Meta-Generalized Gradient
  Approximation}},\ }\href {https://doi.org/10.1021/acs.jpclett.0c02405}
  {\bibfield  {journal} {\bibinfo  {journal} {The Journal of Physical Chemistry
  Letters}\ }\textbf {\bibinfo {volume} {11}},\ \bibinfo {pages} {8208}
  (\bibinfo {year} {2020})},\ \bibinfo {note} {pMID: 32876454}\BibitemShut
  {NoStop}%
\bibitem [{\citenamefont {Kingsbury}\ \emph {et~al.}(2022)\citenamefont
  {Kingsbury}, \citenamefont {Gupta}, \citenamefont {Bartel}, \citenamefont
  {Munro}, \citenamefont {Dwaraknath}, \citenamefont {Horton},\ and\
  \citenamefont {Persson}}]{scan-r2scan}%
  \BibitemOpen
  \bibfield  {author} {\bibinfo {author} {\bibfnamefont {R.}~\bibnamefont
  {Kingsbury}}, \bibinfo {author} {\bibfnamefont {A.~S.}\ \bibnamefont
  {Gupta}}, \bibinfo {author} {\bibfnamefont {C.~J.}\ \bibnamefont {Bartel}},
  \bibinfo {author} {\bibfnamefont {J.~M.}\ \bibnamefont {Munro}}, \bibinfo
  {author} {\bibfnamefont {S.}~\bibnamefont {Dwaraknath}}, \bibinfo {author}
  {\bibfnamefont {M.}~\bibnamefont {Horton}},\ and\ \bibinfo {author}
  {\bibfnamefont {K.~A.}\ \bibnamefont {Persson}},\ }\bibfield  {title}
  {\bibinfo {title} {{Performance comparison of ${r}^{2}\mathrm{SCAN}$ and SCAN
  metaGGA density functionals for solid materials via an automated,
  high-throughput computational workflow}},\ }\href
  {https://doi.org/10.1103/PhysRevMaterials.6.013801} {\bibfield  {journal}
  {\bibinfo  {journal} {Phys. Rev. Mater.}\ }\textbf {\bibinfo {volume} {6}},\
  \bibinfo {pages} {013801} (\bibinfo {year} {2022})}\BibitemShut {NoStop}%
\bibitem [{\citenamefont {Cohen}\ \emph {et~al.}(2012)\citenamefont {Cohen},
  \citenamefont {Mori-Sánchez},\ and\ \citenamefont {Yang}}]{Yang2012}%
  \BibitemOpen
  \bibfield  {author} {\bibinfo {author} {\bibfnamefont {A.~J.}\ \bibnamefont
  {Cohen}}, \bibinfo {author} {\bibfnamefont {P.}~\bibnamefont
  {Mori-Sánchez}},\ and\ \bibinfo {author} {\bibfnamefont {W.}~\bibnamefont
  {Yang}},\ }\bibfield  {title} {\bibinfo {title} {{Challenges for Density
  Functional Theory}},\ }\href {https://doi.org/10.1021/cr200107z} {\bibfield
  {journal} {\bibinfo  {journal} {Chemical Reviews}\ }\textbf {\bibinfo
  {volume} {112}},\ \bibinfo {pages} {289} (\bibinfo {year} {2012})},\ \bibinfo
  {note} {pMID: 22191548}\BibitemShut {NoStop}%
\bibitem [{\citenamefont {Mej\'{\i}a-Rodr\'{\i}guez}\ and\ \citenamefont
  {Trickey}(2019)}]{SCAN_Fe1}%
  \BibitemOpen
  \bibfield  {author} {\bibinfo {author} {\bibfnamefont {D.}~\bibnamefont
  {Mej\'{\i}a-Rodr\'{\i}guez}}\ and\ \bibinfo {author} {\bibfnamefont {S.~B.}\
  \bibnamefont {Trickey}},\ }\bibfield  {title} {\bibinfo {title} {{Analysis of
  over-magnetization of elemental transition metal solids from the SCAN density
  functional}},\ }\href {https://doi.org/10.1103/PhysRevB.100.041113}
  {\bibfield  {journal} {\bibinfo  {journal} {Phys. Rev. B}\ }\textbf {\bibinfo
  {volume} {100}},\ \bibinfo {pages} {041113} (\bibinfo {year}
  {2019})}\BibitemShut {NoStop}%
\bibitem [{\citenamefont {Fu}\ and\ \citenamefont {Singh}(2019)}]{SCAN_Fe2}%
  \BibitemOpen
  \bibfield  {author} {\bibinfo {author} {\bibfnamefont {Y.}~\bibnamefont
  {Fu}}\ and\ \bibinfo {author} {\bibfnamefont {D.~J.}\ \bibnamefont {Singh}},\
  }\bibfield  {title} {\bibinfo {title} {{Density functional methods for the
  magnetism of transition metals: SCAN in relation to other functionals}},\
  }\href {https://doi.org/10.1103/PhysRevB.100.045126} {\bibfield  {journal}
  {\bibinfo  {journal} {Phys. Rev. B}\ }\textbf {\bibinfo {volume} {100}},\
  \bibinfo {pages} {045126} (\bibinfo {year} {2019})}\BibitemShut {NoStop}%
\bibitem [{\citenamefont {Liu}\ \emph {et~al.}(2024)\citenamefont {Liu},
  \citenamefont {Bai}, \citenamefont {Ning}, \citenamefont {Hou}, \citenamefont
  {Song}, \citenamefont {Ramasamy}, \citenamefont {Zhang}, \citenamefont {Li},
  \citenamefont {Sun},\ and\ \citenamefont {Xiao}}]{r2scan_3d_4d_5d}%
  \BibitemOpen
  \bibfield  {author} {\bibinfo {author} {\bibfnamefont {H.}~\bibnamefont
  {Liu}}, \bibinfo {author} {\bibfnamefont {X.}~\bibnamefont {Bai}}, \bibinfo
  {author} {\bibfnamefont {J.}~\bibnamefont {Ning}}, \bibinfo {author}
  {\bibfnamefont {Y.}~\bibnamefont {Hou}}, \bibinfo {author} {\bibfnamefont
  {Z.}~\bibnamefont {Song}}, \bibinfo {author} {\bibfnamefont {A.}~\bibnamefont
  {Ramasamy}}, \bibinfo {author} {\bibfnamefont {R.}~\bibnamefont {Zhang}},
  \bibinfo {author} {\bibfnamefont {Y.}~\bibnamefont {Li}}, \bibinfo {author}
  {\bibfnamefont {J.}~\bibnamefont {Sun}},\ and\ \bibinfo {author}
  {\bibfnamefont {B.}~\bibnamefont {Xiao}},\ }\bibfield  {title} {\bibinfo
  {title} {{Assessing r2SCAN meta-GGA functional for structural parameters,
  cohesive energy, mechanical modulus, and thermophysical properties of 3d, 4d,
  and 5d transition metals}},\ }\href {https://doi.org/10.1063/5.0176415}
  {\bibfield  {journal} {\bibinfo  {journal} {The Journal of Chemical Physics}\
  }\textbf {\bibinfo {volume} {160}},\ \bibinfo {pages} {024102} (\bibinfo
  {year} {2024})}\BibitemShut {NoStop}%
\bibitem [{\citenamefont {Swathilakshmi}\ \emph {et~al.}(2023)\citenamefont
  {Swathilakshmi}, \citenamefont {Devi},\ and\ \citenamefont
  {Sai~Gautam}}]{SCAN_TMO}%
  \BibitemOpen
  \bibfield  {author} {\bibinfo {author} {\bibfnamefont {S.}~\bibnamefont
  {Swathilakshmi}}, \bibinfo {author} {\bibfnamefont {R.}~\bibnamefont
  {Devi}},\ and\ \bibinfo {author} {\bibfnamefont {G.}~\bibnamefont
  {Sai~Gautam}},\ }\bibfield  {title} {\bibinfo {title} {{Performance of the
  r2SCAN Functional in Transition Metal Oxides}},\ }\href
  {https://doi.org/10.1021/acs.jctc.3c00030} {\bibfield  {journal} {\bibinfo
  {journal} {Journal of Chemical Theory and Computation}\ }\textbf {\bibinfo
  {volume} {19}},\ \bibinfo {pages} {4202} (\bibinfo {year} {2023})},\ \bibinfo
  {note} {pMID: 37329316}\BibitemShut {NoStop}%
\bibitem [{\citenamefont {Mosleh}\ and\ \citenamefont
  {Alaei}(2023)}]{Mosleh2023}%
  \BibitemOpen
  \bibfield  {author} {\bibinfo {author} {\bibfnamefont {Z.}~\bibnamefont
  {Mosleh}}\ and\ \bibinfo {author} {\bibfnamefont {M.}~\bibnamefont {Alaei}},\
  }\bibfield  {title} {\bibinfo {title} {{Benchmarking density functional
  theory on the prediction of antiferromagnetic transition temperatures}},\
  }\href {https://doi.org/10.1103/PhysRevB.108.144413} {\bibfield  {journal}
  {\bibinfo  {journal} {Phys. Rev. B}\ }\textbf {\bibinfo {volume} {108}},\
  \bibinfo {pages} {144413} (\bibinfo {year} {2023})}\BibitemShut {NoStop}%
\bibitem [{\citenamefont {Perdew}\ and\ \citenamefont
  {Schmidt}(2001)}]{Jacob-ladder}%
  \BibitemOpen
  \bibfield  {author} {\bibinfo {author} {\bibfnamefont {J.~P.}\ \bibnamefont
  {Perdew}}\ and\ \bibinfo {author} {\bibfnamefont {K.}~\bibnamefont
  {Schmidt}},\ }\bibfield  {title} {\bibinfo {title} {{Jacob’s ladder of
  density functional approximations for the exchange-correlation energy}},\
  }\href {https://doi.org/10.1063/1.1390175} {\bibfield  {journal} {\bibinfo
  {journal} {AIP Conference Proceedings}\ }\textbf {\bibinfo {volume} {577}},\
  \bibinfo {pages} {1} (\bibinfo {year} {2001})}\BibitemShut {NoStop}%
\bibitem [{\citenamefont {Alaei}\ and\ \citenamefont
  {Oganov}(2024)}]{superhex}%
  \BibitemOpen
  \bibfield  {author} {\bibinfo {author} {\bibfnamefont {M.}~\bibnamefont
  {Alaei}}\ and\ \bibinfo {author} {\bibfnamefont {A.~R.}\ \bibnamefont
  {Oganov}},\ }\href {https://arxiv.org/abs/2410.14356} {\bibinfo {title}
  {{Optimizing Supercell Structures for Heisenberg Exchange Interaction
  Calculations}}} (\bibinfo {year} {2024}),\ \Eprint
  {https://arxiv.org/abs/2410.14356} {arXiv:2410.14356 [cond-mat.mtrl-sci]}
  \BibitemShut {NoStop}%
\bibitem [{\citenamefont {Kresse}\ and\ \citenamefont {Joubert}(1999)}]{vasp}%
  \BibitemOpen
  \bibfield  {author} {\bibinfo {author} {\bibfnamefont {G.}~\bibnamefont
  {Kresse}}\ and\ \bibinfo {author} {\bibfnamefont {D.}~\bibnamefont
  {Joubert}},\ }\bibfield  {title} {\bibinfo {title} {{From ultrasoft
  pseudopotentials to the projector augmented-wave method}},\ }\href
  {https://doi.org/10.1103/PhysRevB.59.1758} {\bibfield  {journal} {\bibinfo
  {journal} {Phys. Rev. B}\ }\textbf {\bibinfo {volume} {59}},\ \bibinfo
  {pages} {1758} (\bibinfo {year} {1999})}\BibitemShut {NoStop}%
\bibitem [{\citenamefont {Rezaei}\ \emph {et~al.}(2019)\citenamefont {Rezaei},
  \citenamefont {Hashemifar}, \citenamefont {Alaei}, \citenamefont {Shahbazi},
  \citenamefont {Hashemifar},\ and\ \citenamefont {Akbarzadeh}}]{Nafise2019}%
  \BibitemOpen
  \bibfield  {author} {\bibinfo {author} {\bibfnamefont {N.}~\bibnamefont
  {Rezaei}}, \bibinfo {author} {\bibfnamefont {T.}~\bibnamefont {Hashemifar}},
  \bibinfo {author} {\bibfnamefont {M.}~\bibnamefont {Alaei}}, \bibinfo
  {author} {\bibfnamefont {F.}~\bibnamefont {Shahbazi}}, \bibinfo {author}
  {\bibfnamefont {S.~J.}\ \bibnamefont {Hashemifar}},\ and\ \bibinfo {author}
  {\bibfnamefont {H.}~\bibnamefont {Akbarzadeh}},\ }\bibfield  {title}
  {\bibinfo {title} {{Ab initio investigation of magnetic ordering in the
  double perovskite ${\mathrm{Sr}}_{2}{\mathrm{NiWO}}_{6}$}},\ }\href
  {https://doi.org/10.1103/PhysRevB.99.104411} {\bibfield  {journal} {\bibinfo
  {journal} {Phys. Rev. B}\ }\textbf {\bibinfo {volume} {99}},\ \bibinfo
  {pages} {104411} (\bibinfo {year} {2019})}\BibitemShut {NoStop}%
\bibitem [{\citenamefont {Alaei}\ and\ \citenamefont
  {Karimi}(2023)}]{Alaei2023}%
  \BibitemOpen
  \bibfield  {author} {\bibinfo {author} {\bibfnamefont {M.}~\bibnamefont
  {Alaei}}\ and\ \bibinfo {author} {\bibfnamefont {H.}~\bibnamefont {Karimi}},\
  }\bibfield  {title} {\bibinfo {title} {{A deep investigation of NiO and MnO
  through the first principle calculations and Monte Carlo simulations}},\
  }\href {https://doi.org/10.1088/2516-1075/acbff8} {\bibfield  {journal}
  {\bibinfo  {journal} {Electronic Structure}\ }\textbf {\bibinfo {volume}
  {5}},\ \bibinfo {pages} {025001} (\bibinfo {year} {2023})}\BibitemShut
  {NoStop}%
\bibitem [{\citenamefont {Rezaei}\ \emph {et~al.}(2022)\citenamefont {Rezaei},
  \citenamefont {Alaei},\ and\ \citenamefont {Akbarzadeh}}]{ESpinS}%
  \BibitemOpen
  \bibfield  {author} {\bibinfo {author} {\bibfnamefont {N.}~\bibnamefont
  {Rezaei}}, \bibinfo {author} {\bibfnamefont {M.}~\bibnamefont {Alaei}},\ and\
  \bibinfo {author} {\bibfnamefont {H.}~\bibnamefont {Akbarzadeh}},\ }\bibfield
   {title} {\bibinfo {title} {{ESpinS: A program for classical Monte-Carlo
  simulations of spin systems}},\ }\href
  {https://doi.org/https://doi.org/10.1016/j.commatsci.2021.110947} {\bibfield
  {journal} {\bibinfo  {journal} {Computational Materials Science}\ }\textbf
  {\bibinfo {volume} {202}},\ \bibinfo {pages} {110947} (\bibinfo {year}
  {2022})}\BibitemShut {NoStop}%
\bibitem [{\citenamefont {Blake}\ \emph {et~al.}(2002)\citenamefont {Blake},
  \citenamefont {Palstra}, \citenamefont {Ren}, \citenamefont {Nugroho},\ and\
  \citenamefont {Menovsky}}]{YVO3-S}%
  \BibitemOpen
  \bibfield  {author} {\bibinfo {author} {\bibfnamefont {G.~R.}\ \bibnamefont
  {Blake}}, \bibinfo {author} {\bibfnamefont {T.~T.~M.}\ \bibnamefont
  {Palstra}}, \bibinfo {author} {\bibfnamefont {Y.}~\bibnamefont {Ren}},
  \bibinfo {author} {\bibfnamefont {A.~A.}\ \bibnamefont {Nugroho}},\ and\
  \bibinfo {author} {\bibfnamefont {A.~A.}\ \bibnamefont {Menovsky}},\
  }\bibfield  {title} {\bibinfo {title} {{Neutron diffraction, x-ray
  diffraction, and specific heat studies of orbital ordering in
  ${\mathrm{YVO}}_{3}$}},\ }\href {https://doi.org/10.1103/PhysRevB.65.174112}
  {\bibfield  {journal} {\bibinfo  {journal} {Phys. Rev. B}\ }\textbf {\bibinfo
  {volume} {65}},\ \bibinfo {pages} {174112} (\bibinfo {year}
  {2002})}\BibitemShut {NoStop}%
\bibitem [{\citenamefont {Kjekshus}\ \emph {et~al.}(1979)\citenamefont
  {Kjekshus}, \citenamefont {Peterzens}, \citenamefont {Rakke},\ and\
  \citenamefont {Andresen}}]{CrSb2-S}%
  \BibitemOpen
  \bibfield  {author} {\bibinfo {author} {\bibfnamefont {A.}~\bibnamefont
  {Kjekshus}}, \bibinfo {author} {\bibfnamefont {P.~G.}\ \bibnamefont
  {Peterzens}}, \bibinfo {author} {\bibfnamefont {T.}~\bibnamefont {Rakke}},\
  and\ \bibinfo {author} {\bibfnamefont {A.~F.}\ \bibnamefont {Andresen}},\
  }\bibfield  {title} {\bibinfo {title} {{Compounds with the marcasite type
  crystal structure. XIII. Structural and magnetic properties of
  Cr\textsubscript{t}Fe\textsubscript{1-t}As\textsubscript{2},
  Cr\textsubscript{t}Fe\textsubscript{1-t}Sb\textsubscript{2},
  Fe\textsubscript{1-t}Ni\textsubscript{t}As\textsubscript{2}, and
  Fe\textsubscript{1-t}Ni\textsubscript{t}Sb\textsubscript{2}}},\ }\href
  {http://actachemscand.org/doi/10.3891/acta.chem.scand.33a-0469} {\bibfield
  {journal} {\bibinfo  {journal} {Acta Chemica Scandinavica A}\ }\textbf
  {\bibinfo {volume} {33}},\ \bibinfo {pages} {469} (\bibinfo {year}
  {1979})}\BibitemShut {NoStop}%
\bibitem [{\citenamefont {Tracy}\ \emph {et~al.}(1961)\citenamefont {Tracy},
  \citenamefont {Gregory}, \citenamefont {Lingafelter}, \citenamefont {Dunitz},
  \citenamefont {Mez}, \citenamefont {Rundle}, \citenamefont {Scheringer},
  \citenamefont {Yakel},\ and\ \citenamefont {Wilkinson}}]{CrCl2-S}%
  \BibitemOpen
  \bibfield  {author} {\bibinfo {author} {\bibfnamefont {J.~W.}\ \bibnamefont
  {Tracy}}, \bibinfo {author} {\bibfnamefont {N.}~\bibnamefont {Gregory}},
  \bibinfo {author} {\bibfnamefont {E.}~\bibnamefont {Lingafelter}}, \bibinfo
  {author} {\bibfnamefont {J.}~\bibnamefont {Dunitz}}, \bibinfo {author}
  {\bibfnamefont {H.-C.}\ \bibnamefont {Mez}}, \bibinfo {author} {\bibfnamefont
  {R.}~\bibnamefont {Rundle}}, \bibinfo {author} {\bibfnamefont
  {C.}~\bibnamefont {Scheringer}}, \bibinfo {author} {\bibfnamefont
  {H.}~\bibnamefont {Yakel}},\ and\ \bibinfo {author} {\bibfnamefont
  {M.}~\bibnamefont {Wilkinson}},\ }\bibfield  {title} {\bibinfo {title} {{The
  crystal structure of chromium (II) chloride}},\ }\href
  {https://doi.org/10.1107/S0365110X61002710} {\bibfield  {journal} {\bibinfo
  {journal} {Acta Crystallographica}\ }\textbf {\bibinfo {volume} {14}},\
  \bibinfo {pages} {927} (\bibinfo {year} {1961})}\BibitemShut {NoStop}%
\bibitem [{\citenamefont {Wyckoff}(1963)}]{CrF2-S}%
  \BibitemOpen
  \bibfield  {author} {\bibinfo {author} {\bibfnamefont {R.~W.~G.}\
  \bibnamefont {Wyckoff}},\ }\href@noop {} {\emph {\bibinfo {title} {{Crystal
  Structures}}}},\ \bibinfo {edition} {2nd}\ ed.\ (\bibinfo  {publisher}
  {Interscience Publishers},\ \bibinfo {address} {New York},\ \bibinfo {year}
  {1963})\BibitemShut {NoStop}%
\bibitem [{Cr2(1962)}]{Cr2O3-S}%
  \BibitemOpen
  \bibfield  {title} {\bibinfo {title} {{Refinement of the $\alpha$ Al2O3,
  Ti2O3, V2O3 and Cr2O3 structures*}},\ }\href
  {https://doi.org/doi:10.1524/zkri.1962.117.2-3.235} {\bibfield  {journal}
  {\bibinfo  {journal} {Zeitschrift für Kristallographie}\ }\textbf {\bibinfo
  {volume} {117}},\ \bibinfo {pages} {235} (\bibinfo {year}
  {1962})}\BibitemShut {NoStop}%
\bibitem [{\citenamefont {López-Paz}\ \emph {et~al.}(2022)\citenamefont
  {López-Paz}, \citenamefont {Guguchia}, \citenamefont {Pomjakushin} \emph
  {et~al.}}]{CrSBr-ST}%
  \BibitemOpen
  \bibfield  {author} {\bibinfo {author} {\bibfnamefont {S.~A.}\ \bibnamefont
  {López-Paz}}, \bibinfo {author} {\bibfnamefont {Z.}~\bibnamefont
  {Guguchia}}, \bibinfo {author} {\bibfnamefont {V.~Y.}\ \bibnamefont
  {Pomjakushin}}, \emph {et~al.},\ }\bibfield  {title} {\bibinfo {title}
  {{Dynamic magnetic crossover at the origin of the hidden-order in van der
  Waals antiferromagnet CrSBr}},\ }\href
  {https://doi.org/10.1038/s41467-022-32290-4} {\bibfield  {journal} {\bibinfo
  {journal} {Nature Communications}\ }\textbf {\bibinfo {volume} {13}},\
  \bibinfo {pages} {4745} (\bibinfo {year} {2022})}\BibitemShut {NoStop}%
\bibitem [{\citenamefont {Kunnmann}\ \emph {et~al.}(1968)\citenamefont
  {Kunnmann}, \citenamefont {{La Placa}}, \citenamefont {Corliss},
  \citenamefont {Hastings},\ and\ \citenamefont {Banks}}]{Cr(Fe)2Te(W)O6-S}%
  \BibitemOpen
  \bibfield  {author} {\bibinfo {author} {\bibfnamefont {W.}~\bibnamefont
  {Kunnmann}}, \bibinfo {author} {\bibfnamefont {S.}~\bibnamefont {{La
  Placa}}}, \bibinfo {author} {\bibfnamefont {L.}~\bibnamefont {Corliss}},
  \bibinfo {author} {\bibfnamefont {J.}~\bibnamefont {Hastings}},\ and\
  \bibinfo {author} {\bibfnamefont {E.}~\bibnamefont {Banks}},\ }\bibfield
  {title} {\bibinfo {title} {{Magnetic structures of the ordered trirutiles
  Cr2WO6, Cr2TeO6 and Fe2TeO6}},\ }\href
  {https://doi.org/https://doi.org/10.1016/0022-3697(68)90187-X} {\bibfield
  {journal} {\bibinfo  {journal} {Journal of Physics and Chemistry of Solids}\
  }\textbf {\bibinfo {volume} {29}},\ \bibinfo {pages} {1359} (\bibinfo {year}
  {1968})}\BibitemShut {NoStop}%
\bibitem [{\citenamefont {Zhang}(1999)}]{MnO-S}%
  \BibitemOpen
  \bibfield  {author} {\bibinfo {author} {\bibfnamefont {J.}~\bibnamefont
  {Zhang}},\ }\bibfield  {title} {\bibinfo {title} {{Room-temperature
  compressibilities of MnO and CdO: further examination of the role of cation
  type in bulk modulus systematics}},\ }\href
  {https://doi.org/10.1007/s002690050229} {\bibfield  {journal} {\bibinfo
  {journal} {Physics and Chemistry of Minerals}\ }\textbf {\bibinfo {volume}
  {26}},\ \bibinfo {pages} {644} (\bibinfo {year} {1999})}\BibitemShut
  {NoStop}%
\bibitem [{\citenamefont {Clark}\ \emph {et~al.}(2021)\citenamefont {Clark},
  \citenamefont {Yannello}, \citenamefont {Samarakoon}, \citenamefont {Ross},
  \citenamefont {Uible}, \citenamefont {Garlea},\ and\ \citenamefont
  {Shatruk}}]{MnS-ST}%
  \BibitemOpen
  \bibfield  {author} {\bibinfo {author} {\bibfnamefont {J.~K.}\ \bibnamefont
  {Clark}}, \bibinfo {author} {\bibfnamefont {V.}~\bibnamefont {Yannello}},
  \bibinfo {author} {\bibfnamefont {A.~M.}\ \bibnamefont {Samarakoon}},
  \bibinfo {author} {\bibfnamefont {C.}~\bibnamefont {Ross}}, \bibinfo {author}
  {\bibfnamefont {M.~C.}\ \bibnamefont {Uible}}, \bibinfo {author}
  {\bibfnamefont {V.~O.}\ \bibnamefont {Garlea}},\ and\ \bibinfo {author}
  {\bibfnamefont {M.}~\bibnamefont {Shatruk}},\ }\bibfield  {title} {\bibinfo
  {title} {{Inelastic Neutron Scattering Study of Magnetic Exchange Pathways in
  MnS}},\ }\href {https://doi.org/10.1021/acs.jpcc.1c02956} {\bibfield
  {journal} {\bibinfo  {journal} {The Journal of Physical Chemistry C}\
  }\textbf {\bibinfo {volume} {125}},\ \bibinfo {pages} {16183} (\bibinfo
  {year} {2021})}\BibitemShut {NoStop}%
\bibitem [{\citenamefont {Peng}\ \emph {et~al.}(2002)\citenamefont {Peng},
  \citenamefont {Dong}, \citenamefont {Deng}, \citenamefont {Kou},
  \citenamefont {Gao},\ and\ \citenamefont {Li}}]{MnSe-S}%
  \BibitemOpen
  \bibfield  {author} {\bibinfo {author} {\bibfnamefont {Q.}~\bibnamefont
  {Peng}}, \bibinfo {author} {\bibfnamefont {Y.}~\bibnamefont {Dong}}, \bibinfo
  {author} {\bibfnamefont {Z.}~\bibnamefont {Deng}}, \bibinfo {author}
  {\bibfnamefont {H.}~\bibnamefont {Kou}}, \bibinfo {author} {\bibfnamefont
  {S.}~\bibnamefont {Gao}},\ and\ \bibinfo {author} {\bibfnamefont
  {Y.}~\bibnamefont {Li}},\ }\bibfield  {title} {\bibinfo {title} {{Selective
  Synthesis and Magnetic Properties of $\alpha$-MnSe and MnSe2 Uniform
  Microcrystals}},\ }\href {https://doi.org/10.1021/jp020635f} {\bibfield
  {journal} {\bibinfo  {journal} {The Journal of Physical Chemistry B}\
  }\textbf {\bibinfo {volume} {106}},\ \bibinfo {pages} {9261} (\bibinfo {year}
  {2002})}\BibitemShut {NoStop}%
\bibitem [{\citenamefont {Grønvold}\ \emph {et~al.}(1972)\citenamefont
  {Grønvold}, \citenamefont {Kveseth}, \citenamefont {Marques},\ and\
  \citenamefont {Tichy}}]{MnTe-S}%
  \BibitemOpen
  \bibfield  {author} {\bibinfo {author} {\bibfnamefont {F.}~\bibnamefont
  {Grønvold}}, \bibinfo {author} {\bibfnamefont {N.~J.}\ \bibnamefont
  {Kveseth}}, \bibinfo {author} {\bibfnamefont {F.~D.~S.}\ \bibnamefont
  {Marques}},\ and\ \bibinfo {author} {\bibfnamefont {J.}~\bibnamefont
  {Tichy}},\ }\bibfield  {title} {\bibinfo {title} {{Thermophysical properties
  of manganese monotelluride from 298 to 700 K. Lattice constants, magnetic
  susceptibility, and antiferromagnetic transition}},\ }\href
  {https://doi.org/https://doi.org/10.1016/0021-9614(72)90001-8} {\bibfield
  {journal} {\bibinfo  {journal} {The Journal of Chemical Thermodynamics}\
  }\textbf {\bibinfo {volume} {4}},\ \bibinfo {pages} {795} (\bibinfo {year}
  {1972})}\BibitemShut {NoStop}%
\bibitem [{\citenamefont {Kondrashev}\ and\ \citenamefont
  {Zaslavskii}(1951)}]{MnO2-S}%
  \BibitemOpen
  \bibfield  {author} {\bibinfo {author} {\bibfnamefont {Y.~D.}\ \bibnamefont
  {Kondrashev}}\ and\ \bibinfo {author} {\bibfnamefont {A.}~\bibnamefont
  {Zaslavskii}},\ }\bibfield  {title} {\bibinfo {title} {{The structure of the
  modifications of manganese (IV) oxide}},\ }\href@noop {} {\bibfield
  {journal} {\bibinfo  {journal} {Izvestiya Akademii Nauk SSSR, Seriya
  Fizicheskaya}\ }\textbf {\bibinfo {volume} {15}},\ \bibinfo {pages} {179}
  (\bibinfo {year} {1951})}\BibitemShut {NoStop}%
\bibitem [{\citenamefont {Li}\ \emph {et~al.}(2009{\natexlab{a}})\citenamefont
  {Li}, \citenamefont {Lu}, \citenamefont {Peng}, \citenamefont {Jin},\ and\
  \citenamefont {Wei}}]{MnF2-S}%
  \BibitemOpen
  \bibfield  {author} {\bibinfo {author} {\bibfnamefont {X.}~\bibnamefont
  {Li}}, \bibinfo {author} {\bibfnamefont {J.}~\bibnamefont {Lu}}, \bibinfo
  {author} {\bibfnamefont {G.}~\bibnamefont {Peng}}, \bibinfo {author}
  {\bibfnamefont {L.}~\bibnamefont {Jin}},\ and\ \bibinfo {author}
  {\bibfnamefont {S.}~\bibnamefont {Wei}},\ }\bibfield  {title} {\bibinfo
  {title} {{Solvothermal synthesis of MnF2 nanocrystals and the first-principle
  study of its electronic structure}},\ }\href
  {https://doi.org/https://doi.org/10.1016/j.jpcs.2009.01.004} {\bibfield
  {journal} {\bibinfo  {journal} {Journal of Physics and Chemistry of Solids}\
  }\textbf {\bibinfo {volume} {70}},\ \bibinfo {pages} {609} (\bibinfo {year}
  {2009}{\natexlab{a}})}\BibitemShut {NoStop}%
\bibitem [{MnS(1992)}]{MnS2-S}%
  \BibitemOpen
  \bibfield  {title} {\bibinfo {title} {{X-ray and neutron diffraction study of
  the crystal structure of MnS2}},\ }\href
  {https://doi.org/doi:10.1524/zkri.1992.199.14.13} {\bibfield  {journal}
  {\bibinfo  {journal} {Zeitschrift für Kristallographie - Crystalline
  Materials}\ }\textbf {\bibinfo {volume} {199}},\ \bibinfo {pages} {13}
  (\bibinfo {year} {1992})}\BibitemShut {NoStop}%
\bibitem [{\citenamefont {Burlet}\ \emph {et~al.}(1997)\citenamefont {Burlet},
  \citenamefont {Ressouche}, \citenamefont {Malaman}, \citenamefont {Welter},
  \citenamefont {Sanchez},\ and\ \citenamefont {Vulliet}}]{MnTe2-ST1}%
  \BibitemOpen
  \bibfield  {author} {\bibinfo {author} {\bibfnamefont {P.}~\bibnamefont
  {Burlet}}, \bibinfo {author} {\bibfnamefont {E.}~\bibnamefont {Ressouche}},
  \bibinfo {author} {\bibfnamefont {B.}~\bibnamefont {Malaman}}, \bibinfo
  {author} {\bibfnamefont {R.}~\bibnamefont {Welter}}, \bibinfo {author}
  {\bibfnamefont {J.~P.}\ \bibnamefont {Sanchez}},\ and\ \bibinfo {author}
  {\bibfnamefont {P.}~\bibnamefont {Vulliet}},\ }\bibfield  {title} {\bibinfo
  {title} {{Noncollinear magnetic structure of ${\mathrm{MnTe}}_{2}$}},\ }\href
  {https://doi.org/10.1103/PhysRevB.56.14013} {\bibfield  {journal} {\bibinfo
  {journal} {Phys. Rev. B}\ }\textbf {\bibinfo {volume} {56}},\ \bibinfo
  {pages} {14013} (\bibinfo {year} {1997})}\BibitemShut {NoStop}%
\bibitem [{\citenamefont {Galakhov}\ \emph {et~al.}(2000)\citenamefont
  {Galakhov}, \citenamefont {Korotin}, \citenamefont {Ovechkina}, \citenamefont
  {Kurmaev}, \citenamefont {Gorshkov}, \citenamefont {Kellerman}, \citenamefont
  {Bartkowski},\ and\ \citenamefont {Neumann}}]{LiMnO2-S}%
  \BibitemOpen
  \bibfield  {author} {\bibinfo {author} {\bibfnamefont {V.}~\bibnamefont
  {Galakhov}}, \bibinfo {author} {\bibfnamefont {M.}~\bibnamefont {Korotin}},
  \bibinfo {author} {\bibfnamefont {N.}~\bibnamefont {Ovechkina}}, \bibinfo
  {author} {\bibfnamefont {E.}~\bibnamefont {Kurmaev}}, \bibinfo {author}
  {\bibfnamefont {V.}~\bibnamefont {Gorshkov}}, \bibinfo {author}
  {\bibfnamefont {D.}~\bibnamefont {Kellerman}}, \bibinfo {author}
  {\bibfnamefont {S.}~\bibnamefont {Bartkowski}},\ and\ \bibinfo {author}
  {\bibfnamefont {M.}~\bibnamefont {Neumann}},\ }\bibfield  {title} {\bibinfo
  {title} {{Electronic structure of LiMnO: X-ray emission and photoelectron
  spectra and band structure calculations}},\ }\href
  {https://doi.org/10.1007/s100510050130} {\bibfield  {journal} {\bibinfo
  {journal} {The European Physical Journal B - Condensed Matter and Complex
  Systems}\ }\textbf {\bibinfo {volume} {14}},\ \bibinfo {pages} {281}
  (\bibinfo {year} {2000})}\BibitemShut {NoStop}%
\bibitem [{\citenamefont {Daoud-Aladine}\ \emph {et~al.}(2007)\citenamefont
  {Daoud-Aladine}, \citenamefont {Martin}, \citenamefont {Chapon},
  \citenamefont {Hervieu}, \citenamefont {Knight}, \citenamefont {Brunelli},\
  and\ \citenamefont {Radaelli}}]{SrMnO3-ST}%
  \BibitemOpen
  \bibfield  {author} {\bibinfo {author} {\bibfnamefont {A.}~\bibnamefont
  {Daoud-Aladine}}, \bibinfo {author} {\bibfnamefont {C.}~\bibnamefont
  {Martin}}, \bibinfo {author} {\bibfnamefont {L.~C.}\ \bibnamefont {Chapon}},
  \bibinfo {author} {\bibfnamefont {M.}~\bibnamefont {Hervieu}}, \bibinfo
  {author} {\bibfnamefont {K.~S.}\ \bibnamefont {Knight}}, \bibinfo {author}
  {\bibfnamefont {M.}~\bibnamefont {Brunelli}},\ and\ \bibinfo {author}
  {\bibfnamefont {P.~G.}\ \bibnamefont {Radaelli}},\ }\bibfield  {title}
  {\bibinfo {title} {{Structural phase transition and magnetism in hexagonal
  $\mathrm{Sr}\mathrm{Mn}{\mathrm{O}}_{3}$ by magnetization measurements and by
  electron, x-ray, and neutron diffraction studies}},\ }\href
  {https://doi.org/10.1103/PhysRevB.75.104417} {\bibfield  {journal} {\bibinfo
  {journal} {Phys. Rev. B}\ }\textbf {\bibinfo {volume} {75}},\ \bibinfo
  {pages} {104417} (\bibinfo {year} {2007})}\BibitemShut {NoStop}%
\bibitem [{\citenamefont {Knight}\ \emph {et~al.}(2020)\citenamefont {Knight},
  \citenamefont {Khalyavin}, \citenamefont {Manuel}, \citenamefont {Bull},\
  and\ \citenamefont {McIntyre}}]{KMnF3-ST1}%
  \BibitemOpen
  \bibfield  {author} {\bibinfo {author} {\bibfnamefont {K.~S.}\ \bibnamefont
  {Knight}}, \bibinfo {author} {\bibfnamefont {D.~D.}\ \bibnamefont
  {Khalyavin}}, \bibinfo {author} {\bibfnamefont {P.}~\bibnamefont {Manuel}},
  \bibinfo {author} {\bibfnamefont {C.~L.}\ \bibnamefont {Bull}},\ and\
  \bibinfo {author} {\bibfnamefont {P.}~\bibnamefont {McIntyre}},\ }\bibfield
  {title} {\bibinfo {title} {{Nuclear and magnetic structures of KMnF3
  perovskite in the temperature interval 10 K–105 K}},\ }\href
  {https://doi.org/https://doi.org/10.1016/j.jallcom.2020.155935} {\bibfield
  {journal} {\bibinfo  {journal} {Journal of Alloys and Compounds}\ }\textbf
  {\bibinfo {volume} {842}},\ \bibinfo {pages} {155935} (\bibinfo {year}
  {2020})}\BibitemShut {NoStop}%
\bibitem [{\citenamefont {Ressouche}\ \emph {et~al.}(2010)\citenamefont
  {Ressouche}, \citenamefont {Loire}, \citenamefont {Simonet}, \citenamefont
  {Ballou}, \citenamefont {Stunault},\ and\ \citenamefont {Wildes}}]{MnPS3-S}%
  \BibitemOpen
  \bibfield  {author} {\bibinfo {author} {\bibfnamefont {E.}~\bibnamefont
  {Ressouche}}, \bibinfo {author} {\bibfnamefont {M.}~\bibnamefont {Loire}},
  \bibinfo {author} {\bibfnamefont {V.}~\bibnamefont {Simonet}}, \bibinfo
  {author} {\bibfnamefont {R.}~\bibnamefont {Ballou}}, \bibinfo {author}
  {\bibfnamefont {A.}~\bibnamefont {Stunault}},\ and\ \bibinfo {author}
  {\bibfnamefont {A.}~\bibnamefont {Wildes}},\ }\bibfield  {title} {\bibinfo
  {title} {{Magnetoelectric ${\text{MnPS}}_{3}$ as a candidate for
  ferrotoroidicity}},\ }\href {https://doi.org/10.1103/PhysRevB.82.100408}
  {\bibfield  {journal} {\bibinfo  {journal} {Phys. Rev. B}\ }\textbf {\bibinfo
  {volume} {82}},\ \bibinfo {pages} {100408} (\bibinfo {year}
  {2010})}\BibitemShut {NoStop}%
\bibitem [{\citenamefont {Calder}\ \emph {et~al.}(2021)\citenamefont {Calder},
  \citenamefont {Haglund}, \citenamefont {Kolesnikov},\ and\ \citenamefont
  {Mandrus}}]{MnPSe3-S}%
  \BibitemOpen
  \bibfield  {author} {\bibinfo {author} {\bibfnamefont {S.}~\bibnamefont
  {Calder}}, \bibinfo {author} {\bibfnamefont {A.~V.}\ \bibnamefont {Haglund}},
  \bibinfo {author} {\bibfnamefont {A.~I.}\ \bibnamefont {Kolesnikov}},\ and\
  \bibinfo {author} {\bibfnamefont {D.}~\bibnamefont {Mandrus}},\ }\bibfield
  {title} {\bibinfo {title} {{Magnetic exchange interactions in the van der
  Waals layered antiferromagnet $\mathrm{Mn}\mathrm{P}{\mathrm{Se}}_{3}$}},\
  }\href {https://doi.org/10.1103/PhysRevB.103.024414} {\bibfield  {journal}
  {\bibinfo  {journal} {Phys. Rev. B}\ }\textbf {\bibinfo {volume} {103}},\
  \bibinfo {pages} {024414} (\bibinfo {year} {2021})}\BibitemShut {NoStop}%
\bibitem [{\citenamefont {Lautenschl\"ager}\ \emph {et~al.}(1993)\citenamefont
  {Lautenschl\"ager}, \citenamefont {Weitzel}, \citenamefont {Vogt},
  \citenamefont {Hock}, \citenamefont {B\"ohm}, \citenamefont {Bonnet},\ and\
  \citenamefont {Fuess}}]{MnWO4-ST}%
  \BibitemOpen
  \bibfield  {author} {\bibinfo {author} {\bibfnamefont {G.}~\bibnamefont
  {Lautenschl\"ager}}, \bibinfo {author} {\bibfnamefont {H.}~\bibnamefont
  {Weitzel}}, \bibinfo {author} {\bibfnamefont {T.}~\bibnamefont {Vogt}},
  \bibinfo {author} {\bibfnamefont {R.}~\bibnamefont {Hock}}, \bibinfo {author}
  {\bibfnamefont {A.}~\bibnamefont {B\"ohm}}, \bibinfo {author} {\bibfnamefont
  {M.}~\bibnamefont {Bonnet}},\ and\ \bibinfo {author} {\bibfnamefont
  {H.}~\bibnamefont {Fuess}},\ }\bibfield  {title} {\bibinfo {title} {{Magnetic
  phase transitions of ${\mathrm{MnWO}}_{4}$ studied by the use of neutron
  diffraction}},\ }\href {https://doi.org/10.1103/PhysRevB.48.6087} {\bibfield
  {journal} {\bibinfo  {journal} {Phys. Rev. B}\ }\textbf {\bibinfo {volume}
  {48}},\ \bibinfo {pages} {6087} (\bibinfo {year} {1993})}\BibitemShut
  {NoStop}%
\bibitem [{\citenamefont {Strobel}\ and\ \citenamefont
  {Lambert-Andron}(1988)}]{Li2MnO3-ST}%
  \BibitemOpen
  \bibfield  {author} {\bibinfo {author} {\bibfnamefont {P.}~\bibnamefont
  {Strobel}}\ and\ \bibinfo {author} {\bibfnamefont {B.}~\bibnamefont
  {Lambert-Andron}},\ }\bibfield  {title} {\bibinfo {title} {{Crystallographic
  and magnetic structure of Li2MnO3}},\ }\href
  {https://doi.org/https://doi.org/10.1016/0022-4596(88)90305-2} {\bibfield
  {journal} {\bibinfo  {journal} {Journal of Solid State Chemistry}\ }\textbf
  {\bibinfo {volume} {75}},\ \bibinfo {pages} {90} (\bibinfo {year}
  {1988})}\BibitemShut {NoStop}%
\bibitem [{\citenamefont {Geller}\ and\ \citenamefont
  {Durand}(1960)}]{LiMnPO4-S}%
  \BibitemOpen
  \bibfield  {author} {\bibinfo {author} {\bibfnamefont {S.}~\bibnamefont
  {Geller}}\ and\ \bibinfo {author} {\bibfnamefont {J.~L.}\ \bibnamefont
  {Durand}},\ }\bibfield  {title} {\bibinfo {title} {{Refinement of the
  structure of LiMnPO${\sb 4}$}},\ }\href
  {https://doi.org/10.1107/S0365110X60002521} {\bibfield  {journal} {\bibinfo
  {journal} {Acta Crystallographica}\ }\textbf {\bibinfo {volume} {13}},\
  \bibinfo {pages} {325} (\bibinfo {year} {1960})}\BibitemShut {NoStop}%
\bibitem [{\citenamefont {Pauling}\ and\ \citenamefont
  {Hendricks}(1925)}]{Fe2O3-S}%
  \BibitemOpen
  \bibfield  {author} {\bibinfo {author} {\bibfnamefont {L.}~\bibnamefont
  {Pauling}}\ and\ \bibinfo {author} {\bibfnamefont {S.~B.}\ \bibnamefont
  {Hendricks}},\ }\bibfield  {title} {\bibinfo {title} {{THE CRYSTAL STRUCTURES
  OF HEMATITE AND CORUNDUM}},\ }\href {https://doi.org/10.1021/ja01680a027}
  {\bibfield  {journal} {\bibinfo  {journal} {Journal of the American Chemical
  Society}\ }\textbf {\bibinfo {volume} {47}},\ \bibinfo {pages} {781}
  (\bibinfo {year} {1925})}\BibitemShut {NoStop}%
\bibitem [{\citenamefont {Tsujimoto}\ \emph {et~al.}(2007)\citenamefont
  {Tsujimoto}, \citenamefont {Tassel}, \citenamefont {Hayashi}, \citenamefont
  {Watanabe}, \citenamefont {Kageyama}, \citenamefont {Yoshimura},
  \citenamefont {Takano}, \citenamefont {Ceretti}, \citenamefont {Ritter},\
  and\ \citenamefont {Paulus}}]{SrFeO2-ST}%
  \BibitemOpen
  \bibfield  {author} {\bibinfo {author} {\bibfnamefont {Y.}~\bibnamefont
  {Tsujimoto}}, \bibinfo {author} {\bibfnamefont {C.}~\bibnamefont {Tassel}},
  \bibinfo {author} {\bibfnamefont {N.}~\bibnamefont {Hayashi}}, \bibinfo
  {author} {\bibfnamefont {T.}~\bibnamefont {Watanabe}}, \bibinfo {author}
  {\bibfnamefont {H.}~\bibnamefont {Kageyama}}, \bibinfo {author}
  {\bibfnamefont {K.}~\bibnamefont {Yoshimura}}, \bibinfo {author}
  {\bibfnamefont {M.}~\bibnamefont {Takano}}, \bibinfo {author} {\bibfnamefont
  {M.}~\bibnamefont {Ceretti}}, \bibinfo {author} {\bibfnamefont
  {C.}~\bibnamefont {Ritter}},\ and\ \bibinfo {author} {\bibfnamefont
  {W.}~\bibnamefont {Paulus}},\ }\bibfield  {title} {\bibinfo {title}
  {{Infinite-layer iron oxide with a square-planar coordination}},\ }\href
  {https://doi.org/10.1038/nature06382} {\bibfield  {journal} {\bibinfo
  {journal} {Nature}\ }\textbf {\bibinfo {volume} {450}},\ \bibinfo {pages}
  {1062} (\bibinfo {year} {2007})}\BibitemShut {NoStop}%
\bibitem [{\citenamefont {Selbach}\ \emph {et~al.}(2007)\citenamefont
  {Selbach}, \citenamefont {Einarsrud}, \citenamefont {Tybell},\ and\
  \citenamefont {Grande}}]{BiFeO3-S}%
  \BibitemOpen
  \bibfield  {author} {\bibinfo {author} {\bibfnamefont {S.~M.}\ \bibnamefont
  {Selbach}}, \bibinfo {author} {\bibfnamefont {M.-A.}\ \bibnamefont
  {Einarsrud}}, \bibinfo {author} {\bibfnamefont {T.}~\bibnamefont {Tybell}},\
  and\ \bibinfo {author} {\bibfnamefont {T.}~\bibnamefont {Grande}},\
  }\bibfield  {title} {\bibinfo {title} {{Synthesis of BiFeO3 by Wet Chemical
  Methods}},\ }\href
  {https://doi.org/https://doi.org/10.1111/j.1551-2916.2007.01937.x} {\bibfield
   {journal} {\bibinfo  {journal} {Journal of the American Ceramic Society}\
  }\textbf {\bibinfo {volume} {90}},\ \bibinfo {pages} {3430} (\bibinfo {year}
  {2007})}\BibitemShut {NoStop}%
\bibitem [{\citenamefont {Sangaletti}\ \emph {et~al.}(2001)\citenamefont
  {Sangaletti}, \citenamefont {Depero}, \citenamefont {Allieri}, \citenamefont
  {Nunziante},\ and\ \citenamefont {Traversa}}]{LaFeO3-S}%
  \BibitemOpen
  \bibfield  {author} {\bibinfo {author} {\bibfnamefont {L.}~\bibnamefont
  {Sangaletti}}, \bibinfo {author} {\bibfnamefont {L.~E.}\ \bibnamefont
  {Depero}}, \bibinfo {author} {\bibfnamefont {B.}~\bibnamefont {Allieri}},
  \bibinfo {author} {\bibfnamefont {P.}~\bibnamefont {Nunziante}},\ and\
  \bibinfo {author} {\bibfnamefont {E.}~\bibnamefont {Traversa}},\ }\bibfield
  {title} {\bibinfo {title} {{An X-ray study of the trimetallic LaxSm1-xFeO3
  orthoferrites}},\ }\href
  {https://doi.org/https://doi.org/10.1016/S0955-2219(00)00267-3} {\bibfield
  {journal} {\bibinfo  {journal} {Journal of the European Ceramic Society}\
  }\textbf {\bibinfo {volume} {21}},\ \bibinfo {pages} {719} (\bibinfo {year}
  {2001})}\BibitemShut {NoStop}%
\bibitem [{\citenamefont {du~Boulay}\ \emph {et~al.}(1995)\citenamefont
  {du~Boulay}, \citenamefont {Maslen}, \citenamefont {Streltsov},\ and\
  \citenamefont {Ishizawa}}]{YFeO3-S}%
  \BibitemOpen
  \bibfield  {author} {\bibinfo {author} {\bibfnamefont {D.}~\bibnamefont
  {du~Boulay}}, \bibinfo {author} {\bibfnamefont {E.~N.}\ \bibnamefont
  {Maslen}}, \bibinfo {author} {\bibfnamefont {V.~A.}\ \bibnamefont
  {Streltsov}},\ and\ \bibinfo {author} {\bibfnamefont {N.}~\bibnamefont
  {Ishizawa}},\ }\bibfield  {title} {\bibinfo {title} {{A synchrotron X-ray
  study of the electron density in YFeO${\sb 3}$}},\ }\href
  {https://doi.org/10.1107/S0108768195004010} {\bibfield  {journal} {\bibinfo
  {journal} {Acta Crystallographica Section B}\ }\textbf {\bibinfo {volume}
  {51}},\ \bibinfo {pages} {921} (\bibinfo {year} {1995})}\BibitemShut
  {NoStop}%
\bibitem [{\citenamefont {Lan\ifmmode~\mbox{\c{c}}\else \c{c}\fi{}on}\ \emph
  {et~al.}(2016)\citenamefont {Lan\ifmmode~\mbox{\c{c}}\else \c{c}\fi{}on},
  \citenamefont {Walker}, \citenamefont {Ressouche}, \citenamefont {Ouladdiaf},
  \citenamefont {Rule}, \citenamefont {McIntyre}, \citenamefont {Hicks},
  \citenamefont {R\o{}nnow},\ and\ \citenamefont {Wildes}}]{FePS3-S}%
  \BibitemOpen
  \bibfield  {author} {\bibinfo {author} {\bibfnamefont {D.}~\bibnamefont
  {Lan\ifmmode~\mbox{\c{c}}\else \c{c}\fi{}on}}, \bibinfo {author}
  {\bibfnamefont {H.~C.}\ \bibnamefont {Walker}}, \bibinfo {author}
  {\bibfnamefont {E.}~\bibnamefont {Ressouche}}, \bibinfo {author}
  {\bibfnamefont {B.}~\bibnamefont {Ouladdiaf}}, \bibinfo {author}
  {\bibfnamefont {K.~C.}\ \bibnamefont {Rule}}, \bibinfo {author}
  {\bibfnamefont {G.~J.}\ \bibnamefont {McIntyre}}, \bibinfo {author}
  {\bibfnamefont {T.~J.}\ \bibnamefont {Hicks}}, \bibinfo {author}
  {\bibfnamefont {H.~M.}\ \bibnamefont {R\o{}nnow}},\ and\ \bibinfo {author}
  {\bibfnamefont {A.~R.}\ \bibnamefont {Wildes}},\ }\bibfield  {title}
  {\bibinfo {title} {{Magnetic structure and magnon dynamics of the
  quasi-two-dimensional antiferromagnet ${\mathrm{FePS}}_{3}$}},\ }\href
  {https://doi.org/10.1103/PhysRevB.94.214407} {\bibfield  {journal} {\bibinfo
  {journal} {Phys. Rev. B}\ }\textbf {\bibinfo {volume} {94}},\ \bibinfo
  {pages} {214407} (\bibinfo {year} {2016})}\BibitemShut {NoStop}%
\bibitem [{\citenamefont {Guo}\ \emph {et~al.}(2017)\citenamefont {Guo},
  \citenamefont {Fernández-Díaz}, \citenamefont {Komarek}, \citenamefont
  {Huh}, \citenamefont {Adler},\ and\ \citenamefont {Valldor}}]{SrFe2S2O-S}%
  \BibitemOpen
  \bibfield  {author} {\bibinfo {author} {\bibfnamefont {H.}~\bibnamefont
  {Guo}}, \bibinfo {author} {\bibfnamefont {M.-T.}\ \bibnamefont
  {Fernández-Díaz}}, \bibinfo {author} {\bibfnamefont {A.~C.}\ \bibnamefont
  {Komarek}}, \bibinfo {author} {\bibfnamefont {S.}~\bibnamefont {Huh}},
  \bibinfo {author} {\bibfnamefont {P.}~\bibnamefont {Adler}},\ and\ \bibinfo
  {author} {\bibfnamefont {M.}~\bibnamefont {Valldor}},\ }\bibfield  {title}
  {\bibinfo {title} {{Long-Range Antiferromagnetic Order on Spin Ladders
  SrFe2S2O and SrFe2Se2O As Probed by Neutron Diffraction and Mössbauer
  Spectroscopy}},\ }\href
  {https://doi.org/https://doi.org/10.1002/ejic.201700684} {\bibfield
  {journal} {\bibinfo  {journal} {European Journal of Inorganic Chemistry}\
  }\textbf {\bibinfo {volume} {2017}},\ \bibinfo {pages} {3829} (\bibinfo
  {year} {2017})}\BibitemShut {NoStop}%
\bibitem [{\citenamefont {WEITZEL}(1976)}]{CoWO4-S}%
  \BibitemOpen
  \bibfield  {author} {\bibinfo {author} {\bibfnamefont {H.}~\bibnamefont
  {WEITZEL}},\ }\bibfield  {title} {\bibinfo {title}
  {{Kristallstrukturverfeinerung von Wolframiten und Columbiten}},\ }\href
  {https://doi.org/doi:10.1524/zkri.1976.144.16.238} {\bibfield  {journal}
  {\bibinfo  {journal} {Zeitschrift für Kristallographie - Crystalline
  Materials}\ }\textbf {\bibinfo {volume} {144}},\ \bibinfo {pages} {238}
  (\bibinfo {year} {1976})}\BibitemShut {NoStop}%
\bibitem [{\citenamefont {Bartel}\ and\ \citenamefont {Morosin}(1971)}]{NiO-S}%
  \BibitemOpen
  \bibfield  {author} {\bibinfo {author} {\bibfnamefont {L.~C.}\ \bibnamefont
  {Bartel}}\ and\ \bibinfo {author} {\bibfnamefont {B.}~\bibnamefont
  {Morosin}},\ }\bibfield  {title} {\bibinfo {title} {{Exchange Striction in
  NiO}},\ }\href {https://doi.org/10.1103/PhysRevB.3.1039} {\bibfield
  {journal} {\bibinfo  {journal} {Phys. Rev. B}\ }\textbf {\bibinfo {volume}
  {3}},\ \bibinfo {pages} {1039} (\bibinfo {year} {1971})}\BibitemShut
  {NoStop}%
\bibitem [{\citenamefont {Costa}\ \emph {et~al.}(1993)\citenamefont {Costa},
  \citenamefont {Paix{\~{a}}o}, \citenamefont {de~Almeida},\ and\ \citenamefont
  {Andrade}}]{NiF2-S}%
  \BibitemOpen
  \bibfield  {author} {\bibinfo {author} {\bibfnamefont {M.~M.~R.}\
  \bibnamefont {Costa}}, \bibinfo {author} {\bibfnamefont {J.~A.}\ \bibnamefont
  {Paix{\~{a}}o}}, \bibinfo {author} {\bibfnamefont {M.~J.~M.}\ \bibnamefont
  {de~Almeida}},\ and\ \bibinfo {author} {\bibfnamefont {L.~C.~R.}\
  \bibnamefont {Andrade}},\ }\bibfield  {title} {\bibinfo {title} {{Charge
  densities of two rutile structures: NiF${\sb 2}$ and CoF${\sb 2}$}},\ }\href
  {https://doi.org/10.1107/S0108768193001624} {\bibfield  {journal} {\bibinfo
  {journal} {Acta Crystallographica Section B}\ }\textbf {\bibinfo {volume}
  {49}},\ \bibinfo {pages} {591} (\bibinfo {year} {1993})}\BibitemShut
  {NoStop}%
\bibitem [{\citenamefont {Adam}\ \emph {et~al.}(1980)\citenamefont {Adam},
  \citenamefont {Billerey}, \citenamefont {Terrier}, \citenamefont {Mainard},
  \citenamefont {Regnault}, \citenamefont {Rossat-Mignod},\ and\ \citenamefont
  {Mériel}}]{NiBr2-S}%
  \BibitemOpen
  \bibfield  {author} {\bibinfo {author} {\bibfnamefont {A.}~\bibnamefont
  {Adam}}, \bibinfo {author} {\bibfnamefont {D.}~\bibnamefont {Billerey}},
  \bibinfo {author} {\bibfnamefont {C.}~\bibnamefont {Terrier}}, \bibinfo
  {author} {\bibfnamefont {R.}~\bibnamefont {Mainard}}, \bibinfo {author}
  {\bibfnamefont {L.}~\bibnamefont {Regnault}}, \bibinfo {author}
  {\bibfnamefont {J.}~\bibnamefont {Rossat-Mignod}},\ and\ \bibinfo {author}
  {\bibfnamefont {P.}~\bibnamefont {Mériel}},\ }\bibfield  {title} {\bibinfo
  {title} {{Neutron diffraction study of the commensurate and incommensurate
  magnetic structures of NiBr2}},\ }\href
  {https://doi.org/https://doi.org/10.1016/0038-1098(80)90757-7} {\bibfield
  {journal} {\bibinfo  {journal} {Solid State Communications}\ }\textbf
  {\bibinfo {volume} {35}},\ \bibinfo {pages} {1} (\bibinfo {year}
  {1980})}\BibitemShut {NoStop}%
\bibitem [{\citenamefont {Yano}\ \emph {et~al.}(2016)\citenamefont {Yano},
  \citenamefont {Louca}, \citenamefont {Yang}, \citenamefont {Chatterjee},
  \citenamefont {Bugaris}, \citenamefont {Chung}, \citenamefont {Peng},
  \citenamefont {Grayson},\ and\ \citenamefont {Kanatzidis}}]{NiS2-ST}%
  \BibitemOpen
  \bibfield  {author} {\bibinfo {author} {\bibfnamefont {S.}~\bibnamefont
  {Yano}}, \bibinfo {author} {\bibfnamefont {D.}~\bibnamefont {Louca}},
  \bibinfo {author} {\bibfnamefont {J.}~\bibnamefont {Yang}}, \bibinfo {author}
  {\bibfnamefont {U.}~\bibnamefont {Chatterjee}}, \bibinfo {author}
  {\bibfnamefont {D.~E.}\ \bibnamefont {Bugaris}}, \bibinfo {author}
  {\bibfnamefont {D.~Y.}\ \bibnamefont {Chung}}, \bibinfo {author}
  {\bibfnamefont {L.}~\bibnamefont {Peng}}, \bibinfo {author} {\bibfnamefont
  {M.}~\bibnamefont {Grayson}},\ and\ \bibinfo {author} {\bibfnamefont {M.~G.}\
  \bibnamefont {Kanatzidis}},\ }\bibfield  {title} {\bibinfo {title} {{Magnetic
  structure of ${\mathrm{NiS}}_{2\ensuremath{-}x}{\mathrm{Se}}_{x}$}},\ }\href
  {https://doi.org/10.1103/PhysRevB.93.024409} {\bibfield  {journal} {\bibinfo
  {journal} {Phys. Rev. B}\ }\textbf {\bibinfo {volume} {93}},\ \bibinfo
  {pages} {024409} (\bibinfo {year} {2016})}\BibitemShut {NoStop}%
\bibitem [{\citenamefont {McGuire}(2017)}]{NiCl2-S}%
  \BibitemOpen
  \bibfield  {author} {\bibinfo {author} {\bibfnamefont {M.~A.}\ \bibnamefont
  {McGuire}},\ }\bibfield  {title} {\bibinfo {title} {Crystal and magnetic
  structures in layered, transition metal dihalides and trihalides},\
  }\bibfield  {journal} {\bibinfo  {journal} {Crystals}\ }\textbf {\bibinfo
  {volume} {7}},\ \href {https://doi.org/10.3390/cryst7050121}
  {10.3390/cryst7050121} (\bibinfo {year} {2017})\BibitemShut {NoStop}%
\bibitem [{\citenamefont {Brec}(1986)}]{NiPS3-S}%
  \BibitemOpen
  \bibfield  {author} {\bibinfo {author} {\bibfnamefont {R.}~\bibnamefont
  {Brec}},\ }\bibfield  {title} {\bibinfo {title} {{Review on structural and
  chemical properties of transition metal phosphorous trisulfides MPS3}},\
  }\href {https://doi.org/https://doi.org/10.1016/0167-2738(86)90055-X}
  {\bibfield  {journal} {\bibinfo  {journal} {Solid State Ionics}\ }\textbf
  {\bibinfo {volume} {22}},\ \bibinfo {pages} {3} (\bibinfo {year}
  {1986})}\BibitemShut {NoStop}%
\bibitem [{\citenamefont {Brec}\ \emph {et~al.}(1980)\citenamefont {Brec},
  \citenamefont {Ouvrard}, \citenamefont {Louisy},\ and\ \citenamefont
  {Rouxel}}]{NiPSe3-S1}%
  \BibitemOpen
  \bibfield  {author} {\bibinfo {author} {\bibfnamefont {R.}~\bibnamefont
  {Brec}}, \bibinfo {author} {\bibfnamefont {G.}~\bibnamefont {Ouvrard}},
  \bibinfo {author} {\bibfnamefont {A.}~\bibnamefont {Louisy}},\ and\ \bibinfo
  {author} {\bibfnamefont {J.}~\bibnamefont {Rouxel}},\ }\bibfield  {title}
  {\bibinfo {title} {{Proprietes structurales de phases M(II)PX3 (X = S,
  Se)}},\ }\href@noop {} {\bibfield  {journal} {\bibinfo  {journal} {Annales de
  Chimie (Paris) (Vol=Year)}\ }\textbf {\bibinfo {volume} {5}},\ \bibinfo
  {pages} {499} (\bibinfo {year} {1980})}\BibitemShut {NoStop}%
\bibitem [{\citenamefont {Scatturin}\ \emph {et~al.}(1961)\citenamefont
  {Scatturin}, \citenamefont {Corliss}, \citenamefont {Elliott},\ and\
  \citenamefont {Hastings}}]{KNiF3-ST1}%
  \BibitemOpen
  \bibfield  {author} {\bibinfo {author} {\bibfnamefont {V.}~\bibnamefont
  {Scatturin}}, \bibinfo {author} {\bibfnamefont {L.}~\bibnamefont {Corliss}},
  \bibinfo {author} {\bibfnamefont {N.}~\bibnamefont {Elliott}},\ and\ \bibinfo
  {author} {\bibfnamefont {J.}~\bibnamefont {Hastings}},\ }\bibfield  {title}
  {\bibinfo {title} {{Magnetic structures of 3{\it d} transition metal double
  fluorides, K{\it Me}F${\sb 3}$}},\ }\href
  {https://doi.org/10.1107/S0365110X61000036} {\bibfield  {journal} {\bibinfo
  {journal} {Acta Crystallographica}\ }\textbf {\bibinfo {volume} {14}},\
  \bibinfo {pages} {19} (\bibinfo {year} {1961})}\BibitemShut {NoStop}%
\bibitem [{\citenamefont {Keeling}(1957)}]{NiWO4-S}%
  \BibitemOpen
  \bibfield  {author} {\bibinfo {author} {\bibfnamefont {R.~O.}\ \bibnamefont
  {Keeling}},\ }\bibfield  {title} {\bibinfo {title} {{The structure of
  NiWO4}},\ }\href {https://doi.org/https://doi.org/10.1107/S0365110X57000651}
  {\bibfield  {journal} {\bibinfo  {journal} {Acta Crystallographica}\ }\textbf
  {\bibinfo {volume} {10}},\ \bibinfo {pages} {209} (\bibinfo {year}
  {1957})}\BibitemShut {NoStop}%
\bibitem [{\citenamefont {Rodriguez-Carvajal}\ \emph
  {et~al.}(1991)\citenamefont {Rodriguez-Carvajal}, \citenamefont
  {Fernandez-Diaz},\ and\ \citenamefont {Martinez}}]{La2NiO4-ST}%
  \BibitemOpen
  \bibfield  {author} {\bibinfo {author} {\bibfnamefont {J.}~\bibnamefont
  {Rodriguez-Carvajal}}, \bibinfo {author} {\bibfnamefont {M.~T.}\ \bibnamefont
  {Fernandez-Diaz}},\ and\ \bibinfo {author} {\bibfnamefont {J.~L.}\
  \bibnamefont {Martinez}},\ }\bibfield  {title} {\bibinfo {title} {{Neutron
  diffraction study on structural and magnetic properties of La2NiO4}},\ }\href
  {https://doi.org/10.1088/0953-8984/3/19/002} {\bibfield  {journal} {\bibinfo
  {journal} {Journal of Physics: Condensed Matter}\ }\textbf {\bibinfo {volume}
  {3}},\ \bibinfo {pages} {3215} (\bibinfo {year} {1991})}\BibitemShut
  {NoStop}%
\bibitem [{\citenamefont {{Plumier, R.}}\ and\ \citenamefont {{Legrand,
  E.}}(1962)}]{K2NiF4-S}%
  \BibitemOpen
  \bibfield  {author} {\bibinfo {author} {\bibnamefont {{Plumier, R.}}}\ and\
  \bibinfo {author} {\bibnamefont {{Legrand, E.}}},\ }\bibfield  {title}
  {\bibinfo {title} {{Structure magnétique de K2NiF4}},\ }\href
  {https://doi.org/10.1051/jphysrad:01962002308-9047400} {\bibfield  {journal}
  {\bibinfo  {journal} {J. Phys. Radium}\ }\textbf {\bibinfo {volume} {23}},\
  \bibinfo {pages} {474} (\bibinfo {year} {1962})}\BibitemShut {NoStop}%
\bibitem [{\citenamefont {Lujan}\ \emph {et~al.}(1994)\citenamefont {Lujan},
  \citenamefont {Rivera}, \citenamefont {Kizhaev}, \citenamefont {Schmid},
  \citenamefont {Triscone}, \citenamefont {Muller}, \citenamefont {Ye},
  \citenamefont {Mettout},\ and\ \citenamefont {Bouzerar}}]{KNiPO4-ST1}%
  \BibitemOpen
  \bibfield  {author} {\bibinfo {author} {\bibfnamefont {M.}~\bibnamefont
  {Lujan}}, \bibinfo {author} {\bibfnamefont {J.-P.}\ \bibnamefont {Rivera}},
  \bibinfo {author} {\bibfnamefont {S.}~\bibnamefont {Kizhaev}}, \bibinfo
  {author} {\bibfnamefont {H.}~\bibnamefont {Schmid}}, \bibinfo {author}
  {\bibfnamefont {G.}~\bibnamefont {Triscone}}, \bibinfo {author}
  {\bibfnamefont {J.}~\bibnamefont {Muller}}, \bibinfo {author} {\bibfnamefont
  {Z.-G.}\ \bibnamefont {Ye}}, \bibinfo {author} {\bibfnamefont
  {B.}~\bibnamefont {Mettout}},\ and\ \bibinfo {author} {\bibfnamefont
  {R.}~\bibnamefont {Bouzerar}},\ }\bibfield  {title} {\bibinfo {title}
  {{Magnetic measurements and magnetoelectric effect of pyroelectric KNiPO4
  single crystals}},\ }\href {https://doi.org/10.1080/00150199408213356}
  {\bibfield  {journal} {\bibinfo  {journal} {Ferroelectrics}\ }\textbf
  {\bibinfo {volume} {161}},\ \bibinfo {pages} {77} (\bibinfo {year}
  {1994})}\BibitemShut {NoStop}%
\bibitem [{\citenamefont {Abrahams}\ and\ \citenamefont
  {Easson}(1993)}]{LiNiPO4-S}%
  \BibitemOpen
  \bibfield  {author} {\bibinfo {author} {\bibfnamefont {I.}~\bibnamefont
  {Abrahams}}\ and\ \bibinfo {author} {\bibfnamefont {K.~S.}\ \bibnamefont
  {Easson}},\ }\bibfield  {title} {\bibinfo {title} {{Structure of lithium
  nickel phosphate}},\ }\href {https://doi.org/10.1107/S0108270192013064}
  {\bibfield  {journal} {\bibinfo  {journal} {Acta Crystallographica Section
  C}\ }\textbf {\bibinfo {volume} {49}},\ \bibinfo {pages} {925} (\bibinfo
  {year} {1993})}\BibitemShut {NoStop}%
\bibitem [{\citenamefont {Forsyth}\ \emph {et~al.}(1988)\citenamefont
  {Forsyth}, \citenamefont {Brown},\ and\ \citenamefont {Wanklyn}}]{CuO-ST}%
  \BibitemOpen
  \bibfield  {author} {\bibinfo {author} {\bibfnamefont {J.~B.}\ \bibnamefont
  {Forsyth}}, \bibinfo {author} {\bibfnamefont {P.~J.}\ \bibnamefont {Brown}},\
  and\ \bibinfo {author} {\bibfnamefont {B.~M.}\ \bibnamefont {Wanklyn}},\
  }\bibfield  {title} {\bibinfo {title} {{Magnetism in cupric oxide}},\ }\href
  {https://doi.org/10.1088/0022-3719/21/15/023} {\bibfield  {journal} {\bibinfo
   {journal} {Journal of Physics C: Solid State Physics}\ }\textbf {\bibinfo
  {volume} {21}},\ \bibinfo {pages} {2917} (\bibinfo {year}
  {1988})}\BibitemShut {NoStop}%
\bibitem [{\citenamefont {Fischer}\ \emph {et~al.}(1974)\citenamefont
  {Fischer}, \citenamefont {Hälg}, \citenamefont {Schwarzenbach},\ and\
  \citenamefont {Gamsjäger}}]{CuF2-S}%
  \BibitemOpen
  \bibfield  {author} {\bibinfo {author} {\bibfnamefont {P.}~\bibnamefont
  {Fischer}}, \bibinfo {author} {\bibfnamefont {W.}~\bibnamefont {Hälg}},
  \bibinfo {author} {\bibfnamefont {D.}~\bibnamefont {Schwarzenbach}},\ and\
  \bibinfo {author} {\bibfnamefont {H.}~\bibnamefont {Gamsjäger}},\ }\bibfield
   {title} {\bibinfo {title} {{Magnetic and crystal structure of copper(II)
  fluoride}},\ }\href
  {https://doi.org/https://doi.org/10.1016/S0022-3697(74)80182-4} {\bibfield
  {journal} {\bibinfo  {journal} {Journal of Physics and Chemistry of Solids}\
  }\textbf {\bibinfo {volume} {35}},\ \bibinfo {pages} {1683} (\bibinfo {year}
  {1974})}\BibitemShut {NoStop}%
\bibitem [{\citenamefont {Blum}\ \emph {et~al.}(2009)\citenamefont {Blum},
  \citenamefont {Gehrke}, \citenamefont {Hanke}, \citenamefont {Havu},
  \citenamefont {Havu}, \citenamefont {Ren}, \citenamefont {Reuter},\ and\
  \citenamefont {Scheffler}}]{FHI-aims}%
  \BibitemOpen
  \bibfield  {author} {\bibinfo {author} {\bibfnamefont {V.}~\bibnamefont
  {Blum}}, \bibinfo {author} {\bibfnamefont {R.}~\bibnamefont {Gehrke}},
  \bibinfo {author} {\bibfnamefont {F.}~\bibnamefont {Hanke}}, \bibinfo
  {author} {\bibfnamefont {P.}~\bibnamefont {Havu}}, \bibinfo {author}
  {\bibfnamefont {V.}~\bibnamefont {Havu}}, \bibinfo {author} {\bibfnamefont
  {X.}~\bibnamefont {Ren}}, \bibinfo {author} {\bibfnamefont {K.}~\bibnamefont
  {Reuter}},\ and\ \bibinfo {author} {\bibfnamefont {M.}~\bibnamefont
  {Scheffler}},\ }\bibfield  {title} {\bibinfo {title} {{Ab initio molecular
  simulations with numeric atom-centered orbitals}},\ }\href
  {https://doi.org/https://doi.org/10.1016/j.cpc.2009.06.022} {\bibfield
  {journal} {\bibinfo  {journal} {Computer Physics Communications}\ }\textbf
  {\bibinfo {volume} {180}},\ \bibinfo {pages} {2175} (\bibinfo {year}
  {2009})}\BibitemShut {NoStop}%
\bibitem [{SM()}]{SM}%
  \BibitemOpen
  \href@noop {} {}\bibinfo {note} {See Supplemental Material at
  URL-will-be-inserted-by-publisher for additional data.}\BibitemShut {Stop}%
\bibitem [{\citenamefont {Reehuis}\ \emph {et~al.}(2006)\citenamefont
  {Reehuis}, \citenamefont {Ulrich}, \citenamefont {Pattison}, \citenamefont
  {Ouladdiaf}, \citenamefont {Rheinst\"adter}, \citenamefont {Ohl},
  \citenamefont {Regnault}, \citenamefont {Miyasaka}, \citenamefont {Tokura},\
  and\ \citenamefont {Keimer}}]{YVO3-T}%
  \BibitemOpen
  \bibfield  {author} {\bibinfo {author} {\bibfnamefont {M.}~\bibnamefont
  {Reehuis}}, \bibinfo {author} {\bibfnamefont {C.}~\bibnamefont {Ulrich}},
  \bibinfo {author} {\bibfnamefont {P.}~\bibnamefont {Pattison}}, \bibinfo
  {author} {\bibfnamefont {B.}~\bibnamefont {Ouladdiaf}}, \bibinfo {author}
  {\bibfnamefont {M.~C.}\ \bibnamefont {Rheinst\"adter}}, \bibinfo {author}
  {\bibfnamefont {M.}~\bibnamefont {Ohl}}, \bibinfo {author} {\bibfnamefont
  {L.~P.}\ \bibnamefont {Regnault}}, \bibinfo {author} {\bibfnamefont
  {M.}~\bibnamefont {Miyasaka}}, \bibinfo {author} {\bibfnamefont
  {Y.}~\bibnamefont {Tokura}},\ and\ \bibinfo {author} {\bibfnamefont
  {B.}~\bibnamefont {Keimer}},\ }\bibfield  {title} {\bibinfo {title} {{Neutron
  diffraction study of ${\mathrm{YVO}}_{3}$, ${\mathrm{NdVO}}_{3}$, and
  ${\mathrm{TbVO}}_{3}$}},\ }\href {https://doi.org/10.1103/PhysRevB.73.094440}
  {\bibfield  {journal} {\bibinfo  {journal} {Phys. Rev. B}\ }\textbf {\bibinfo
  {volume} {73}},\ \bibinfo {pages} {094440} (\bibinfo {year}
  {2006})}\BibitemShut {NoStop}%
\bibitem [{\citenamefont {Holseth}\ \emph {et~al.}(1970)\citenamefont
  {Holseth}, \citenamefont {Kjekshus}, \citenamefont {Andresen}, \citenamefont
  {Sch{\"a}fer},\ and\ \citenamefont {Shimizu}}]{CrSb2-T1}%
  \BibitemOpen
  \bibfield  {author} {\bibinfo {author} {\bibfnamefont {H.}~\bibnamefont
  {Holseth}}, \bibinfo {author} {\bibfnamefont {A.}~\bibnamefont {Kjekshus}},
  \bibinfo {author} {\bibfnamefont {A.~F.}\ \bibnamefont {Andresen}}, \bibinfo
  {author} {\bibfnamefont {L.}~\bibnamefont {Sch{\"a}fer}},\ and\ \bibinfo
  {author} {\bibfnamefont {A.}~\bibnamefont {Shimizu}},\ }\bibfield  {title}
  {\bibinfo {title} {{Compounds with the Marcasite Type Crystal Structure. VI.
  Neutron Diffraction Studies of CrSb2 and FeSb2.}},\ }\href
  {http://actachemscand.org/doi/10.3891/acta.chem.scand.24-3309} {\bibfield
  {journal} {\bibinfo  {journal} {Acta Chemica Scandinavica}\ }\textbf
  {\bibinfo {volume} {24}},\ \bibinfo {pages} {3309} (\bibinfo {year}
  {1970})}\BibitemShut {NoStop}%
\bibitem [{\citenamefont {Sales}\ \emph {et~al.}(2012)\citenamefont {Sales},
  \citenamefont {May}, \citenamefont {McGuire}, \citenamefont {Stone},
  \citenamefont {Singh},\ and\ \citenamefont {Mandrus}}]{CrSb2-T2}%
  \BibitemOpen
  \bibfield  {author} {\bibinfo {author} {\bibfnamefont {B.~C.}\ \bibnamefont
  {Sales}}, \bibinfo {author} {\bibfnamefont {A.~F.}\ \bibnamefont {May}},
  \bibinfo {author} {\bibfnamefont {M.~A.}\ \bibnamefont {McGuire}}, \bibinfo
  {author} {\bibfnamefont {M.~B.}\ \bibnamefont {Stone}}, \bibinfo {author}
  {\bibfnamefont {D.~J.}\ \bibnamefont {Singh}},\ and\ \bibinfo {author}
  {\bibfnamefont {D.}~\bibnamefont {Mandrus}},\ }\bibfield  {title} {\bibinfo
  {title} {{Transport, thermal, and magnetic properties of the narrow-gap
  semiconductor CrSb${}_{2}$}},\ }\href
  {https://doi.org/10.1103/PhysRevB.86.235136} {\bibfield  {journal} {\bibinfo
  {journal} {Phys. Rev. B}\ }\textbf {\bibinfo {volume} {86}},\ \bibinfo
  {pages} {235136} (\bibinfo {year} {2012})}\BibitemShut {NoStop}%
\bibitem [{\citenamefont {Stout}\ and\ \citenamefont
  {Chisholm}(1962)}]{CrCl2-T1}%
  \BibitemOpen
  \bibfield  {author} {\bibinfo {author} {\bibfnamefont {J.~W.}\ \bibnamefont
  {Stout}}\ and\ \bibinfo {author} {\bibfnamefont {R.~C.}\ \bibnamefont
  {Chisholm}},\ }\bibfield  {title} {\bibinfo {title} {{Heat Capacity and
  Entropy of CuCl2 and CrCl2 from 11° to 300°K. Magnetic Ordering in Linear
  Chain Crystals}},\ }\href {https://doi.org/10.1063/1.1732699} {\bibfield
  {journal} {\bibinfo  {journal} {The Journal of Chemical Physics}\ }\textbf
  {\bibinfo {volume} {36}},\ \bibinfo {pages} {979} (\bibinfo {year}
  {1962})}\BibitemShut {NoStop}%
\bibitem [{\citenamefont {Winkelmann}\ \emph {et~al.}(1997)\citenamefont
  {Winkelmann}, \citenamefont {Baehr}, \citenamefont {Reehuis}, \citenamefont
  {Steiner}, \citenamefont {Hagiwara},\ and\ \citenamefont
  {Katsumata}}]{CrCl2-T2}%
  \BibitemOpen
  \bibfield  {author} {\bibinfo {author} {\bibfnamefont {M.}~\bibnamefont
  {Winkelmann}}, \bibinfo {author} {\bibfnamefont {M.}~\bibnamefont {Baehr}},
  \bibinfo {author} {\bibfnamefont {M.}~\bibnamefont {Reehuis}}, \bibinfo
  {author} {\bibfnamefont {M.}~\bibnamefont {Steiner}}, \bibinfo {author}
  {\bibfnamefont {M.}~\bibnamefont {Hagiwara}},\ and\ \bibinfo {author}
  {\bibfnamefont {K.}~\bibnamefont {Katsumata}},\ }\bibfield  {title} {\bibinfo
  {title} {{Structural and magnetic characterization of a new phase of
  CrCl2}},\ }\href
  {https://doi.org/https://doi.org/10.1016/S0022-3697(96)00122-9} {\bibfield
  {journal} {\bibinfo  {journal} {Journal of Physics and Chemistry of Solids}\
  }\textbf {\bibinfo {volume} {58}},\ \bibinfo {pages} {481} (\bibinfo {year}
  {1997})}\BibitemShut {NoStop}%
\bibitem [{\citenamefont {Cable}\ \emph {et~al.}(1960)\citenamefont {Cable},
  \citenamefont {Wilkinson},\ and\ \citenamefont {Wollan}}]{CrCl2-T3}%
  \BibitemOpen
  \bibfield  {author} {\bibinfo {author} {\bibfnamefont {J.~W.}\ \bibnamefont
  {Cable}}, \bibinfo {author} {\bibfnamefont {M.~K.}\ \bibnamefont
  {Wilkinson}},\ and\ \bibinfo {author} {\bibfnamefont {E.~O.}\ \bibnamefont
  {Wollan}},\ }\bibfield  {title} {\bibinfo {title} {{Neutron Diffraction
  Studies of Antiferromagnetism in Cr${\mathrm{F}}_{2}$ and
  Cr${\mathrm{Cl}}_{2}$}},\ }\href {https://doi.org/10.1103/PhysRev.118.950}
  {\bibfield  {journal} {\bibinfo  {journal} {Phys. Rev.}\ }\textbf {\bibinfo
  {volume} {118}},\ \bibinfo {pages} {950} (\bibinfo {year}
  {1960})}\BibitemShut {NoStop}%
\bibitem [{\citenamefont {Chatterji}\ and\ \citenamefont
  {Hansen}(2011)}]{CrF2-T}%
  \BibitemOpen
  \bibfield  {author} {\bibinfo {author} {\bibfnamefont {T.}~\bibnamefont
  {Chatterji}}\ and\ \bibinfo {author} {\bibfnamefont {T.~C.}\ \bibnamefont
  {Hansen}},\ }\bibfield  {title} {\bibinfo {title} {{Magnetoelastic effects in
  Jahn–Teller distorted CrF2 and CuF2 studied by neutron powder
  diffraction}},\ }\href {https://doi.org/10.1088/0953-8984/23/27/276007}
  {\bibfield  {journal} {\bibinfo  {journal} {Journal of Physics: Condensed
  Matter}\ }\textbf {\bibinfo {volume} {23}},\ \bibinfo {pages} {276007}
  (\bibinfo {year} {2011})}\BibitemShut {NoStop}%
\bibitem [{\citenamefont {Samuelsen}\ \emph {et~al.}(1970)\citenamefont
  {Samuelsen}, \citenamefont {Hutchings},\ and\ \citenamefont
  {Shirane}}]{Cr2O3-T1}%
  \BibitemOpen
  \bibfield  {author} {\bibinfo {author} {\bibfnamefont {E.}~\bibnamefont
  {Samuelsen}}, \bibinfo {author} {\bibfnamefont {M.}~\bibnamefont
  {Hutchings}},\ and\ \bibinfo {author} {\bibfnamefont {G.}~\bibnamefont
  {Shirane}},\ }\bibfield  {title} {\bibinfo {title} {{Inelastic neutron
  scattering investigation of spin waves and magnetic interactions in Cr2O3}},\
  }\href {https://doi.org/https://doi.org/10.1016/0031-8914(70)90158-8}
  {\bibfield  {journal} {\bibinfo  {journal} {Physica}\ }\textbf {\bibinfo
  {volume} {48}},\ \bibinfo {pages} {13} (\bibinfo {year} {1970})}\BibitemShut
  {NoStop}%
\bibitem [{\citenamefont {Brockhouse}(1953)}]{Cr2O3-T2}%
  \BibitemOpen
  \bibfield  {author} {\bibinfo {author} {\bibfnamefont {B.~N.}\ \bibnamefont
  {Brockhouse}},\ }\bibfield  {title} {\bibinfo {title} {{Antiferromagnetic
  Structure in Cr2O3}},\ }\href {https://doi.org/10.1063/1.1699098} {\bibfield
  {journal} {\bibinfo  {journal} {The Journal of Chemical Physics}\ }\textbf
  {\bibinfo {volume} {21}},\ \bibinfo {pages} {961} (\bibinfo {year}
  {1953})}\BibitemShut {NoStop}%
\bibitem [{\citenamefont {Fiebig}\ \emph {et~al.}(1996)\citenamefont {Fiebig},
  \citenamefont {Fr\"ohlich},\ and\ \citenamefont {Thiele}}]{Cr2O3-T3}%
  \BibitemOpen
  \bibfield  {author} {\bibinfo {author} {\bibfnamefont {M.}~\bibnamefont
  {Fiebig}}, \bibinfo {author} {\bibfnamefont {D.}~\bibnamefont {Fr\"ohlich}},\
  and\ \bibinfo {author} {\bibfnamefont {H.~J.}\ \bibnamefont {Thiele}},\
  }\bibfield  {title} {\bibinfo {title} {{Determination of spin direction in
  the spin-flop phase of ${\mathrm{Cr}}_{2}$${\mathrm{O}}_{3}$}},\ }\href
  {https://doi.org/10.1103/PhysRevB.54.R12681} {\bibfield  {journal} {\bibinfo
  {journal} {Phys. Rev. B}\ }\textbf {\bibinfo {volume} {54}},\ \bibinfo
  {pages} {R12681} (\bibinfo {year} {1996})}\BibitemShut {NoStop}%
\bibitem [{\citenamefont {Telford}\ \emph {et~al.}(2020)\citenamefont
  {Telford}, \citenamefont {Dismukes}, \citenamefont {Lee}, \citenamefont
  {Cheng}, \citenamefont {Wieteska}, \citenamefont {Bartholomew}, \citenamefont
  {Chen}, \citenamefont {Xu}, \citenamefont {Pasupathy}, \citenamefont {Zhu},
  \citenamefont {Dean},\ and\ \citenamefont {Roy}}]{CrSBr-T}%
  \BibitemOpen
  \bibfield  {author} {\bibinfo {author} {\bibfnamefont {E.~J.}\ \bibnamefont
  {Telford}}, \bibinfo {author} {\bibfnamefont {A.~H.}\ \bibnamefont
  {Dismukes}}, \bibinfo {author} {\bibfnamefont {K.}~\bibnamefont {Lee}},
  \bibinfo {author} {\bibfnamefont {M.}~\bibnamefont {Cheng}}, \bibinfo
  {author} {\bibfnamefont {A.}~\bibnamefont {Wieteska}}, \bibinfo {author}
  {\bibfnamefont {A.~K.}\ \bibnamefont {Bartholomew}}, \bibinfo {author}
  {\bibfnamefont {Y.-S.}\ \bibnamefont {Chen}}, \bibinfo {author}
  {\bibfnamefont {X.}~\bibnamefont {Xu}}, \bibinfo {author} {\bibfnamefont
  {A.~N.}\ \bibnamefont {Pasupathy}}, \bibinfo {author} {\bibfnamefont
  {X.}~\bibnamefont {Zhu}}, \bibinfo {author} {\bibfnamefont {C.~R.}\
  \bibnamefont {Dean}},\ and\ \bibinfo {author} {\bibfnamefont
  {X.}~\bibnamefont {Roy}},\ }\bibfield  {title} {\bibinfo {title} {{Layered
  Antiferromagnetism Induces Large Negative Magnetoresistance in the van der
  Waals Semiconductor CrSBr}},\ }\href
  {https://doi.org/https://doi.org/10.1002/adma.202003240} {\bibfield
  {journal} {\bibinfo  {journal} {Advanced Materials}\ }\textbf {\bibinfo
  {volume} {32}},\ \bibinfo {pages} {2003240} (\bibinfo {year}
  {2020})}\BibitemShut {NoStop}%
\bibitem [{\citenamefont {Zhu}\ \emph {et~al.}(2019)\citenamefont {Zhu},
  \citenamefont {Matsumoto}, \citenamefont {Stone}, \citenamefont {Dun},
  \citenamefont {Zhou}, \citenamefont {Hong}, \citenamefont {Zou},
  \citenamefont {Mahanti},\ and\ \citenamefont {Ke}}]{Cr2Te(W)O6-T}%
  \BibitemOpen
  \bibfield  {author} {\bibinfo {author} {\bibfnamefont {M.}~\bibnamefont
  {Zhu}}, \bibinfo {author} {\bibfnamefont {M.}~\bibnamefont {Matsumoto}},
  \bibinfo {author} {\bibfnamefont {M.~B.}\ \bibnamefont {Stone}}, \bibinfo
  {author} {\bibfnamefont {Z.~L.}\ \bibnamefont {Dun}}, \bibinfo {author}
  {\bibfnamefont {H.~D.}\ \bibnamefont {Zhou}}, \bibinfo {author}
  {\bibfnamefont {T.}~\bibnamefont {Hong}}, \bibinfo {author} {\bibfnamefont
  {T.}~\bibnamefont {Zou}}, \bibinfo {author} {\bibfnamefont {S.~D.}\
  \bibnamefont {Mahanti}},\ and\ \bibinfo {author} {\bibfnamefont
  {X.}~\bibnamefont {Ke}},\ }\bibfield  {title} {\bibinfo {title} {{Amplitude
  modes in three-dimensional spin dimers away from quantum critical point}},\
  }\href {https://doi.org/10.1103/PhysRevResearch.1.033111} {\bibfield
  {journal} {\bibinfo  {journal} {Phys. Rev. Res.}\ }\textbf {\bibinfo {volume}
  {1}},\ \bibinfo {pages} {033111} (\bibinfo {year} {2019})}\BibitemShut
  {NoStop}%
\bibitem [{\citenamefont {Kohgi}\ \emph {et~al.}(1972)\citenamefont {Kohgi},
  \citenamefont {Ishikawa},\ and\ \citenamefont {Endoh}}]{MnO-T}%
  \BibitemOpen
  \bibfield  {author} {\bibinfo {author} {\bibfnamefont {M.}~\bibnamefont
  {Kohgi}}, \bibinfo {author} {\bibfnamefont {Y.}~\bibnamefont {Ishikawa}},\
  and\ \bibinfo {author} {\bibfnamefont {Y.}~\bibnamefont {Endoh}},\ }\bibfield
   {title} {\bibinfo {title} {{Inelastic neutron scattering study of spin waves
  in MnO}},\ }\href
  {https://doi.org/https://doi.org/10.1016/0038-1098(72)90255-4} {\bibfield
  {journal} {\bibinfo  {journal} {Solid State Communications}\ }\textbf
  {\bibinfo {volume} {11}},\ \bibinfo {pages} {391} (\bibinfo {year}
  {1972})}\BibitemShut {NoStop}%
\bibitem [{\citenamefont {Grzybowski}\ \emph {et~al.}(2024)\citenamefont
  {Grzybowski}, \citenamefont {Autieri}, \citenamefont {Domagala},
  \citenamefont {Krasucki}, \citenamefont {Kaleta}, \citenamefont {Kret},
  \citenamefont {Gas}, \citenamefont {Sawicki}, \citenamefont {Bożek},
  \citenamefont {Suffczyński},\ and\ \citenamefont {Pacuski}}]{MnSe-T1}%
  \BibitemOpen
  \bibfield  {author} {\bibinfo {author} {\bibfnamefont {M.~J.}\ \bibnamefont
  {Grzybowski}}, \bibinfo {author} {\bibfnamefont {C.}~\bibnamefont {Autieri}},
  \bibinfo {author} {\bibfnamefont {J.}~\bibnamefont {Domagala}}, \bibinfo
  {author} {\bibfnamefont {C.}~\bibnamefont {Krasucki}}, \bibinfo {author}
  {\bibfnamefont {A.}~\bibnamefont {Kaleta}}, \bibinfo {author} {\bibfnamefont
  {S.}~\bibnamefont {Kret}}, \bibinfo {author} {\bibfnamefont {K.}~\bibnamefont
  {Gas}}, \bibinfo {author} {\bibfnamefont {M.}~\bibnamefont {Sawicki}},
  \bibinfo {author} {\bibfnamefont {R.}~\bibnamefont {Bożek}}, \bibinfo
  {author} {\bibfnamefont {J.}~\bibnamefont {Suffczyński}},\ and\ \bibinfo
  {author} {\bibfnamefont {W.}~\bibnamefont {Pacuski}},\ }\bibfield  {title}
  {\bibinfo {title} {Wurtzite vs. rock-salt mnse epitaxy: electronic and
  altermagnetic properties},\ }\href {https://doi.org/10.1039/D3NR04798A}
  {\bibfield  {journal} {\bibinfo  {journal} {Nanoscale}\ }\textbf {\bibinfo
  {volume} {16}},\ \bibinfo {pages} {6259} (\bibinfo {year}
  {2024})}\BibitemShut {NoStop}%
\bibitem [{\citenamefont {Pollard}\ \emph {et~al.}(1983)\citenamefont
  {Pollard}, \citenamefont {McCann},\ and\ \citenamefont {Ward}}]{MnSe-T2}%
  \BibitemOpen
  \bibfield  {author} {\bibinfo {author} {\bibfnamefont {R.~J.}\ \bibnamefont
  {Pollard}}, \bibinfo {author} {\bibfnamefont {V.~H.}\ \bibnamefont
  {McCann}},\ and\ \bibinfo {author} {\bibfnamefont {J.~B.}\ \bibnamefont
  {Ward}},\ }\bibfield  {title} {\bibinfo {title} {{Magnetic structures of
  $\alpha$-MnS and MnSe from 57Fe Mossbauer spectroscopy}},\ }\href
  {https://doi.org/10.1088/0022-3719/16/2/017} {\bibfield  {journal} {\bibinfo
  {journal} {Journal of Physics C: Solid State Physics}\ }\textbf {\bibinfo
  {volume} {16}},\ \bibinfo {pages} {345} (\bibinfo {year} {1983})}\BibitemShut
  {NoStop}%
\bibitem [{\citenamefont {Uchida}\ \emph {et~al.}(1956)\citenamefont {Uchida},
  \citenamefont {Kondoh},\ and\ \citenamefont {Fukuoka}}]{MnTe-T1}%
  \BibitemOpen
  \bibfield  {author} {\bibinfo {author} {\bibfnamefont {E.}~\bibnamefont
  {Uchida}}, \bibinfo {author} {\bibfnamefont {H.}~\bibnamefont {Kondoh}},\
  and\ \bibinfo {author} {\bibfnamefont {N.}~\bibnamefont {Fukuoka}},\
  }\bibfield  {title} {\bibinfo {title} {{Magnetic and Electrical Properties of
  Manganese Telluride}},\ }\href {https://doi.org/10.1143/JPSJ.11.27}
  {\bibfield  {journal} {\bibinfo  {journal} {Journal of the Physical Society
  of Japan}\ }\textbf {\bibinfo {volume} {11}},\ \bibinfo {pages} {27}
  (\bibinfo {year} {1956})}\BibitemShut {NoStop}%
\bibitem [{\citenamefont {Szuszkiewicz}\ \emph {et~al.}(2006)\citenamefont
  {Szuszkiewicz}, \citenamefont {Dynowska}, \citenamefont {Witkowska},\ and\
  \citenamefont {Hennion}}]{MnTe-T2}%
  \BibitemOpen
  \bibfield  {author} {\bibinfo {author} {\bibfnamefont {W.}~\bibnamefont
  {Szuszkiewicz}}, \bibinfo {author} {\bibfnamefont {E.}~\bibnamefont
  {Dynowska}}, \bibinfo {author} {\bibfnamefont {B.}~\bibnamefont
  {Witkowska}},\ and\ \bibinfo {author} {\bibfnamefont {B.}~\bibnamefont
  {Hennion}},\ }\bibfield  {title} {\bibinfo {title} {{Spin-wave measurements
  on hexagonal $\mathrm{MnTe}$ of $\mathrm{NiAs}$-type structure by inelastic
  neutron scattering}},\ }\href {https://doi.org/10.1103/PhysRevB.73.104403}
  {\bibfield  {journal} {\bibinfo  {journal} {Phys. Rev. B}\ }\textbf {\bibinfo
  {volume} {73}},\ \bibinfo {pages} {104403} (\bibinfo {year}
  {2006})}\BibitemShut {NoStop}%
\bibitem [{\citenamefont {Ohama}\ and\ \citenamefont
  {Hamaguchi}(1971)}]{MnO2-T1}%
  \BibitemOpen
  \bibfield  {author} {\bibinfo {author} {\bibfnamefont {N.}~\bibnamefont
  {Ohama}}\ and\ \bibinfo {author} {\bibfnamefont {Y.}~\bibnamefont
  {Hamaguchi}},\ }\bibfield  {title} {\bibinfo {title} {{Determination of the
  Exchange Integrals in $\beta$-MnO2}},\ }\href
  {https://doi.org/10.1143/JPSJ.30.1311} {\bibfield  {journal} {\bibinfo
  {journal} {Journal of the Physical Society of Japan}\ }\textbf {\bibinfo
  {volume} {30}},\ \bibinfo {pages} {1311} (\bibinfo {year}
  {1971})}\BibitemShut {NoStop}%
\bibitem [{\citenamefont {Sato}\ \emph {et~al.}(2001)\citenamefont {Sato},
  \citenamefont {Wakiya}, \citenamefont {Enoki}, \citenamefont {Kiyama},
  \citenamefont {Wakabayashi}, \citenamefont {Nakao},\ and\ \citenamefont
  {Murakami}}]{MnO2-T2}%
  \BibitemOpen
  \bibfield  {author} {\bibinfo {author} {\bibfnamefont {H.}~\bibnamefont
  {Sato}}, \bibinfo {author} {\bibfnamefont {K.}~\bibnamefont {Wakiya}},
  \bibinfo {author} {\bibfnamefont {T.}~\bibnamefont {Enoki}}, \bibinfo
  {author} {\bibfnamefont {T.}~\bibnamefont {Kiyama}}, \bibinfo {author}
  {\bibfnamefont {Y.}~\bibnamefont {Wakabayashi}}, \bibinfo {author}
  {\bibfnamefont {H.}~\bibnamefont {Nakao}},\ and\ \bibinfo {author}
  {\bibfnamefont {Y.}~\bibnamefont {Murakami}},\ }\bibfield  {title} {\bibinfo
  {title} {{Magnetic Structure of $\beta$-MnO 2: X-ray Magnetic Scattering
  Study}},\ }\href {https://doi.org/10.1143/JPSJ.70.37} {\bibfield  {journal}
  {\bibinfo  {journal} {Journal of the Physical Society of Japan}\ }\textbf
  {\bibinfo {volume} {70}},\ \bibinfo {pages} {37} (\bibinfo {year}
  {2001})}\BibitemShut {NoStop}%
\bibitem [{\citenamefont {Nordblad}\ \emph {et~al.}(1981)\citenamefont
  {Nordblad}, \citenamefont {Lundgren}, \citenamefont {Figueroa},\ and\
  \citenamefont {Beckman}}]{MnF2-T}%
  \BibitemOpen
  \bibfield  {author} {\bibinfo {author} {\bibfnamefont {P.}~\bibnamefont
  {Nordblad}}, \bibinfo {author} {\bibfnamefont {L.}~\bibnamefont {Lundgren}},
  \bibinfo {author} {\bibfnamefont {E.}~\bibnamefont {Figueroa}},\ and\
  \bibinfo {author} {\bibfnamefont {O.}~\bibnamefont {Beckman}},\ }\bibfield
  {title} {\bibinfo {title} {{Specific heat and magnetic susceptibility of MnF2
  and Mn0.98Fe0.02F2 near TN}},\ }\href
  {https://doi.org/https://doi.org/10.1016/0304-8853(81)90056-1} {\bibfield
  {journal} {\bibinfo  {journal} {Journal of Magnetism and Magnetic Materials}\
  }\textbf {\bibinfo {volume} {23}},\ \bibinfo {pages} {333} (\bibinfo {year}
  {1981})}\BibitemShut {NoStop}%
\bibitem [{\citenamefont {Akutsu}\ and\ \citenamefont
  {Ikeda}(1981)}]{MnF2-KMnF3-T2}%
  \BibitemOpen
  \bibfield  {author} {\bibinfo {author} {\bibfnamefont {N.}~\bibnamefont
  {Akutsu}}\ and\ \bibinfo {author} {\bibfnamefont {H.}~\bibnamefont {Ikeda}},\
  }\bibfield  {title} {\bibinfo {title} {{Specific Heat Capacities of CoF$_2$,
  MnF$_2$ and KMnF$_3$ near the N\'eel Temperatures}},\ }\href
  {https://doi.org/10.1143/JPSJ.50.2865} {\bibfield  {journal} {\bibinfo
  {journal} {Journal of the Physical Society of Japan}\ }\textbf {\bibinfo
  {volume} {50}},\ \bibinfo {pages} {2865} (\bibinfo {year}
  {1981})}\BibitemShut {NoStop}%
\bibitem [{\citenamefont {Chattopadhyay}\ \emph {et~al.}(1991)\citenamefont
  {Chattopadhyay}, \citenamefont {Br\"uckel},\ and\ \citenamefont
  {Burlet}}]{MnS2-T1}%
  \BibitemOpen
  \bibfield  {author} {\bibinfo {author} {\bibfnamefont {T.}~\bibnamefont
  {Chattopadhyay}}, \bibinfo {author} {\bibfnamefont {T.}~\bibnamefont
  {Br\"uckel}},\ and\ \bibinfo {author} {\bibfnamefont {P.}~\bibnamefont
  {Burlet}},\ }\bibfield  {title} {\bibinfo {title} {{Spin correlation in the
  frustrated antiferromagnet ${\mathrm{MnS}}_{2}$ above the N\'eel
  temperature}},\ }\href {https://doi.org/10.1103/PhysRevB.44.7394} {\bibfield
  {journal} {\bibinfo  {journal} {Phys. Rev. B}\ }\textbf {\bibinfo {volume}
  {44}},\ \bibinfo {pages} {7394} (\bibinfo {year} {1991})}\BibitemShut
  {NoStop}%
\bibitem [{\citenamefont {Chattopadhyay}\ \emph {et~al.}(1984)\citenamefont
  {Chattopadhyay}, \citenamefont {Schnering},\ and\ \citenamefont
  {Graf}}]{MnS2-T2}%
  \BibitemOpen
  \bibfield  {author} {\bibinfo {author} {\bibfnamefont {T.}~\bibnamefont
  {Chattopadhyay}}, \bibinfo {author} {\bibfnamefont {H.}~\bibnamefont
  {Schnering}},\ and\ \bibinfo {author} {\bibfnamefont {H.}~\bibnamefont
  {Graf}},\ }\bibfield  {title} {\bibinfo {title} {{First order
  antiferromagnetic phase transition in MnS2}},\ }\href
  {https://doi.org/https://doi.org/10.1016/0038-1098(84)90348-X} {\bibfield
  {journal} {\bibinfo  {journal} {Solid State Communications}\ }\textbf
  {\bibinfo {volume} {50}},\ \bibinfo {pages} {865} (\bibinfo {year}
  {1984})}\BibitemShut {NoStop}%
\bibitem [{\citenamefont {Westrum}\ and\ \citenamefont
  {Gro/nvold}(1970)}]{MnS(Te)2-T3}%
  \BibitemOpen
  \bibfield  {author} {\bibinfo {author} {\bibfnamefont {J.}~\bibnamefont
  {Westrum}, \bibfnamefont {Edgar~F.}}\ and\ \bibinfo {author} {\bibfnamefont
  {F.}~\bibnamefont {Gro/nvold}},\ }\bibfield  {title} {\bibinfo {title}
  {{Manganese Disulfide (Hauerite) and Manganese Ditelluride. Thermal
  Properties from 5 to 350°K and Antiferromagnetic Transitions}},\ }\href
  {https://doi.org/10.1063/1.1673563} {\bibfield  {journal} {\bibinfo
  {journal} {The Journal of Chemical Physics}\ }\textbf {\bibinfo {volume}
  {52}},\ \bibinfo {pages} {3820} (\bibinfo {year} {1970})}\BibitemShut
  {NoStop}%
\bibitem [{\citenamefont {Pasternak}(1969)}]{MnTe2-T2}%
  \BibitemOpen
  \bibfield  {author} {\bibinfo {author} {\bibfnamefont {M.}~\bibnamefont
  {Pasternak}},\ }\bibfield  {title} {\bibinfo {title} {Magnetization near an
  iodine impurity in antiferromagnetic mn${\mathrm{te}}_{2}$},\ }\href
  {https://doi.org/10.1103/PhysRev.184.523} {\bibfield  {journal} {\bibinfo
  {journal} {Phys. Rev.}\ }\textbf {\bibinfo {volume} {184}},\ \bibinfo {pages}
  {523} (\bibinfo {year} {1969})}\BibitemShut {NoStop}%
\bibitem [{\citenamefont {Greedan}\ \emph {et~al.}(1997)\citenamefont
  {Greedan}, \citenamefont {Raju},\ and\ \citenamefont {Davidson}}]{LiMnO2-T1}%
  \BibitemOpen
  \bibfield  {author} {\bibinfo {author} {\bibfnamefont {J.}~\bibnamefont
  {Greedan}}, \bibinfo {author} {\bibfnamefont {N.}~\bibnamefont {Raju}},\ and\
  \bibinfo {author} {\bibfnamefont {I.}~\bibnamefont {Davidson}},\ }\bibfield
  {title} {\bibinfo {title} {{Long Range and Short Range Magnetic Order in
  Orthorhombic LiMnO2}},\ }\href
  {https://doi.org/https://doi.org/10.1006/jssc.1996.7189} {\bibfield
  {journal} {\bibinfo  {journal} {Journal of Solid State Chemistry}\ }\textbf
  {\bibinfo {volume} {128}},\ \bibinfo {pages} {209} (\bibinfo {year}
  {1997})}\BibitemShut {NoStop}%
\bibitem [{\citenamefont {Kellerman}\ \emph {et~al.}(2007)\citenamefont
  {Kellerman}, \citenamefont {Medvedeva}, \citenamefont {Gorshkov},
  \citenamefont {Kurbakov}, \citenamefont {Zubkov}, \citenamefont
  {Tyutyunnik},\ and\ \citenamefont {Trunov}}]{LiMnO2-T2}%
  \BibitemOpen
  \bibfield  {author} {\bibinfo {author} {\bibfnamefont {D.}~\bibnamefont
  {Kellerman}}, \bibinfo {author} {\bibfnamefont {J.}~\bibnamefont
  {Medvedeva}}, \bibinfo {author} {\bibfnamefont {V.}~\bibnamefont {Gorshkov}},
  \bibinfo {author} {\bibfnamefont {A.}~\bibnamefont {Kurbakov}}, \bibinfo
  {author} {\bibfnamefont {V.}~\bibnamefont {Zubkov}}, \bibinfo {author}
  {\bibfnamefont {A.}~\bibnamefont {Tyutyunnik}},\ and\ \bibinfo {author}
  {\bibfnamefont {V.}~\bibnamefont {Trunov}},\ }\bibfield  {title} {\bibinfo
  {title} {{Structural and magnetic properties of orthorhombic LixMnO2}},\
  }\href
  {https://doi.org/https://doi.org/10.1016/j.solidstatesciences.2006.11.013}
  {\bibfield  {journal} {\bibinfo  {journal} {Solid State Sciences}\ }\textbf
  {\bibinfo {volume} {9}},\ \bibinfo {pages} {196} (\bibinfo {year}
  {2007})}\BibitemShut {NoStop}%
\bibitem [{\citenamefont {Battle}\ \emph {et~al.}(1988)\citenamefont {Battle},
  \citenamefont {Gibb},\ and\ \citenamefont {Jones}}]{SrMnO3-T}%
  \BibitemOpen
  \bibfield  {author} {\bibinfo {author} {\bibfnamefont {P.}~\bibnamefont
  {Battle}}, \bibinfo {author} {\bibfnamefont {T.}~\bibnamefont {Gibb}},\ and\
  \bibinfo {author} {\bibfnamefont {C.}~\bibnamefont {Jones}},\ }\bibfield
  {title} {\bibinfo {title} {{The structural and magnetic properties of SrMnO3:
  A reinvestigation}},\ }\href
  {https://doi.org/https://doi.org/10.1016/0022-4596(88)90331-3} {\bibfield
  {journal} {\bibinfo  {journal} {Journal of Solid State Chemistry}\ }\textbf
  {\bibinfo {volume} {74}},\ \bibinfo {pages} {60} (\bibinfo {year}
  {1988})}\BibitemShut {NoStop}%
\bibitem [{\citenamefont {Okazaki}\ and\ \citenamefont
  {Suemune}(1961)}]{KNi(Mn)F3-T2}%
  \BibitemOpen
  \bibfield  {author} {\bibinfo {author} {\bibfnamefont {A.}~\bibnamefont
  {Okazaki}}\ and\ \bibinfo {author} {\bibfnamefont {Y.}~\bibnamefont
  {Suemune}},\ }\bibfield  {title} {\bibinfo {title} {{The Crystal Structures
  of KMnF3, KFeF3, KCoF3, KNiF3 and KCuF3 above and below their Néel
  Temperatures}},\ }\href {https://doi.org/10.1143/JPSJ.16.671} {\bibfield
  {journal} {\bibinfo  {journal} {Journal of the Physical Society of Japan}\
  }\textbf {\bibinfo {volume} {16}},\ \bibinfo {pages} {671} (\bibinfo {year}
  {1961})}\BibitemShut {NoStop}%
\bibitem [{\citenamefont {Hirakawa}\ \emph {et~al.}(1960)\citenamefont
  {Hirakawa}, \citenamefont {Hirakawa},\ and\ \citenamefont
  {Hashimoto}}]{KNi(Mn)F3-T3}%
  \BibitemOpen
  \bibfield  {author} {\bibinfo {author} {\bibfnamefont {K.}~\bibnamefont
  {Hirakawa}}, \bibinfo {author} {\bibfnamefont {K.}~\bibnamefont {Hirakawa}},\
  and\ \bibinfo {author} {\bibfnamefont {T.}~\bibnamefont {Hashimoto}},\
  }\bibfield  {title} {\bibinfo {title} {{Magnetic Properties of Potassium Iron
  Group Fluorides KMF$_3$}},\ }\href {https://doi.org/10.1143/JPSJ.15.2063}
  {\bibfield  {journal} {\bibinfo  {journal} {Journal of the Physical Society
  of Japan}\ }\textbf {\bibinfo {volume} {15}},\ \bibinfo {pages} {2063}
  (\bibinfo {year} {1960})}\BibitemShut {NoStop}%
\bibitem [{\citenamefont {Suemune}\ and\ \citenamefont
  {Ikawa}(1964)}]{KMn(Ni)F3-T3}%
  \BibitemOpen
  \bibfield  {author} {\bibinfo {author} {\bibfnamefont {Y.}~\bibnamefont
  {Suemune}}\ and\ \bibinfo {author} {\bibfnamefont {H.}~\bibnamefont
  {Ikawa}},\ }\bibfield  {title} {\bibinfo {title} {{Thermal Conductivity of
  KMnF3, KCoF3, KNiF3, and KZnF3 Single Crystals}},\ }\href
  {https://doi.org/10.1143/JPSJ.19.1686} {\bibfield  {journal} {\bibinfo
  {journal} {Journal of the Physical Society of Japan}\ }\textbf {\bibinfo
  {volume} {19}},\ \bibinfo {pages} {1686} (\bibinfo {year}
  {1964})}\BibitemShut {NoStop}%
\bibitem [{\citenamefont {Ogawa}(1959)}]{KMnF3-T4}%
  \BibitemOpen
  \bibfield  {author} {\bibinfo {author} {\bibfnamefont {S.}~\bibnamefont
  {Ogawa}},\ }\bibfield  {title} {\bibinfo {title} {{Antiferromagnetism in
  KMnF$_3$}},\ }\href {https://doi.org/10.1143/JPSJ.14.1115} {\bibfield
  {journal} {\bibinfo  {journal} {Journal of the Physical Society of Japan}\
  }\textbf {\bibinfo {volume} {14}},\ \bibinfo {pages} {1115} (\bibinfo {year}
  {1959})}\BibitemShut {NoStop}%
\bibitem [{\citenamefont {{Le Flem}}\ \emph {et~al.}(1982)\citenamefont {{Le
  Flem}}, \citenamefont {Brec}, \citenamefont {Ouvard}, \citenamefont
  {Louisy},\ and\ \citenamefont {Segransan}}]{MPX3-T}%
  \BibitemOpen
  \bibfield  {author} {\bibinfo {author} {\bibfnamefont {G.}~\bibnamefont {{Le
  Flem}}}, \bibinfo {author} {\bibfnamefont {R.}~\bibnamefont {Brec}}, \bibinfo
  {author} {\bibfnamefont {G.}~\bibnamefont {Ouvard}}, \bibinfo {author}
  {\bibfnamefont {A.}~\bibnamefont {Louisy}},\ and\ \bibinfo {author}
  {\bibfnamefont {P.}~\bibnamefont {Segransan}},\ }\bibfield  {title} {\bibinfo
  {title} {{Magnetic interactions in the layer compounds MPX3 (M = Mn, Fe, Ni;
  X = S, Se)}},\ }\href
  {https://doi.org/https://doi.org/10.1016/0022-3697(82)90156-1} {\bibfield
  {journal} {\bibinfo  {journal} {Journal of Physics and Chemistry of Solids}\
  }\textbf {\bibinfo {volume} {43}},\ \bibinfo {pages} {455} (\bibinfo {year}
  {1982})}\BibitemShut {NoStop}%
\bibitem [{\citenamefont {Joy}\ and\ \citenamefont {Vasudevan}(1992)}]{MPS3-T}%
  \BibitemOpen
  \bibfield  {author} {\bibinfo {author} {\bibfnamefont {P.~A.}\ \bibnamefont
  {Joy}}\ and\ \bibinfo {author} {\bibfnamefont {S.}~\bibnamefont
  {Vasudevan}},\ }\bibfield  {title} {\bibinfo {title} {{Magnetism in the
  layered transition-metal thiophosphates M${\mathrm{PS}}_{3}$ (M=Mn, Fe, and
  Ni)}},\ }\href {https://doi.org/10.1103/PhysRevB.46.5425} {\bibfield
  {journal} {\bibinfo  {journal} {Phys. Rev. B}\ }\textbf {\bibinfo {volume}
  {46}},\ \bibinfo {pages} {5425} (\bibinfo {year} {1992})}\BibitemShut
  {NoStop}%
\bibitem [{\citenamefont {Wiedenmann}\ \emph {et~al.}(1981)\citenamefont
  {Wiedenmann}, \citenamefont {Rossat-Mignod}, \citenamefont {Louisy},
  \citenamefont {Brec},\ and\ \citenamefont {Rouxel}}]{MnPSe3-T}%
  \BibitemOpen
  \bibfield  {author} {\bibinfo {author} {\bibfnamefont {A.}~\bibnamefont
  {Wiedenmann}}, \bibinfo {author} {\bibfnamefont {J.}~\bibnamefont
  {Rossat-Mignod}}, \bibinfo {author} {\bibfnamefont {A.}~\bibnamefont
  {Louisy}}, \bibinfo {author} {\bibfnamefont {R.}~\bibnamefont {Brec}},\ and\
  \bibinfo {author} {\bibfnamefont {J.}~\bibnamefont {Rouxel}},\ }\bibfield
  {title} {\bibinfo {title} {{Neutron diffraction study of the layered
  compounds MnPSe3 and FePSe3}},\ }\href
  {https://doi.org/https://doi.org/10.1016/0038-1098(81)90253-2} {\bibfield
  {journal} {\bibinfo  {journal} {Solid State Communications}\ }\textbf
  {\bibinfo {volume} {40}},\ \bibinfo {pages} {1067} (\bibinfo {year}
  {1981})}\BibitemShut {NoStop}%
\bibitem [{\citenamefont {Mays}(1963)}]{LiMnPO4-T1}%
  \BibitemOpen
  \bibfield  {author} {\bibinfo {author} {\bibfnamefont {J.~M.}\ \bibnamefont
  {Mays}},\ }\bibfield  {title} {\bibinfo {title} {{Nuclear Magnetic Resonances
  and Mn-O-P-O-Mn Superexchange Linkages in Paramagnetic and Antiferromagnetic
  LiMnP${\mathrm{O}}_{4}$}},\ }\href {https://doi.org/10.1103/PhysRev.131.38}
  {\bibfield  {journal} {\bibinfo  {journal} {Phys. Rev.}\ }\textbf {\bibinfo
  {volume} {131}},\ \bibinfo {pages} {38} (\bibinfo {year} {1963})}\BibitemShut
  {NoStop}%
\bibitem [{\citenamefont {Gnewuch}\ and\ \citenamefont
  {Rodriguez}(2020)}]{LiMnPO4-T2}%
  \BibitemOpen
  \bibfield  {author} {\bibinfo {author} {\bibfnamefont {S.}~\bibnamefont
  {Gnewuch}}\ and\ \bibinfo {author} {\bibfnamefont {E.~E.}\ \bibnamefont
  {Rodriguez}},\ }\bibfield  {title} {\bibinfo {title} {{Distinguishing the
  Intrinsic Antiferromagnetism in Polycrystalline LiCoPO\(_4\) and LiMnPO\(_4\)
  Olivines}},\ }\href {https://doi.org/10.1021/acs.inorgchem.9b03545}
  {\bibfield  {journal} {\bibinfo  {journal} {Inorganic Chemistry}\ }\textbf
  {\bibinfo {volume} {59}},\ \bibinfo {pages} {5883} (\bibinfo {year}
  {2020})},\ \bibinfo {note} {epub 2020 Apr 22}\BibitemShut {NoStop}%
\bibitem [{\citenamefont {Arčon}\ \emph {et~al.}(2004)\citenamefont {Arčon},
  \citenamefont {Zorko}, \citenamefont {Dominko},\ and\ \citenamefont
  {Jagličič}}]{LiMnPO4-T3}%
  \BibitemOpen
  \bibfield  {author} {\bibinfo {author} {\bibfnamefont {D.}~\bibnamefont
  {Arčon}}, \bibinfo {author} {\bibfnamefont {A.}~\bibnamefont {Zorko}},
  \bibinfo {author} {\bibfnamefont {R.}~\bibnamefont {Dominko}},\ and\ \bibinfo
  {author} {\bibfnamefont {Z.}~\bibnamefont {Jagličič}},\ }\bibfield  {title}
  {\bibinfo {title} {{A comparative study of magnetic properties of LiFePO4 and
  LiMnPO4}},\ }\href {https://doi.org/10.1088/0953-8984/16/30/014} {\bibfield
  {journal} {\bibinfo  {journal} {Journal of Physics: Condensed Matter}\
  }\textbf {\bibinfo {volume} {16}},\ \bibinfo {pages} {5531} (\bibinfo {year}
  {2004})}\BibitemShut {NoStop}%
\bibitem [{\citenamefont {Li}\ \emph {et~al.}(2009{\natexlab{b}})\citenamefont
  {Li}, \citenamefont {Tian}, \citenamefont {Chen}, \citenamefont {Zarestky},
  \citenamefont {Lynn},\ and\ \citenamefont {Vaknin}}]{LiMnPO4-T4}%
  \BibitemOpen
  \bibfield  {author} {\bibinfo {author} {\bibfnamefont {J.}~\bibnamefont
  {Li}}, \bibinfo {author} {\bibfnamefont {W.}~\bibnamefont {Tian}}, \bibinfo
  {author} {\bibfnamefont {Y.}~\bibnamefont {Chen}}, \bibinfo {author}
  {\bibfnamefont {J.~L.}\ \bibnamefont {Zarestky}}, \bibinfo {author}
  {\bibfnamefont {J.~W.}\ \bibnamefont {Lynn}},\ and\ \bibinfo {author}
  {\bibfnamefont {D.}~\bibnamefont {Vaknin}},\ }\bibfield  {title} {\bibinfo
  {title} {{Antiferromagnetism in the magnetoelectric effect single crystal
  ${\text{LiMnPO}}_{4}$}},\ }\href {https://doi.org/10.1103/PhysRevB.79.144410}
  {\bibfield  {journal} {\bibinfo  {journal} {Phys. Rev. B}\ }\textbf {\bibinfo
  {volume} {79}},\ \bibinfo {pages} {144410} (\bibinfo {year}
  {2009}{\natexlab{b}})}\BibitemShut {NoStop}%
\bibitem [{\citenamefont {Gilad}\ \emph {et~al.}(1963)\citenamefont {Gilad},
  \citenamefont {Greenshpan}, \citenamefont {Hillman},\ and\ \citenamefont
  {Shechter}}]{Fe2O30-T1}%
  \BibitemOpen
  \bibfield  {author} {\bibinfo {author} {\bibfnamefont {P.}~\bibnamefont
  {Gilad}}, \bibinfo {author} {\bibfnamefont {M.}~\bibnamefont {Greenshpan}},
  \bibinfo {author} {\bibfnamefont {P.}~\bibnamefont {Hillman}},\ and\ \bibinfo
  {author} {\bibfnamefont {H.}~\bibnamefont {Shechter}},\ }\bibfield  {title}
  {\bibinfo {title} {On the curie temperature of $\alpha$-fe2o3},\ }\href
  {https://doi.org/https://doi.org/10.1016/0031-9163(63)90311-1} {\bibfield
  {journal} {\bibinfo  {journal} {Physics Letters}\ }\textbf {\bibinfo {volume}
  {7}},\ \bibinfo {pages} {239} (\bibinfo {year} {1963})}\BibitemShut {NoStop}%
\bibitem [{\citenamefont {Oravova}\ \emph {et~al.}(2013)\citenamefont
  {Oravova}, \citenamefont {Zhang}, \citenamefont {Church}, \citenamefont
  {Harrison}, \citenamefont {Howard},\ and\ \citenamefont
  {Carpenter}}]{Fe2O3-T3}%
  \BibitemOpen
  \bibfield  {author} {\bibinfo {author} {\bibfnamefont {L.}~\bibnamefont
  {Oravova}}, \bibinfo {author} {\bibfnamefont {Z.}~\bibnamefont {Zhang}},
  \bibinfo {author} {\bibfnamefont {N.}~\bibnamefont {Church}}, \bibinfo
  {author} {\bibfnamefont {R.~J.}\ \bibnamefont {Harrison}}, \bibinfo {author}
  {\bibfnamefont {C.~J.}\ \bibnamefont {Howard}},\ and\ \bibinfo {author}
  {\bibfnamefont {M.~A.}\ \bibnamefont {Carpenter}},\ }\bibfield  {title}
  {\bibinfo {title} {{Elastic and anelastic relaxations accompanying magnetic
  ordering and spin-flop transitions in hematite, Fe2O3}},\ }\href
  {https://doi.org/10.1088/0953-8984/25/11/116006} {\bibfield  {journal}
  {\bibinfo  {journal} {Journal of Physics: Condensed Matter}\ }\textbf
  {\bibinfo {volume} {25}},\ \bibinfo {pages} {116006} (\bibinfo {year}
  {2013})}\BibitemShut {NoStop}%
\bibitem [{\citenamefont {Morin}(1950)}]{Fe2O3-T4}%
  \BibitemOpen
  \bibfield  {author} {\bibinfo {author} {\bibfnamefont {F.~J.}\ \bibnamefont
  {Morin}},\ }\bibfield  {title} {\bibinfo {title} {{Magnetic Susceptibility of
  $\ensuremath{\alpha}{\mathrm{Fe}}_{2}{\mathrm{O}}_{3}$ and
  $\ensuremath{\alpha}{\mathrm{Fe}}_{2}{\mathrm{O}}_{3}$ with Added
  Titanium}},\ }\href {https://doi.org/10.1103/PhysRev.78.819.2} {\bibfield
  {journal} {\bibinfo  {journal} {Phys. Rev.}\ }\textbf {\bibinfo {volume}
  {78}},\ \bibinfo {pages} {819} (\bibinfo {year} {1950})}\BibitemShut
  {NoStop}%
\bibitem [{\citenamefont {{Nešković, N. B. and Babić, B. and
  Konstantinović, J.}}(1977)}]{FE2O3-T5}%
  \BibitemOpen
  \bibfield  {author} {\bibinfo {author} {\bibnamefont {{Nešković, N. B. and
  Babić, B. and Konstantinović, J.}}},\ }\bibfield  {title} {\bibinfo {title}
  {High temperature anomalous behaviour of the crystal lattice of hematite},\
  }\href {https://doi.org/https://doi.org/10.1002/pssa.2210410252} {\bibfield
  {journal} {\bibinfo  {journal} {physica status solidi (a)}\ }\textbf
  {\bibinfo {volume} {41}},\ \bibinfo {pages} {K133} (\bibinfo {year}
  {1977})}\BibitemShut {NoStop}%
\bibitem [{\citenamefont {Białek}\ \emph {et~al.}(2022)\citenamefont
  {Białek}, \citenamefont {Zhang}, \citenamefont {Yu},\ and\ \citenamefont
  {Ansermet}}]{Fe2O3-T6}%
  \BibitemOpen
  \bibfield  {author} {\bibinfo {author} {\bibfnamefont {M.}~\bibnamefont
  {Białek}}, \bibinfo {author} {\bibfnamefont {J.}~\bibnamefont {Zhang}},
  \bibinfo {author} {\bibfnamefont {H.}~\bibnamefont {Yu}},\ and\ \bibinfo
  {author} {\bibfnamefont {J.-P.}\ \bibnamefont {Ansermet}},\ }\bibfield
  {title} {\bibinfo {title} {{Antiferromagnetic resonance in $\alpha$-Fe2O3 up
  to its N\'eel temperature}},\ }\href {https://doi.org/10.1063/5.0094868}
  {\bibfield  {journal} {\bibinfo  {journal} {Applied Physics Letters}\
  }\textbf {\bibinfo {volume} {121}},\ \bibinfo {pages} {032401} (\bibinfo
  {year} {2022})}\BibitemShut {NoStop}%
\bibitem [{\citenamefont {Sosnowska}\ \emph {et~al.}(1982)\citenamefont
  {Sosnowska}, \citenamefont {Neumaier},\ and\ \citenamefont
  {Steichele}}]{BiFeO3-T1}%
  \BibitemOpen
  \bibfield  {author} {\bibinfo {author} {\bibfnamefont {I.}~\bibnamefont
  {Sosnowska}}, \bibinfo {author} {\bibfnamefont {T.~P.}\ \bibnamefont
  {Neumaier}},\ and\ \bibinfo {author} {\bibfnamefont {E.}~\bibnamefont
  {Steichele}},\ }\bibfield  {title} {\bibinfo {title} {{Spiral magnetic
  ordering in bismuth ferrite}},\ }\href
  {https://doi.org/10.1088/0022-3719/15/23/020} {\bibfield  {journal} {\bibinfo
   {journal} {Journal of Physics C: Solid State Physics}\ }\textbf {\bibinfo
  {volume} {15}},\ \bibinfo {pages} {4835} (\bibinfo {year}
  {1982})}\BibitemShut {NoStop}%
\bibitem [{\citenamefont {Fischer}\ \emph {et~al.}(1980)\citenamefont
  {Fischer}, \citenamefont {Polomska}, \citenamefont {Sosnowska},\ and\
  \citenamefont {Szymanski}}]{BiFeO3-T2}%
  \BibitemOpen
  \bibfield  {author} {\bibinfo {author} {\bibfnamefont {P.}~\bibnamefont
  {Fischer}}, \bibinfo {author} {\bibfnamefont {M.}~\bibnamefont {Polomska}},
  \bibinfo {author} {\bibfnamefont {I.}~\bibnamefont {Sosnowska}},\ and\
  \bibinfo {author} {\bibfnamefont {M.}~\bibnamefont {Szymanski}},\ }\bibfield
  {title} {\bibinfo {title} {{Temperature dependence of the crystal and
  magnetic structures of BiFeO3}},\ }\href
  {https://doi.org/10.1088/0022-3719/13/10/012} {\bibfield  {journal} {\bibinfo
   {journal} {Journal of Physics C: Solid State Physics}\ }\textbf {\bibinfo
  {volume} {13}},\ \bibinfo {pages} {1931} (\bibinfo {year}
  {1980})}\BibitemShut {NoStop}%
\bibitem [{\citenamefont {Koehler}\ and\ \citenamefont
  {Wollan}(1957)}]{LaFeO3-T1}%
  \BibitemOpen
  \bibfield  {author} {\bibinfo {author} {\bibfnamefont {W.}~\bibnamefont
  {Koehler}}\ and\ \bibinfo {author} {\bibfnamefont {E.}~\bibnamefont
  {Wollan}},\ }\bibfield  {title} {\bibinfo {title} {{Neutron-diffraction study
  of the magnetic properties of perovskite-like compounds LaBO3}},\ }\href
  {https://doi.org/https://doi.org/10.1016/0022-3697(57)90095-1} {\bibfield
  {journal} {\bibinfo  {journal} {Journal of Physics and Chemistry of Solids}\
  }\textbf {\bibinfo {volume} {2}},\ \bibinfo {pages} {100} (\bibinfo {year}
  {1957})}\BibitemShut {NoStop}%
\bibitem [{\citenamefont {Park}\ \emph {et~al.}(2018)\citenamefont {Park},
  \citenamefont {Sim}, \citenamefont {Leiner}, \citenamefont {Yoshida},
  \citenamefont {Jeong}, \citenamefont {ichiro Yano}, \citenamefont {Gardner},
  \citenamefont {Bourges}, \citenamefont {Klicpera}, \citenamefont
  {Sechovský}, \citenamefont {Boehm},\ and\ \citenamefont {Park}}]{LaFeO3-T2}%
  \BibitemOpen
  \bibfield  {author} {\bibinfo {author} {\bibfnamefont {K.}~\bibnamefont
  {Park}}, \bibinfo {author} {\bibfnamefont {H.}~\bibnamefont {Sim}}, \bibinfo
  {author} {\bibfnamefont {J.~C.}\ \bibnamefont {Leiner}}, \bibinfo {author}
  {\bibfnamefont {Y.}~\bibnamefont {Yoshida}}, \bibinfo {author} {\bibfnamefont
  {J.}~\bibnamefont {Jeong}}, \bibinfo {author} {\bibfnamefont
  {S.}~\bibnamefont {ichiro Yano}}, \bibinfo {author} {\bibfnamefont
  {J.}~\bibnamefont {Gardner}}, \bibinfo {author} {\bibfnamefont
  {P.}~\bibnamefont {Bourges}}, \bibinfo {author} {\bibfnamefont
  {M.}~\bibnamefont {Klicpera}}, \bibinfo {author} {\bibfnamefont
  {V.}~\bibnamefont {Sechovský}}, \bibinfo {author} {\bibfnamefont
  {M.}~\bibnamefont {Boehm}},\ and\ \bibinfo {author} {\bibfnamefont {J.-G.}\
  \bibnamefont {Park}},\ }\bibfield  {title} {\bibinfo {title} {{Low-energy
  spin dynamics of orthoferrites AFeO3 (A=Y, La, Bi)}},\ }\href
  {https://doi.org/10.1088/1361-648X/aac06b} {\bibfield  {journal} {\bibinfo
  {journal} {Journal of Physics: Condensed Matter}\ }\textbf {\bibinfo {volume}
  {30}},\ \bibinfo {pages} {235802} (\bibinfo {year} {2018})}\BibitemShut
  {NoStop}%
\bibitem [{\citenamefont {Shen}\ \emph {et~al.}(2009)\citenamefont {Shen},
  \citenamefont {Xu}, \citenamefont {Wu}, \citenamefont {Zhao},\ and\
  \citenamefont {Shi}}]{YFeO3-T}%
  \BibitemOpen
  \bibfield  {author} {\bibinfo {author} {\bibfnamefont {H.}~\bibnamefont
  {Shen}}, \bibinfo {author} {\bibfnamefont {J.}~\bibnamefont {Xu}}, \bibinfo
  {author} {\bibfnamefont {A.}~\bibnamefont {Wu}}, \bibinfo {author}
  {\bibfnamefont {J.}~\bibnamefont {Zhao}},\ and\ \bibinfo {author}
  {\bibfnamefont {M.}~\bibnamefont {Shi}},\ }\bibfield  {title} {\bibinfo
  {title} {{Magnetic and thermal properties of perovskite YFeO3 single
  crystals}},\ }\href
  {https://doi.org/https://doi.org/10.1016/j.mseb.2008.12.020} {\bibfield
  {journal} {\bibinfo  {journal} {Materials Science and Engineering: B}\
  }\textbf {\bibinfo {volume} {157}},\ \bibinfo {pages} {77} (\bibinfo {year}
  {2009})}\BibitemShut {NoStop}%
\bibitem [{\citenamefont {Rule}\ \emph {et~al.}(2007)\citenamefont {Rule},
  \citenamefont {McIntyre}, \citenamefont {Kennedy},\ and\ \citenamefont
  {Hicks}}]{FePS3-T}%
  \BibitemOpen
  \bibfield  {author} {\bibinfo {author} {\bibfnamefont {K.~C.}\ \bibnamefont
  {Rule}}, \bibinfo {author} {\bibfnamefont {G.~J.}\ \bibnamefont {McIntyre}},
  \bibinfo {author} {\bibfnamefont {S.~J.}\ \bibnamefont {Kennedy}},\ and\
  \bibinfo {author} {\bibfnamefont {T.~J.}\ \bibnamefont {Hicks}},\ }\bibfield
  {title} {\bibinfo {title} {{Single-crystal and powder neutron diffraction
  experiments on $\mathrm{Fe}\mathrm{P}{\mathrm{S}}_{3}$: Search for the
  magnetic structure}},\ }\href {https://doi.org/10.1103/PhysRevB.76.134402}
  {\bibfield  {journal} {\bibinfo  {journal} {Phys. Rev. B}\ }\textbf {\bibinfo
  {volume} {76}},\ \bibinfo {pages} {134402} (\bibinfo {year}
  {2007})}\BibitemShut {NoStop}%
\bibitem [{\citenamefont {Dehn}\ \emph {et~al.}(1968)\citenamefont {Dehn},
  \citenamefont {Newnham},\ and\ \citenamefont {Mulay}}]{Fe2TeO6-1}%
  \BibitemOpen
  \bibfield  {author} {\bibinfo {author} {\bibfnamefont {J.~T.}\ \bibnamefont
  {Dehn}}, \bibinfo {author} {\bibfnamefont {R.~E.}\ \bibnamefont {Newnham}},\
  and\ \bibinfo {author} {\bibfnamefont {L.~N.}\ \bibnamefont {Mulay}},\
  }\bibfield  {title} {\bibinfo {title} {{Mixed Magnetic Ordering: Magnetic
  Susceptibility and Mössbauer Studies on Iron(III) Tellurate (Fe2TeO6)}},\
  }\href {https://doi.org/10.1063/1.1670571} {\bibfield  {journal} {\bibinfo
  {journal} {The Journal of Chemical Physics}\ }\textbf {\bibinfo {volume}
  {49}},\ \bibinfo {pages} {3201} (\bibinfo {year} {1968})}\BibitemShut
  {NoStop}%
\bibitem [{\citenamefont {Yamaguchi}\ and\ \citenamefont
  {Ishikawa}(1994)}]{Fe2TeO6-2}%
  \BibitemOpen
  \bibfield  {author} {\bibinfo {author} {\bibfnamefont {M.}~\bibnamefont
  {Yamaguchi}}\ and\ \bibinfo {author} {\bibfnamefont {M.}~\bibnamefont
  {Ishikawa}},\ }\bibfield  {title} {\bibinfo {title} {{Magnetic Phase
  Transitions in Inverse Trirutile-Type Compounds}},\ }\href
  {https://doi.org/10.1143/JPSJ.63.1666} {\bibfield  {journal} {\bibinfo
  {journal} {Journal of the Physical Society of Japan}\ }\textbf {\bibinfo
  {volume} {63}},\ \bibinfo {pages} {1666} (\bibinfo {year}
  {1994})}\BibitemShut {NoStop}%
\bibitem [{\citenamefont {Buksphan}\ \emph {et~al.}(1972)\citenamefont
  {Buksphan}, \citenamefont {Fischer},\ and\ \citenamefont
  {Hornreich}}]{Fe2TeO6-3}%
  \BibitemOpen
  \bibfield  {author} {\bibinfo {author} {\bibfnamefont {S.}~\bibnamefont
  {Buksphan}}, \bibinfo {author} {\bibfnamefont {E.}~\bibnamefont {Fischer}},\
  and\ \bibinfo {author} {\bibfnamefont {R.}~\bibnamefont {Hornreich}},\
  }\bibfield  {title} {\bibinfo {title} {{Magnetoelectric and Mössbauer
  studies of Fe2TeO6}},\ }\href
  {https://doi.org/https://doi.org/10.1016/0038-1098(72)90580-7} {\bibfield
  {journal} {\bibinfo  {journal} {Solid State Communications}\ }\textbf
  {\bibinfo {volume} {10}},\ \bibinfo {pages} {657} (\bibinfo {year}
  {1972})}\BibitemShut {NoStop}%
\bibitem [{\citenamefont {Huh}\ \emph {et~al.}(2015)\citenamefont {Huh},
  \citenamefont {Prots}, \citenamefont {Adler}, \citenamefont {Tjeng},\ and\
  \citenamefont {Valldor}}]{SrFe2S2O-T}%
  \BibitemOpen
  \bibfield  {author} {\bibinfo {author} {\bibfnamefont {S.}~\bibnamefont
  {Huh}}, \bibinfo {author} {\bibfnamefont {Y.}~\bibnamefont {Prots}}, \bibinfo
  {author} {\bibfnamefont {P.}~\bibnamefont {Adler}}, \bibinfo {author}
  {\bibfnamefont {L.~H.}\ \bibnamefont {Tjeng}},\ and\ \bibinfo {author}
  {\bibfnamefont {M.}~\bibnamefont {Valldor}},\ }\bibfield  {title} {\bibinfo
  {title} {{Synthesis and Characterization of Frustrated Spin Ladders SrFe2S2O
  and SrFe2Se2O}},\ }\href
  {https://doi.org/https://doi.org/10.1002/ejic.201500385} {\bibfield
  {journal} {\bibinfo  {journal} {European Journal of Inorganic Chemistry}\
  }\textbf {\bibinfo {volume} {2015}},\ \bibinfo {pages} {2982} (\bibinfo
  {year} {2015})}\BibitemShut {NoStop}%
\bibitem [{\citenamefont {Deng}\ \emph {et~al.}(2012)\citenamefont {Deng},
  \citenamefont {Chang}, \citenamefont {Wang}, \citenamefont {Zhang},
  \citenamefont {Ma},\ and\ \citenamefont {Wang}}]{CoWO4-T}%
  \BibitemOpen
  \bibfield  {author} {\bibinfo {author} {\bibfnamefont {J.}~\bibnamefont
  {Deng}}, \bibinfo {author} {\bibfnamefont {L.}~\bibnamefont {Chang}},
  \bibinfo {author} {\bibfnamefont {P.}~\bibnamefont {Wang}}, \bibinfo {author}
  {\bibfnamefont {E.}~\bibnamefont {Zhang}}, \bibinfo {author} {\bibfnamefont
  {J.}~\bibnamefont {Ma}},\ and\ \bibinfo {author} {\bibfnamefont
  {T.}~\bibnamefont {Wang}},\ }\bibfield  {title} {\bibinfo {title}
  {{Preparation and magnetic properties of CoWO4 nanocrystals}},\ }\href
  {https://doi.org/https://doi.org/10.1002/crat.201200130} {\bibfield
  {journal} {\bibinfo  {journal} {Crystal Research and Technology}\ }\textbf
  {\bibinfo {volume} {47}},\ \bibinfo {pages} {1004} (\bibinfo {year}
  {2012})}\BibitemShut {NoStop}%
\bibitem [{\citenamefont {Tomlinson}\ \emph {et~al.}(1955)\citenamefont
  {Tomlinson}, \citenamefont {Domash}, \citenamefont {Hay},\ and\ \citenamefont
  {Montgomery}}]{NiO-T}%
  \BibitemOpen
  \bibfield  {author} {\bibinfo {author} {\bibfnamefont {J.~R.}\ \bibnamefont
  {Tomlinson}}, \bibinfo {author} {\bibfnamefont {L.}~\bibnamefont {Domash}},
  \bibinfo {author} {\bibfnamefont {R.~G.}\ \bibnamefont {Hay}},\ and\ \bibinfo
  {author} {\bibfnamefont {C.~W.}\ \bibnamefont {Montgomery}},\ }\bibfield
  {title} {\bibinfo {title} {{The High Temperature Heat Content of Nickel
  Oxide}},\ }\href {https://doi.org/10.1021/ja01609a032} {\bibfield  {journal}
  {\bibinfo  {journal} {Journal of the American Chemical Society}\ }\textbf
  {\bibinfo {volume} {77}},\ \bibinfo {pages} {909} (\bibinfo {year}
  {1955})}\BibitemShut {NoStop}%
\bibitem [{\citenamefont {Cooke}\ \emph {et~al.}(1965)\citenamefont {Cooke},
  \citenamefont {Gehring},\ and\ \citenamefont {Lazenby}}]{NiF2-T1}%
  \BibitemOpen
  \bibfield  {author} {\bibinfo {author} {\bibfnamefont {A.~H.}\ \bibnamefont
  {Cooke}}, \bibinfo {author} {\bibfnamefont {K.~A.}\ \bibnamefont {Gehring}},\
  and\ \bibinfo {author} {\bibfnamefont {R.}~\bibnamefont {Lazenby}},\
  }\bibfield  {title} {\bibinfo {title} {{The magnetic properties of NiF2}},\
  }\href {https://doi.org/10.1088/0370-1328/85/5/315} {\bibfield  {journal}
  {\bibinfo  {journal} {Proceedings of the Physical Society}\ }\textbf
  {\bibinfo {volume} {85}},\ \bibinfo {pages} {967} (\bibinfo {year}
  {1965})}\BibitemShut {NoStop}%
\bibitem [{\citenamefont {Fleury}(1969)}]{NiF2-T2}%
  \BibitemOpen
  \bibfield  {author} {\bibinfo {author} {\bibfnamefont {P.~A.}\ \bibnamefont
  {Fleury}},\ }\bibfield  {title} {\bibinfo {title} {Paramagnetic spin waves
  and correlation functions in ni${\mathrm{f}}_{2}$},\ }\href
  {https://doi.org/10.1103/PhysRev.180.591} {\bibfield  {journal} {\bibinfo
  {journal} {Phys. Rev.}\ }\textbf {\bibinfo {volume} {180}},\ \bibinfo {pages}
  {591} (\bibinfo {year} {1969})}\BibitemShut {NoStop}%
\bibitem [{\citenamefont {Day}\ \emph {et~al.}(1976)\citenamefont {Day},
  \citenamefont {Dinsdale}, \citenamefont {Krausz},\ and\ \citenamefont
  {Robbins}}]{NiBr2-T}%
  \BibitemOpen
  \bibfield  {author} {\bibinfo {author} {\bibfnamefont {P.}~\bibnamefont
  {Day}}, \bibinfo {author} {\bibfnamefont {A.}~\bibnamefont {Dinsdale}},
  \bibinfo {author} {\bibfnamefont {E.~R.}\ \bibnamefont {Krausz}},\ and\
  \bibinfo {author} {\bibfnamefont {D.~J.}\ \bibnamefont {Robbins}},\
  }\bibfield  {title} {\bibinfo {title} {{Optical and neutron diffraction study
  of the magnetic phase diagram of NiBr2}},\ }\href
  {https://doi.org/10.1088/0022-3719/9/13/008} {\bibfield  {journal} {\bibinfo
  {journal} {Journal of Physics C: Solid State Physics}\ }\textbf {\bibinfo
  {volume} {9}},\ \bibinfo {pages} {2481} (\bibinfo {year} {1976})}\BibitemShut
  {NoStop}%
\bibitem [{\citenamefont {Busey}\ and\ \citenamefont
  {Giauque}(1952)}]{NiCl2-T1}%
  \BibitemOpen
  \bibfield  {author} {\bibinfo {author} {\bibfnamefont {R.~H.}\ \bibnamefont
  {Busey}}\ and\ \bibinfo {author} {\bibfnamefont {W.~F.}\ \bibnamefont
  {Giauque}},\ }\bibfield  {title} {\bibinfo {title} {{The Heat Capacity of
  Anhydrous NiCl2 from 15 to 300°K. The Antiferromagnetic Anomaly near 52°K.
  Entropy and Free Energy1}},\ }\href {https://doi.org/10.1021/ja01137a062}
  {\bibfield  {journal} {\bibinfo  {journal} {Journal of the American Chemical
  Society}\ }\textbf {\bibinfo {volume} {74}},\ \bibinfo {pages} {4443}
  (\bibinfo {year} {1952})}\BibitemShut {NoStop}%
\bibitem [{\citenamefont {Lindgard}\ \emph {et~al.}(1975)\citenamefont
  {Lindgard}, \citenamefont {Birgeneau}, \citenamefont {Guggenheim},\ and\
  \citenamefont {Als-Nielsen}}]{NiCl2-T2}%
  \BibitemOpen
  \bibfield  {author} {\bibinfo {author} {\bibfnamefont {P.~A.}\ \bibnamefont
  {Lindgard}}, \bibinfo {author} {\bibfnamefont {R.~J.}\ \bibnamefont
  {Birgeneau}}, \bibinfo {author} {\bibfnamefont {H.~J.}\ \bibnamefont
  {Guggenheim}},\ and\ \bibinfo {author} {\bibfnamefont {J.}~\bibnamefont
  {Als-Nielsen}},\ }\bibfield  {title} {\bibinfo {title} {{Spin-wave dispersion
  and sublattice magnetization in NiCl2}},\ }\href
  {https://doi.org/10.1088/0022-3719/8/7/021} {\bibfield  {journal} {\bibinfo
  {journal} {Journal of Physics C: Solid State Physics}\ }\textbf {\bibinfo
  {volume} {8}},\ \bibinfo {pages} {1059} (\bibinfo {year} {1975})}\BibitemShut
  {NoStop}%
\bibitem [{\citenamefont {Prosnikov}\ \emph {et~al.}(2017)\citenamefont
  {Prosnikov}, \citenamefont {Davydov}, \citenamefont {Smirnov}, \citenamefont
  {Volkov}, \citenamefont {Pisarev}, \citenamefont {Becker},\ and\
  \citenamefont {Bohat\'y}}]{NiWO4-T}%
  \BibitemOpen
  \bibfield  {author} {\bibinfo {author} {\bibfnamefont {M.~A.}\ \bibnamefont
  {Prosnikov}}, \bibinfo {author} {\bibfnamefont {V.~Y.}\ \bibnamefont
  {Davydov}}, \bibinfo {author} {\bibfnamefont {A.~N.}\ \bibnamefont
  {Smirnov}}, \bibinfo {author} {\bibfnamefont {M.~P.}\ \bibnamefont {Volkov}},
  \bibinfo {author} {\bibfnamefont {R.~V.}\ \bibnamefont {Pisarev}}, \bibinfo
  {author} {\bibfnamefont {P.}~\bibnamefont {Becker}},\ and\ \bibinfo {author}
  {\bibfnamefont {L.}~\bibnamefont {Bohat\'y}},\ }\bibfield  {title} {\bibinfo
  {title} {{Lattice and spin dynamics in a low-symmetry antiferromagnet
  ${\mathrm{NiWO}}_{4}$}},\ }\href {https://doi.org/10.1103/PhysRevB.96.014428}
  {\bibfield  {journal} {\bibinfo  {journal} {Phys. Rev. B}\ }\textbf {\bibinfo
  {volume} {96}},\ \bibinfo {pages} {014428} (\bibinfo {year}
  {2017})}\BibitemShut {NoStop}%
\bibitem [{\citenamefont {Nakajima}\ \emph {et~al.}(1993)\citenamefont
  {Nakajima}, \citenamefont {Yamada}, \citenamefont {Hosoya}, \citenamefont
  {Omata},\ and\ \citenamefont {Endoh}}]{La2NiO4-T2}%
  \BibitemOpen
  \bibfield  {author} {\bibinfo {author} {\bibfnamefont {K.}~\bibnamefont
  {Nakajima}}, \bibinfo {author} {\bibfnamefont {K.}~\bibnamefont {Yamada}},
  \bibinfo {author} {\bibfnamefont {S.}~\bibnamefont {Hosoya}}, \bibinfo
  {author} {\bibfnamefont {T.}~\bibnamefont {Omata}},\ and\ \bibinfo {author}
  {\bibfnamefont {Y.}~\bibnamefont {Endoh}},\ }\bibfield  {title} {\bibinfo
  {title} {{Spin-Wave Excitations in Two Dimensional Antiferromagnet of
  Stoichiometric La2NiO4}},\ }\href {https://doi.org/10.1143/JPSJ.62.4438}
  {\bibfield  {journal} {\bibinfo  {journal} {Journal of the Physical Society
  of Japan}\ }\textbf {\bibinfo {volume} {62}},\ \bibinfo {pages} {4438}
  (\bibinfo {year} {1993})}\BibitemShut {NoStop}%
\bibitem [{\citenamefont {Birgeneau}\ \emph {et~al.}(1970)\citenamefont
  {Birgeneau}, \citenamefont {Skalyo},\ and\ \citenamefont
  {Shirane}}]{K2NiF4-T1}%
  \BibitemOpen
  \bibfield  {author} {\bibinfo {author} {\bibfnamefont {R.~J.}\ \bibnamefont
  {Birgeneau}}, \bibinfo {author} {\bibfnamefont {J.}~\bibnamefont {Skalyo},
  \bibfnamefont {J.}},\ and\ \bibinfo {author} {\bibfnamefont {G.}~\bibnamefont
  {Shirane}},\ }\bibfield  {title} {\bibinfo {title} {{Phase Transitions and
  Magnetic Correlations in Two‐Dimensional Antiferromagnets}},\ }\href
  {https://doi.org/10.1063/1.1658917} {\bibfield  {journal} {\bibinfo
  {journal} {Journal of Applied Physics}\ }\textbf {\bibinfo {volume} {41}},\
  \bibinfo {pages} {1303} (\bibinfo {year} {1970})}\BibitemShut {NoStop}%
\bibitem [{\citenamefont {Salamon}\ and\ \citenamefont
  {Ikeda}(1973)}]{K2NiF4-T2}%
  \BibitemOpen
  \bibfield  {author} {\bibinfo {author} {\bibfnamefont {M.~B.}\ \bibnamefont
  {Salamon}}\ and\ \bibinfo {author} {\bibfnamefont {H.}~\bibnamefont
  {Ikeda}},\ }\bibfield  {title} {\bibinfo {title} {{Specific Heat of
  Two-Dimensional Heisenberg Antiferromagnets:
  ${\mathrm{K}}_{2}$Mn${\mathrm{F}}_{4}$ and
  ${\mathrm{K}}_{2}$Ni${\mathrm{F}}_{4}$}},\ }\href
  {https://doi.org/10.1103/PhysRevB.7.2017} {\bibfield  {journal} {\bibinfo
  {journal} {Phys. Rev. B}\ }\textbf {\bibinfo {volume} {7}},\ \bibinfo {pages}
  {2017} (\bibinfo {year} {1973})}\BibitemShut {NoStop}%
\bibitem [{\citenamefont {Legrand}\ and\ \citenamefont
  {Plumier}(1962)}]{K2NiF4-T3}%
  \BibitemOpen
  \bibfield  {author} {\bibinfo {author} {\bibfnamefont {E.}~\bibnamefont
  {Legrand}}\ and\ \bibinfo {author} {\bibfnamefont {R.}~\bibnamefont
  {Plumier}},\ }\bibfield  {title} {\bibinfo {title} {{Neutron diffraction
  investigation of antiferromagnetic K2NiF4}},\ }\href
  {https://doi.org/https://doi.org/10.1002/pssb.19620020306} {\bibfield
  {journal} {\bibinfo  {journal} {physica status solidi (b)}\ }\textbf
  {\bibinfo {volume} {2}},\ \bibinfo {pages} {317} (\bibinfo {year}
  {1962})}\BibitemShut {NoStop}%
\bibitem [{\citenamefont {Pages}\ \emph {et~al.}(2022)\citenamefont {Pages},
  \citenamefont {Soh}, \citenamefont {Fesharaki}, \citenamefont {Ronnow},\ and\
  \citenamefont {Ahmadvand}}]{KNiPO4-T2}%
  \BibitemOpen
  \bibfield  {author} {\bibinfo {author} {\bibfnamefont {A.}~\bibnamefont
  {Pages}}, \bibinfo {author} {\bibfnamefont {J.-R.}\ \bibnamefont {Soh}},
  \bibinfo {author} {\bibfnamefont {M.~J.}\ \bibnamefont {Fesharaki}}, \bibinfo
  {author} {\bibfnamefont {H.~M.}\ \bibnamefont {Ronnow}},\ and\ \bibinfo
  {author} {\bibfnamefont {H.}~\bibnamefont {Ahmadvand}},\ }\href
  {https://arxiv.org/abs/2207.06969} {\bibinfo {title} {{Revisiting the
  magnetic and crystal structure of multiferroic KNiPO$_4$}}} (\bibinfo {year}
  {2022}),\ \Eprint {https://arxiv.org/abs/2207.06969} {arXiv:2207.06969}
  \BibitemShut {NoStop}%
\bibitem [{\citenamefont {Kharchenko}\ \emph {et~al.}(2003)\citenamefont
  {Kharchenko}, \citenamefont {Kharcheno}, \citenamefont {Baran},\ and\
  \citenamefont {Szymczak}}]{LiNiPO4-T1}%
  \BibitemOpen
  \bibfield  {author} {\bibinfo {author} {\bibfnamefont {Y.~N.}\ \bibnamefont
  {Kharchenko}}, \bibinfo {author} {\bibfnamefont {N.~F.}\ \bibnamefont
  {Kharcheno}}, \bibinfo {author} {\bibfnamefont {M.}~\bibnamefont {Baran}},\
  and\ \bibinfo {author} {\bibfnamefont {R.}~\bibnamefont {Szymczak}},\
  }\bibfield  {title} {\bibinfo {title} {{Weak ferromagnetism and an
  intermediate incommensurate antiferromagnetic phase in LiNiPO4}},\ }\href
  {https://doi.org/10.1063/1.1596583} {\bibfield  {journal} {\bibinfo
  {journal} {Low Temperature Physics}\ }\textbf {\bibinfo {volume} {29}},\
  \bibinfo {pages} {579} (\bibinfo {year} {2003})}\BibitemShut {NoStop}%
\bibitem [{\citenamefont {Vaknin}\ \emph {et~al.}(2004)\citenamefont {Vaknin},
  \citenamefont {Zarestky}, \citenamefont {Rivera},\ and\ \citenamefont
  {Schmid}}]{LiNiPO4-T2}%
  \BibitemOpen
  \bibfield  {author} {\bibinfo {author} {\bibfnamefont {D.}~\bibnamefont
  {Vaknin}}, \bibinfo {author} {\bibfnamefont {J.~L.}\ \bibnamefont
  {Zarestky}}, \bibinfo {author} {\bibfnamefont {J.-P.}\ \bibnamefont
  {Rivera}},\ and\ \bibinfo {author} {\bibfnamefont {H.}~\bibnamefont
  {Schmid}},\ }\bibfield  {title} {\bibinfo {title}
  {{Commensurate-Incommensurate Magnetic Phase Transition in Magnetoelectric
  Single Crystal ${\mathrm{LiNiPO}}_{4}$}},\ }\href
  {https://doi.org/10.1103/PhysRevLett.92.207201} {\bibfield  {journal}
  {\bibinfo  {journal} {Phys. Rev. Lett.}\ }\textbf {\bibinfo {volume} {92}},\
  \bibinfo {pages} {207201} (\bibinfo {year} {2004})}\BibitemShut {NoStop}%
\bibitem [{\citenamefont {Hu}\ and\ \citenamefont
  {Johnston}(1953)}]{CuO-T(220)}%
  \BibitemOpen
  \bibfield  {author} {\bibinfo {author} {\bibfnamefont {J.-H.}\ \bibnamefont
  {Hu}}\ and\ \bibinfo {author} {\bibfnamefont {H.~L.}\ \bibnamefont
  {Johnston}},\ }\bibfield  {title} {\bibinfo {title} {{Low Temperature Heat
  Capacities of Inorganic Solids. XVI. Heat Capacity of Cupric Oxide from 15 to
  300 °K.1}},\ }\href {https://doi.org/10.1021/ja01106a056} {\bibfield
  {journal} {\bibinfo  {journal} {Journal of the American Chemical Society}\
  }\textbf {\bibinfo {volume} {75}},\ \bibinfo {pages} {2471} (\bibinfo {year}
  {1953})}\BibitemShut {NoStop}%
\bibitem [{\citenamefont {Yang}\ \emph {et~al.}(1988)\citenamefont {Yang},
  \citenamefont {Tranquada},\ and\ \citenamefont {Shirane}}]{CuO-T(225)}%
  \BibitemOpen
  \bibfield  {author} {\bibinfo {author} {\bibfnamefont {B.~X.}\ \bibnamefont
  {Yang}}, \bibinfo {author} {\bibfnamefont {J.~M.}\ \bibnamefont
  {Tranquada}},\ and\ \bibinfo {author} {\bibfnamefont {G.}~\bibnamefont
  {Shirane}},\ }\bibfield  {title} {\bibinfo {title} {{Neutron scattering
  studies of the magnetic structure of cupric oxide}},\ }\href
  {https://doi.org/10.1103/PhysRevB.38.174} {\bibfield  {journal} {\bibinfo
  {journal} {Phys. Rev. B}\ }\textbf {\bibinfo {volume} {38}},\ \bibinfo
  {pages} {174} (\bibinfo {year} {1988})}\BibitemShut {NoStop}%
\bibitem [{\citenamefont {Joenk}\ and\ \citenamefont {Bozorth}(1965)}]{CuF2-T}%
  \BibitemOpen
  \bibfield  {author} {\bibinfo {author} {\bibfnamefont {R.~J.}\ \bibnamefont
  {Joenk}}\ and\ \bibinfo {author} {\bibfnamefont {R.~M.}\ \bibnamefont
  {Bozorth}},\ }\bibfield  {title} {\bibinfo {title} {Magnetic properties of
  cuf2},\ }\href {https://doi.org/10.1063/1.1714152} {\bibfield  {journal}
  {\bibinfo  {journal} {Journal of Applied Physics}\ }\textbf {\bibinfo
  {volume} {36}},\ \bibinfo {pages} {1167} (\bibinfo {year}
  {1965})}\BibitemShut {NoStop}%
\end{thebibliography}%


\pagebreak
\onecolumngrid
\begin{center}

        \textbf{\large Supplemental Material for: Evaluating SCAN and r$^2$SCAN meta-GGA functionals for predicting transition temperatures in antiferromagnetic materials}\\[.2cm]
	Nafise Rezaei$^{1}$, Mojtaba Alaei$^{1,2}$, Artem R. Oganov$^{1}$  \\[.1cm]
	{\itshape $^{1}$Skolkovo Institute of Science and Technology, 121205, Bolshoy Boulevard 30, bld. 1, Moscow, Russia.}\\
	{\itshape $^{2}$Department of Physics, Isfahan University of Technology, Isfahan 84156-83111, Iran.}
\end{center}

\section*{Heisenberg exchange parameters}
{\Huge
\begin{longtable}{>{\centering\hspace{2pt}}m{0.1\linewidth}>
{\centering\hspace{0pt}}m{0.2\linewidth}>{\centering\arraybackslash\hspace{0pt}}m{0.2\linewidth}}
\caption{Heisenberg exchange parameters in SCAN and r$^2$SCAN functionals}
\\
\hline\hline
Compound                               &  $J_{\text{SCAN}}$ (meV)   &   $J_{\text{r}^2\text{SCAN}}$ (meV)  \\* 
\hline
\multirow{7}{*}{\Centering{}YVO$_3$}      & $J_1$=   -6.567 & $J_1$=   -3.219                    \\*
                                        & $J_2$=   -4.890 & $J_2$=   -5.903                    \\*
                                        & $J_3$=   -0.519 & $J_3$=   -2.543                    \\*
                                       & $J_4$=   -0.096 & $J_4$=   -0.464                    \\*
                                       & $J_5$=   -0.882 & $J_5$=   -0.919                    \\*
                                       & $J_6$=    0.032 & $J_6$=    0.350                    \\*
                                       & $J_7$=   -0.034 & $J_7$=   -0.699                    \\* 
\hline
\multirow{7}{*}{\Centering{}CrSb$_2$}  & $J_1$= -55.770   & $J_1$= -59.119                      \\*
                                       & $J_2$= 6.562     & $J_2$= 6.811                        \\*
                                       & $J_3$= -4.203    & $J_3$= -4.635                       \\*
                                       & $J_4$= 10.151    & $J_4$= 10.250                       \\*
                                       & $J_5$= -0.869    & $J_5$= -1.299                       \\*
                                       & $J_6$= -5.115    & $J_6$= -6.055                       \\*
                                       & $J_7$= 0.865     & $J_7$= 0.451                        \\* 
\hline
\multirow{6}{*}{\Centering{}CrCl$_2$}     & $J_1$=   -5.151 & $J_1$=   -6.796                    \\*
                                       & $J_2$=    0.080 & $J_2$=    1.391                    \\*
                                       & $J_3$=    0.063 & $J_3$=    0.191                    \\*
                                       & $J_4$=   -0.035 & $J_4$=   -0.013                    \\*
                                       & $J_5$=   -0.022 & $J_5$=    0.036                    \\*
                                       & $J_6$=   -0.070 & $J_6$=   -0.026                    \\* 
\hline
\multirow{11}{*}{\Centering{}CrF$_2$}     & $J_1$= 4.333     & $J_1$= 5.019                        \\*
                                       & $J_2$= -1.456    & $J_2$= 0.076                        \\*
                                       & $J_3$= 1.028     & $J_3$= 1.581                        \\*
                                       & $J_4$= 0.086     & $J_4$= 0.092                        \\*
                                       & $J_5$= -0.001    & $J_5$= 0.017                        \\*
                                       & $J_6$= -0.011    & $J_6$= -0.043                       \\*
                                       & $J_7$= -0.005    & $J_7$= 0.007                        \\*
                                       & $J_8$= -0.068    & $J_8$= 0.035                        \\*
                                       & $J_9$= 0.001     & $J_9$= 0.062                        \\*
                                       & $J_{10}$= 0.050    & $J_{10}$= 0.029                       \\*
                                       & $J_{11}$= 0.072    & $J_{11}$= 0.062                       \\* 
\hline
\multirow{9}{*}{\Centering{}Cr$_2$O$_3$}     & $J_1$=  -29.210 & $J_1$=  -29.765                    \\*
                                       & $J_2$=  -17.628 & $J_2$=  -18.370                    \\*
                                       & $J_3$=   -1.058 & $J_3$=    1.328                    \\*
                                       & $J_4$=    0.293 & $J_4$=    2.331                    \\*
                                       & $J_5$=    0.582 & $J_5$=    0.927                    \\*
                                       & $J_6$=   -0.339 & $J_6$=   -0.240                    \\*
                                       & $J_7$=    0.096 & $J_7$=    0.030                    \\*
                                       & $J_8$=   -0.180 & $J_8$=   -0.142                    \\*
                                       & $J_9$=   -1.404 & $J_9$=   -1.522                    \\* 
\hline
\multirow{8}{*}{\Centering{}CrSBr}     & $J_1$= 5.350     & $J_1$= 7.721                        \\*
                                       & $J_2$= 9.743     & $J_2$= 12.150                       \\*
                                       & $J_3$= 5.956     & $J_3$= 7.883                        \\*
                                       & $J_4$= 0.058     & $J_4$= 0.100                        \\*
                                       & $J_5$= -0.008    & $J_5$= 0.219                        \\*
                                       & $J_6$= -0.134    & $J_6$= -0.146                       \\*
                                       & $J_7$= -1.247    & $J_7$= -1.468                       \\*
                                       & $J_8$= 0.582     & $J_8$= 0.980                        \\* 
\hline
\multirow{5}{*}{\Centering{}Cr$_2$TeO$_6$}   & $J_1$=  -13.030 & $J_1$=  -13.362                    \\*
                                       & $J_2$=   -6.252 & $J_2$=   -5.752                    \\*
                                       & $J_3$=   -0.726 & $J_3$=   -0.716                    \\*
                                       & $J_4$=   -0.459 & $J_4$=   -0.442                    \\*
                                       & $J_5$=   -0.134 & $J_5$=   -0.097                    \\* 
\hline
\multirow{6}{*}{\Centering{}Cr$_2$WO$_6$}    & $J_1$=   -3.962 & $J_1$=   -3.123                    \\*
                                       & $J_2$=   -0.325 & $J_2$=    1.331                    \\*
                                       & $J_3$=   -2.084 & $J_3$=   -2.226                    \\*
                                       & $J_4$=   -2.859 & $J_4$=   -2.912                    \\*
                                       & $J_5$=   -0.594 & $J_5$=   -0.615                    \\*
                                       & $J_6$=   -0.557 & $J_6$=   -0.665                    \\* 
\hline
\multirow{4}{*}{\Centering{}MnO}       & $J_1$=   -7.851 & $J_1$=   -7.156                    \\*
                                       & $J_2$=  -10.127 & $J_2$=  -10.870                    \\*
                                       & $J_3$=   -0.085 & $J_3$=   -0.092                    \\*
                                       & $J_4$=   -0.622 & $J_4$=   -0.582                    \\* 
\hline
\multirow{4}{*}{\Centering{}MnS}       & $J_1$=   -2.002 & $J_1$=   -2.070                    \\*
                                       & $J_2$=  -13.788 & $J_2$=  -14.872                    \\*
                                       & $J_3$=   -0.481 & $J_3$=   -0.468                    \\*
                                       & $J_4$=   -1.835 & $J_4$=   -1.750                    \\* 
\hline
\multirow{4}{*}{\Centering{}MnSe}      & $J_1$=    0.011 & $J_1$=   -0.060                    \\*
                                       & $J_2$=  -13.308 & $J_2$=  -14.387                    \\*
                                       & $J_3$=   -0.352 & $J_3$=   -0.324                    \\*
                                       & $J_4$=   -2.635 & $J_4$=   -2.505                    \\* 
\hline
\multirow{7}{*}{\Centering{}MnTe}      & $J_1$= -44.203   & $J_1$= -34.403                      \\*
                                       & $J_2$= -4.147    & $J_2$= -1.936                       \\*
                                       & $J_3$= -5.252    & $J_3$= -7.484                       \\*
                                       & $J_4$= -2.123    & $J_4$= -2.421                       \\*
                                       & $J_5$= 0.452     & $J_5$= 0.758                        \\*
                                       & $J_6$= -1.086    & $J_6$= -1.116                       \\*
                                       & $J_7$= 0.231     & $J_7$= 0.472                        \\* 
\hline
\multirow{7}{*}{\Centering{}MnO$_2$}      & $J_1$= -5.462    & $J_1$= -5.080                       \\*
                                       & $J_2$= -8.179    & $J_2$= -7.095                       \\*
                                       & $J_3$= -1.521    & $J_3$= -1.118                       \\*
                                       & $J_4$= -0.747    & $J_4$= -0.881                       \\*
                                       & $J_5$= 0.001     & $J_5$= -0.097                       \\*
                                       & $J_6$= -0.209    & $J_6$= -0.445                       \\*
                                       & $J_7$= -1.200    & $J_7$= -1.302                       \\* 
\hline
\multirow{5}{*}{\Centering{}MnF$_2$}      & $J_1$=    0.338 & $J_1$=    0.341                    \\*
                                       & $J_2$=   -4.069 & $J_2$=   -4.033                    \\*
                                       & $J_3$=   -0.189 & $J_3$=   -0.201                    \\*
                                       & $J_4$=   -0.094 & $J_4$=   -0.086                    \\*
                                       & $J_5$=    0.046 & $J_5$=    0.043                    \\* 
\hline
\multirow{5}{*}{\Centering{}MnS$_2$}      & $J_1$= -12.347   & $J_1$= -12.469                      \\*
                                       & $J_2$= -1.860    & $J_2$= -1.925                       \\*
                                       & $J_3$= -0.260    & $J_3$= -0.264                       \\*
                                       & $J_4$= -0.409    & $J_4$= -0.409                       \\*
                                       & $J_5$= -0.253    & $J_5$= -0.258                       \\* 
\hline
\multirow{5}{*}{\Centering{}MnTe$_2$}     & $J_1$= -10.750   & $J_1$= -11.469                      \\*
                                       & $J_2$= -0.823    & $J_2$= -0.626                       \\*
                                       & $J_3$= -0.973    & $J_3$= -0.943                       \\*
                                       & $J_4$= -0.493    & $J_4$= -0.478                       \\*
                                       & $J_5$= -0.462    & $J_5$= -0.417                       \\* 
\hline
\multirow{6}{*}{\Centering{}LiMnO$_2$}    & $J_1$=  -31.396 & $J_1$=  -32.556                    \\*
                                       & $J_2$=   -1.688 & $J_2$=   -1.221                    \\*
                                       & $J_3$=  -31.198 & $J_3$=  -30.534                    \\*
                                       & $J_4$=    0.308 & $J_4$=    0.251                    \\*
                                       & $J_5$=   -1.293 & $J_5$=   -1.065                    \\*
                                       & $J_6$=   -1.065 & $J_6$=   -0.923                    \\* 
\hline
\multirow{11}{*}{\Centering{}SrMnO$_3$}   & $J_1$= -18.370   & $J_1$= -20.107                      \\*
                                       & $J_2$= -27.321   & $J_2$= -28.244                      \\*
                                       & $J_3$= 0.933     & $J_3$= 1.036                        \\*
                                       & $J_4$= -1.103    & $J_4$= -1.123                       \\*
                                       & $J_5$= -1.004    & $J_5$= -1.015                       \\*
                                       & $J_6$= -0.371    & $J_6$= -0.240                       \\*
                                       & $J_7$= -0.446    & $J_7$= -0.341                       \\*
                                       & $J_8$= -0.459    & $J_8$= -0.378                       \\*
                                       & $J_9$= -0.182    & $J_9$= -0.109                       \\*
                                       & $J_{10}$= -0.070   & $J_{10}$= -0.024                      \\*
                                       & $J_{11}$= 0.010    & $J_{11}$= 0.040                       \\* 
\hline
\multirow{9}{*}{\Centering{}KMnF$_3$}     & $J_1$= -8.263    & $J_1$= -8.320                       \\*
                                       & $J_2$= -8.413    & $J_2$= -8.489                       \\*
                                       & $J_3$= -0.137    & $J_3$= -0.124                       \\*
                                       & $J_4$= -0.179    & $J_4$= -0.165                       \\*
                                       & $J_5$= -0.001    & $J_5$= -0.004                       \\*
                                       & $J_6$= -0.010    & $J_6$= -0.010                       \\*
                                       & $J_7$= -0.019    & $J_7$= -0.018                       \\*
                                       & $J_8$= -0.006    & $J_8$= -0.004                       \\*
                                       & $J_9$= -0.015    & $J_9$= -0.017                       \\* 
\hline
\multirow{8}{*}{\Centering{}MnPS$_3$}     & $J_1$= -12.390   & $J_1$= -11.978                      \\*
                                       & $J_2$= -12.242   & $J_2$= -11.836                      \\*
                                       & $J_3$= -0.966    & $J_3$= -0.936                       \\*
                                       & $J_4$= -0.542    & $J_4$= -0.513                       \\*
                                       & $J_5$= -5.981    & $J_5$= -6.026                       \\*
                                       & $J_6$= -5.580    & $J_6$= -5.609                       \\*
                                       & $J_7$= -0.155    & $J_7$= -0.158                       \\*
                                       & $J_8$= -0.213    & $J_8$= -0.206                       \\* 
\hline
\multirow{6}{*}{\Centering{}MnPSe$_3$}    & $J_1$= -8.873    & $J_1$= -8.554                       \\*
                                       & $J_2$= -0.771    & $J_2$= -0.731                       \\*
                                       & $J_3$= -0.243    & $J_3$= -0.224                       \\*
                                       & $J_4$= -4.880    & $J_4$= -5.116                       \\*
                                       & $J_5$= -0.205    & $J_5$= -0.194                       \\*
                                       & $J_6$= -0.546    & $J_6$= -0.551                       \\* 
\hline
\multirow{9}{*}{\Centering{}MnWO$_4$}  & $J_1$=  -10.331 & $J_1$=   -9.682                    \\*
                                       & $J_2$=   -0.245 & $J_2$=   -0.198                    \\*
                                       & $J_3$=   -3.632 & $J_3$=   -3.328                    \\*
                                       & $J_4$=   -2.139 & $J_4$=   -2.064                    \\*
                                       & $J_5$=   -4.531 & $J_5$=   -4.592                    \\*
                                       & $J_6$=   -2.989 & $J_6$=   -3.072                    \\*
                                       & $J_7$=   -0.905 & $J_7$=   -1.009                    \\*
                                       & $J_8$=   -0.112 & $J_8$=   -0.136                    \\*
                                       & $J_9$=   -0.840 & $J_9$=   -0.802                    \\* 
\hline
\multirow{9}{*}{\Centering{}Li$_2$MnO$_3$}   & $J_1$=   -0.518 & $J_1$=   -0.311                    \\*
                                       & $J_2$=    0.0292 & $J_2$=    0.034                    \\*
                                       & $J_3$=    0.345 & $J_3$=    0.562                    \\*
                                       & $J_4$=    0.361 & $J_4$=    0.588                   \\*
                                       & $J_5$=   -0.058 & $J_5$=    0.006                    \\*
                                       & $J_6$=   -0.061 & $J_6$=   -0.078                    \\*
                                       & $J_7$=   -0.902 & $J_7$=   -0.627                    \\*
                                       & $J_8$=   -0.075 & $J_8$=   -0.392                    \\*
                                       & $J_9$=   -0.157 & $J_9$=   -0.183                    \\* 
\hline
\multirow{6}{*}{\Centering{}LiMnPO$_4$} & $J_1$=   -3.636 & $J_1$=   -3.641                    \\*
                                       & $J_2$=   -0.115 & $J_2$=   -0.139                    \\*
                                       & $J_3$=   -1.541 & $J_3$=   -1.549                    \\*
                                       & $J_4$=   -0.091 & $J_4$=   -0.074                    \\*
                                       & $J_5$=   -0.139 & $J_5$=   -0.133                    \\*
                                       & $J_6$=   -0.599 & $J_6$=   -0.634                    \\* 
\hline
\multirow{9}{*}{\Centering{}Fe$_2$O$_3$}     & $J_1$= -4.795    & $J_1$= -4.190                       \\*
                                       & $J_2$= -9.239    & $J_2$= -8.779                       \\*
                                       & $J_3$= -102.946  & $J_3$= -104.406                     \\*
                                       & $J_4$= -27.956   & $J_4$= -28.102                      \\*
                                       & $J_5$= 6.362     & $J_5$= 6.003                        \\*
                                       & $J_6$= -3.569    & $J_6$= -3.105                       \\*
                                       & $J_7$= 1.313     & $J_7$= 0.947                        \\*
                                       & $J_8$= -2.560    & $J_8$= -2.653                       \\*
                                       & $J_9$= -5.144    & $J_9$= -5.249                       \\* 
\hline
\multirow{8}{*}{\Centering{}SrFeO$_2$}    & $J_1$= -5.913    & $J_1$= -6.686                       \\*
                                       & $J_2$= -40.632   & $J_2$= -41.214                      \\*
                                       & $J_3$= 2.301     & $J_3$= 2.365                        \\*
                                       & $J_4$= -5.108    & $J_4$= -4.801                       \\*
                                       & $J_5$= 0.139     & $J_5$= 0.137                        \\*
                                       & $J_6$= 5.070     & $J_6$= 5.431                        \\*
                                       & $J_7$= -2.772    & $J_7$= -2.824                       \\*
                                       & $J_8$= 0.321     & $J_8$= 0.478                        \\* 
\hline
\multirow{4}{*}{\Centering{}BiFeO$_3$}    & $J_1$=  -50.324 & $J_1$=  -50.932                    \\*
                                       & $J_2$=   -5.105 & $J_2$=   -5.131                    \\*
                                       & $J_3$=    1.447 & $J_3$=    1.512                    \\*
                                       & $J_4$=   -0.435 & $J_4$=   -0.430                    \\* 
\hline
\multirow{6}{*}{\Centering{}LaFeO$_3$}    & $J_1$= -53.042   & $J_1$= -53.850                      \\*
                                       & $J_2$= -55.861   & $J_2$= -56.636                      \\*
                                       & $J_3$= -1.550    & $J_3$= -1.529                       \\*
                                       & $J_4$= -6.604    & $J_4$= -6.650                       \\*
                                       & $J_5$= -0.123    & $J_5$= -0.119                       \\*
                                       & $J_6$= -0.456    & $J_6$= -0.442                       \\* 
\hline
\multirow{7}{*}{\Centering{}YFeO$_3$}  & $J_1$=  -46.628 & $J_1$=  -47.157                    \\*
                                       & $J_2$=  -48.972 & $J_2$=  -49.491                    \\*
                                       & $J_3$=   -4.484 & $J_3$=   -4.448                    \\*
                                       & $J_4$=   -1.545 & $J_4$=   -1.531                    \\*
                                       & $J_5$=   -2.784 & $J_5$=   -2.716                    \\*
                                       & $J_6$=    0.056 & $J_6$=    0.047                    \\*
                                       & $J_7$=   -0.046 & $J_7$=   -0.074                    \\* 
\hline
\multirow{8}{*}{\Centering{}FePS$_3$}     & $J_1$= 8.506     & $J_1$= 12.323                       \\*
                                       & $J_2$= -11.313   & $J_2$= -10.795                      \\*
                                       & $J_3$= -2.981    & $J_3$= -2.793                       \\*
                                       & $J_4$= -2.274    & $J_4$= -1.718                       \\*
                                       & $J_5$= -2.327    & $J_5$= -1.738                       \\*
                                       & $J_6$= -0.763    & $J_6$= -0.734                       \\*
                                       & $J_7$= -10.008   & $J_7$= -9.343                       \\*
                                       & $J_8$= -11.087   & $J_8$= -11.464                      \\* 
\hline
\multirow{6}{*}{\Centering{}Fe$_2$TeO$_6$}   & $J_1$= -14.857   & $J_1$= -14.861                      \\*
                                       & $J_2$= -28.964   & $J_2$= -29.393                      \\*
                                       & $J_3$= -8.553    & $J_3$= -8.546                       \\*
                                       & $J_4$= -4.094    & $J_4$= -4.015                       \\*
                                       & $J_5$= -2.108    & $J_5$= -2.081                       \\*
                                       & $J_6$= -2.399    & $J_6$= -2.533                       \\* 
\hline
\multirow{15}{*}{\Centering{}SrFe$_2$S$_2$O} & $J_1$= -20.005   & $J_1$= -13.473                      \\*
                                       & $J_2$= -72.220   & $J_2$= -74.900                      \\*
                                       & $J_3$= -18.405   & $J_3$= -16.191                      \\*
                                       & $J_4$= 4.476     & $J_4$= 3.973                        \\*
                                       & $J_5$= 4.142     & $J_5$= 6.547                        \\*
                                       & $J_6$= 2.843     & $J_6$= -1.147                       \\*
                                       & $J_7$= -11.299   & $J_7$= -10.957                      \\*
                                       & $J_8$= -1.721    & $J_8$= -3.748                       \\*
                                       & $J_9$= 1.243     & $J_9$= -0.489                       \\*
                                       & $J_{10}$= -2.510   & $J_{10}$= -3.961                      \\*
                                       & $J_{11}$= -1.913   & $J_{11}$= -0.007                      \\*
                                       & $J_{12}$= 1.461    & $J_{12}$= 3.124                       \\*
                                       & $J_{13}$= -2.408   & $J_{13}$= -0.867                      \\*
                                       & $J_{14}$= 0.247    & $J_{14}$= 0.773                       \\*
                                       & $J_{15}$= 2.335    & $J_{15}$= 1.293                       \\* 
\hline
\multirow{10}{*}{\Centering{}CoWO$_4$}    & $J_1$= 2.180     & $J_1$= 2.827                        \\*
                                       & $J_2$= -0.274    & $J_2$= -0.109                       \\*
                                       & $J_3$= 0.077     & $J_3$= 0.186                        \\*
                                       & $J_4$= -1.100    & $J_4$= -0.898                       \\*
                                       & $J_5$= -1.039    & $J_5$= -0.667                       \\*
                                       & $J_6$= 0.060     & $J_6$= 0.078                        \\*
                                       & $J_7$= -1.107    & $J_7$= -1.136                       \\*
                                       & $J_8$= -0.380    & $J_8$= -0.484                       \\*
                                       & $J_9$= -0.150    & $J_9$= -0.223                       \\*
                                       & $J_{10}$= 0.040    & $J_{10}$= -0.063                      \\* 
\hline
\multirow{4}{*}{\Centering{}NiO}       & $J_1$=    2.290 & $J_1$=    2.611                    \\*
                                       & $J_2$=  -28.296 & $J_2$=  -31.200                    \\*
                                       & $J_3$=   -0.034 & $J_3$=    0.084                    \\*
                                       & $J_4$=   -1.098 & $J_4$=   -1.387                    \\* 
\hline
\multirow{5}{*}{\Centering{}NiF$_2$}      & $J_1$=   -0.797 & $J_1$=   -0.708                    \\*
                                       & $J_2$=   -3.842 & $J_2$=   -3.888                    \\*
                                       & $J_3$=   -0.675 & $J_3$=   -0.597                    \\*
                                       & $J_4$=    0.0118 & $J_4$=   -0.052                    \\*
                                       & $J_5$=    0.018 & $J_5$=    0.008                    \\* 
\hline
\multirow{6}{*}{\Centering{}NiBr$_2$}     & $J_1$=    4.290 & $J_1$=    4.923                    \\*
                                       & $J_2$=    0.053 & $J_2$=    0.059                    \\*
                                       & $J_3$=    0.027 & $J_3$=    0.063                    \\*
                                       & $J_4$=   -1.273 & $J_4$=   -1.489                    \\*
                                       & $J_5$=   -0.530 & $J_5$=   -0.646                    \\*
                                       & $J_6$=   -0.050 & $J_6$=   -0.064                    \\* 
\hline
\multirow{5}{*}{\Centering{}NiS$_2$}      & $J_1$= -6.569    & $J_1$= -8.763                       \\*
                                       & $J_2$= -2.386    & $J_2$= -3.507                       \\*
                                       & $J_3$= -0.420    & $J_3$= -0.557                       \\*
                                       & $J_4$= -0.604    & $J_4$= -1.325                       \\*
                                       & $J_5$= -0.127    & $J_5$= -0.182                       \\* 
\hline
\multirow{8}{*}{\Centering{}NiCl$_2$}     & $J_1$= 4.250     & $J_1$= 4.786                        \\*
                                       & $J_2$= 0.055     & $J_2$= 0.080                        \\*
                                       & $J_3$= 0.025     & $J_3$= 0.020                        \\*
                                       & $J_4$= -0.836    & $J_4$= -0.952                       \\*
                                       & $J_5$= -0.309    & $J_5$= -0.342                       \\*
                                       & $J_6$= -0.029    & $J_6$= -0.029                       \\*
                                       & $J_7$= 0.014     & $J_7$= 0.017                        \\*
                                       & $J_8$= 0.009     & $J_8$= -0.007                       \\* 
\hline
\multirow{8}{*}{\Centering{}NiPS$_3$}     & $J_1$= 1.619     & $J_1$= 3.141                        \\*
                                       & $J_2$= 2.011     & $J_2$= 3.480                        \\*
                                       & $J_3$= 2.170     & $J_3$= 2.841                        \\*
                                       & $J_4$= -0.243    & $J_4$= -0.552                       \\*
                                       & $J_5$= -0.251    & $J_5$= -0.334                       \\*
                                       & $J_6$= -0.049    & $J_6$= -0.026                       \\*
                                       & $J_7$= -18.312   & $J_7$= -22.235                      \\*
                                       & $J_8$= -16.347   & $J_8$= -19.880                      \\* 
\hline
\multirow{10}{*}{\Centering{}NiPSe$_3$}   & $J_1$= 4.376     & $J_1$= 6.892                        \\*
                                       & $J_2$= 3.917     & $J_2$= 6.280                        \\*
                                       & $J_3$= 0.741     & $J_3$= 0.467                        \\*
                                       & $J_4$= -0.497    & $J_4$= -1.033                       \\*
                                       & $J_5$= -0.223    & $J_5$= -0.375                       \\*
                                       & $J_6$= -0.581    & $J_6$= -0.832                       \\*
                                       & $J_7$= -18.774   & $J_7$= -22.873                      \\*
                                       & $J_8$= -20.046   & $J_8$= -24.377                      \\*
                                       & $J_9$= -1.558    & $J_9$= -2.005                       \\*
                                       & $J_{10}$= -2.402   & $J_{10}$= -2.923                      \\* 
\hline
\multirow{6}{*}{\Centering{}KNiF$_3$}     & $J_1$= -18.126   & $J_1$= -18.724                      \\*
                                       & $J_2$= -0.205    & $J_2$= -0.226                       \\*
                                       & $J_3$= -0.038    & $J_3$= -0.037                       \\*
                                       & $J_4$= -0.284    & $J_4$= -0.187                       \\*
                                       & $J_5$= -0.083    & $J_5$= -0.040                       \\*
                                       & $J_6$= 0.044     & $J_6$= 0.014                        \\* 
\hline
\multirow{9}{*}{\Centering{}NiWO$_4$}     & $J_1$=    3.166 & $J_1$=    3.464                    \\*
                                       & $J_2$=   -0.480 & $J_2$=   -0.497                    \\*
                                       & $J_3$=   -1.832 & $J_3$=   -1.778                    \\*
                                       & $J_4$=   -1.897 & $J_4$=   -1.989                    \\*
                                       & $J_5$=   -2.301 & $J_5$=   -2.487                    \\*
                                       & $J_6$=   -0.069 & $J_6$=    0.019                    \\*
                                       & $J_7$=   -0.206 & $J_7$=   -0.175                    \\*
                                       & $J_8$=    0.507 & $J_8$=    0.547                    \\*
                                       & $J_9$=   -3.972 & $J_9$=   -4.335                    \\* 
\hline
\multirow{3}{*}{\Centering{}La$_2$NiO$_4$}   & $J_1$=  -44.296 & $J_1$= -50.227                      \\*
                                       & $J_2$=   -3.546 & $J_2$= -4.181                       \\*
                                       & $J_3$=   -0.017 & $J_3$= -0.007                       \\* 
\hline
\multirow{6}{*}{\Centering{}K$_2$NiF$_4$}    & $J_1$= -18.762   & $J_1$= -19.397                      \\*
                                       & $J_2$= -0.239    & $J_2$= -0.227                       \\*
                                       & $J_3$= -0.016    & $J_3$= -0.016                       \\*
                                       & $J_4$= -0.197    & $J_4$= -0.188                       \\*
                                       & $J_5$= -0.024    & $J_5$= -0.005                       \\*
                                       & $J_6$= 0.002     & $J_6$= -0.006                       \\* 
\hline
\multirow{4}{*}{\Centering{}KNiPO$_4$}    & $J_1$=   -3.270 & $J_1$=   -3.372                    \\*
                                       & $J_2$=    0.383 & $J_2$=    0.425                    \\*
                                       & $J_3$=    0.160 & $J_3$=    0.183                    \\*
                                       & $J_4$=   -0.770 & $J_4$=   -0.837                    \\* 
\hline
\multirow{8}{*}{\Centering{}LiNiPO$_4$}   & $J_1$=   -2.331 & $J_1$=   -2.321                    \\*
                                       & $J_2$=    0.138 & $J_2$=    0.159                    \\*
                                       & $J_3$=   -1.805 & $J_3$=   -1.968                    \\*
                                       & $J_4$=    0.025 & $J_4$=    0.019                    \\*
                                       & $J_5$=    0.268 & $J_5$=    0.281                    \\*
                                       & $J_6$=   -1.849 & $J_6$=   -1.998                    \\*
                                       & $J_7$=   -0.034 & $J_7$=   -0.033                    \\*
                                       & $J_8$=    0.001 & $J_8$=    0.013                    \\* 
\hline
\multirow{13}{*}{\Centering{}CuO}      & $J_1$= -0.328    & $J_1$= -0.632                       \\*
                                       & $J_2$= 1.017     & $J_2$= 0.892                        \\*
                                       & $J_3$= 1.830     & $J_3$= 2.223                        \\*
                                       & $J_4$= -0.987    & $J_4$= -1.859                       \\*
                                       & $J_5$= -32.971   & $J_5$= -37.712                      \\*
                                       & $J_6$= 1.216     & $J_6$= 0.806                        \\*
                                       & $J_7$= -0.070    & $J_7$= -0.232                       \\*
                                       & $J_8$= 0.199     & $J_8$= -0.019                       \\*
                                       & $J_9$= -6.733    & $J_9$= -7.792                       \\*
                                       & $J_{10}$= 0.420    & $J_{10}$= 0.259                       \\*
                                       & $J_{11}$= -0.372   & $J_{11}$= -0.468                      \\*
                                       & $J_{12}$= -0.398   & $J_{12}$= -0.159                      \\*
                                       & $J_{13}$= -0.155   & $J_{13}$= -0.332                      \\* 
\hline
\multirow{10}{*}{\Centering{}CuF$_2$}     & $J_1$= 1.065     & $J_1$= 1.511                        \\*
                                       & $J_2$= -6.053    & $J_2$= -6.725                       \\*
                                       & $J_3$= 1.046     & $J_3$= 1.506                        \\*
                                       & $J_4$= 0.061     & $J_4$= 0.111                        \\*
                                       & $J_5$= 0.015     & $J_5$= -0.008                       \\*
                                       & $J_6$= -0.328    & $J_6$= -0.346                       \\*
                                       & $J_7$= -0.432    & $J_7$= -0.537                       \\*
                                       & $J_8$= 0.046     & $J_8$= 0.046                        \\*
                                       & $J_9$= -0.094    & $J_9$= -0.098                       \\*
                                       & $J_{10}$= -0.010   & $J_{10}$= 0.019  
                                                 \\* 
\hline
\hline
\end{longtable}

}

\end{document}